\documentclass[prd,preprint,superscriptaddress,amsmath,amssymb,nofootinbib]{revtex4}

\usepackage{slashed,bbold,bbm}
\usepackage{soul}

\usepackage{graphicx,color}
\usepackage{hyperref}

\newcommand{\oovt}{\frac{1}{3}}

\begin{document}

\title{Phenomenological study of a gauged ${L_\mu -L_\tau}$ model with a scalar leptoquark}

\author{Chuan-Hung Chen}
\email[E-mail: ]{physchen@mail.ncku.edu.tw}
\affiliation{Department of Physics, National Cheng-Kung University, Tainan 70101, Taiwan}
\affiliation{Physics Division, National Center for Theoretical Sciences, Taipei 10617, Taiwan}

\author{Cheng-Wei Chiang}
\email[E-mail: ]{chengwei@phys.ntu.edu.tw}
\affiliation{Department of Physics and Center for Theoretical Physics, National Taiwan University, Taipei 10617, Taiwan}
\affiliation{Physics Division, National Center for Theoretical Sciences, Taipei 10617, Taiwan}

\author{Chun-Wei Su}
\email[E-mail: ]{r10222026@ntu.edu.tw}
\affiliation{Department of Physics and Center for Theoretical Physics, National Taiwan University, Taipei 10617, Taiwan}

\date{\today}

\begin{abstract}

A $Z'$ gauge boson with sub-GeV mass has acquired a significant interest in phenomenology, particularly in view of the muon $g-2$ anomaly and coherent elastic neutrino-nucleon scattering. The latter is challenged by the nuclear recoil energy of a few tens of keV but has been observed by the COHERENT experiment. To further reconcile the observed excesses in $R(D^{(*)})$ from semileptonic charmful $B$ decays and in the $W$ boson mass, we investigate a model with a gauged $U(1)_{L_\mu-L_\tau}$ symmetry and a scalar leptoquark.  In contrast to the mechanism that involves kinetic mixing between the gauge bosons of $U(1)_{\rm em}$ and $U(1)_{L_\mu-L_\tau}$, we adopt a dynamical symmetry breaking of $U(1)_{L_\mu-L_\tau}$ by incorporating an additional Higgs doublet.  Through mixing with the $U(1)_{L_\mu-L_\tau}$-charged Higgs doublet, new Higgs decay channels $h\to Z_1 Z_1/Z_1 Z_2$ occur at percent-level branching ratios, which could be accessible at the LHC. The $W$-mass anomaly observed by CDF II can be potentially resolved through the enhancement in the oblique parameter $T$. Due to the flavored gauge symmetry, the introduced scalar leptoquark $S^{\frac{1}{3}}=(\bar{3},1,2/3)$ exhibits a unique coupling to the $\tau$-lepton, offering an explanation for the excesses observed in $R(D^{(*)})$. Moreover, $\tau \to \mu (Z_1\to ) e^- e^+$ via the resonant light gauge boson decay can reach the sensitivity of Belle II at an integrated luminosity of 50 ab$^{-1}$.

\end{abstract}
\maketitle

\newpage

\section{Introduction}

The possible existence of a sub-GeV $Z'$ gauge boson has attracted much attention in recent years in addressing unresolved problems in particle physics phenomenology, particularly in flavor physics and dark matter (DM).  The former includes the long-lasting puzzle in the anomalous magnetic dipole moment of muon (muon $g-2$), while the latter could lead to establishing a portal between the visible and dark sectors.

Furthermore, a light $Z'$ can play a substantial role in coherent elastic neutrino-nucleon scattering (CE$\nu$NS). Since the proposal of measuring CE$\nu$NS~\cite{Freedman:1973yd}, conducting such experiments has been quite challenging, not just because of its tiny cross section but mostly due to the fact that the maximum nuclear recoil energy would be only several tens of keV.  
Nevertheless, the CE$\nu$NS has finally been observed by the COHERENT experiment using CsI and Ar targets~\cite{COHERENT:2017ipa,COHERENT:2021xmm,COHERENT:2020iec}. The obtained total cross sections averaged over neutrino fluxes are
\begin{align}
\langle \sigma \rangle_\phi &= (16.5 ^{+3.0}_{-2.5}) \times 10^{-39}\, {\rm cm^2}~~[{\rm CsI}]\,, 
\nonumber \\
\langle \sigma \rangle_\phi &= (2.2\pm 0.7 ) \times 10^{-39}\, {\rm cm^2}~[{\rm Ar}]\,.
\end{align}
The standard model (SM) predictions are $18.9\times 10^{-39}$ cm$^2$~\cite{COHERENT:2021xmm} and $1.8 \times 10^{-39}$ cm$^2$~\cite{COHERENT:2020iec}, respectively. 
Additionally, besides improving our understanding of atomic nuclei and neutrinos, precision measurements of CE$\nu$NS can be used to explore or constrain physics beyond the SM~\cite{Coloma:2017ncl,Liao:2017uzy,Giunti:2019xpr,Coloma:2019mbs,Denton:2020hop,Khan:2021wzy,Hoferichter:2020osn,Liao:2022hno,Abdullah:2022zue,Calabrese:2022mnp}. Since the momentum transfer to the nucleus is at the sub-MeV level, a light $Z'$ gauge boson stands out as one of the most appealing extensions of the SM~\cite{Papoulias:2017qdn,Abdullah:2018ykz,Denton:2018xmq,CONNIE:2019xid,Miranda:2020tif,Cadeddu:2020nbr,Coloma:2020gfv,delaVega:2021wpx,CONUS:2021dwh,Coloma:2022avw,AtzoriCorona:2022moj}.

Meanwhile, several deviations from the SM predictions have emerged in experiments, such as the muon $g-2$, $R(D^{(*)})$ in the $B\to D^{(*)} \tau \bar\nu$ decays, and the mass of $W$ gauge boson. It would be intriguing to build a light $Z'$ model that can not only resolve these observed anomalies but also have a significant impact on the CE$\nu$NS phenomenon.  To fulfill these objectives within one coherent framework, we consider extending the SM with a new local $U(1)$ gauge symmetry. Numerous potential candidates of such a $U(1)$ gauge symmetry exist in the literature, including universal $U(1)$, $B-L$, $B_y+L_\mu+L_\tau$, $B-3 L_{\ell}$, $B-L_e-2L_\mu$, and $L_{\ell-\ell'}$~\cite{AtzoriCorona:2022moj}, where $B_i$ and $L_\ell$ denote the quantum numbers of quark and lepton, respectively.  Among these, we find that in addition to satisfying gauge anomaly-free conditions without introducing new chiral fermions, the gauged $U(1)_{L_\mu-L_\tau}\equiv U(1)_{\mu-\tau}$ model can effectively address the above-mentioned concerns.

Having $U(1)_{\mu-\tau}$ symmetry as a gauge extension of the SM has many advantages from a phenomenological viewpoint~\cite{He:1990pn,He:1991qd}.  As mentioned earlier, the gauge coupling $g_{Z'}$ of ${\cal O}(10^{-4})$ with $Z'$ mass of ${\cal O}(10-200)$~MeV can explain muon $g-2$~\cite{Chen:2017cic}, where the discrepancy between experimental measurements and theoretical calculations, which use the data-driven approach to evaluate the hadronic vacuum polarization, reaches $\sim 4\sigma$~\cite{Aoyama:2020ynm}:
\begin{equation}
\Delta a_\mu=a^{\rm exp}_\mu-a^{\rm SM}_\mu = (2.51\pm 0.59)\times 10^{-9}\,. \label{eq:Damu}
\end{equation}

Instead of kinetic mixing between $U(1)_{\rm em}$ and $U(1)_{\mu-\tau}$~\cite{Abdullah:2018ykz,Cadeddu:2020nbr,AtzoriCorona:2022moj}, we examine the $Z'-Z$ mass mixing scenario, which arises from the consideration of spontaneous $U(1)_{\mu-\tau}$ symmetry breakdown by a new scalar field carrying the $U(1)_{\mu-\tau}$ charge~\cite{Heeck:2011wj,Chen:2017cic}.  It would be useful if the new scalar field could also resolve any potential anomalies from a phenomenological perspective.  Interestingly,  the CDF~II Collaboration used the full dataset from proton-antiproton collisions at $\sqrt{s}=1.96$~TeV to determine the mass of the $W$ boson as~\cite{CDF:2022hxs}:
  \begin{equation}
  m_W = 80.4335\pm 0.0094~\text{GeV}\,.
  \end{equation}
The newly observed value differs from earlier measurements of $m_W=80.385\pm 0.015$~GeV from the combined results of  LEP and Tevatron~\cite{CDF:2013dpa} and $m_W=80.360\pm 0.016$~GeV from the updated ATLAS result~\cite{ATLAS:2023fsi}. Moreover, it deviates from the SM prediction of $m_W=80.361$~GeV~\cite{Heinemeyer:2013dia} by $\sim 7\sigma$.  If the anomaly in the $W$-mass measurement is confirmed with more data from the LHC, it would provide another piece of strong evidence for new physics.~\cite{Fan:2022dck,Strumia:2022qkt,Bagnaschi:2022whn,Bahl:2022xzi,Cheng:2022jyi,Asadi:2022xiy,Heckman:2022the,Crivellin:2022fdf,FileviezPerez:2022lxp,Kanemura:2022ahw,Kim:2022hvh,Li:2022gwc,Dcruz:2022dao,Chowdhury:2022dps,Gao:2022wxk,Han:2022juu,Cheng:2022hbo,Bandyopadhyay:2022bgx,Chen:2023eof}.  This anomaly also motivates the introduction of a Higgs doublet charged under the $U(1)_{\mu-\tau}$ symmetry~\cite{Heeck:2014qea,Heeck:2016xkh}.

The observed anomalies in the ratio of branching ratios (BRs) in semileptonic charmed $B$ decays are defined by 
   \begin{equation}
   R(M)= \frac{BR(B\to M \tau \bar\nu)}{ BR(B\to M \ell \bar\nu)}\,,
   \end{equation}
with $M=D, D^{*}$. The SM predictions are $R(D) = 0.298\pm 0.004$ and $R(D^*) = 0.254\pm 0.005$~\cite{MILC:2015uhg,Na:2015kha,Bigi:2016mdz,Bernlochner:2017jka,Jaiswal:2017rve,BaBar:2019vpl,Bordone:2019vic,Martinelli:2021onb} while the current experimental values are $R(D)=0.356\pm 0.029$ and $R(D^*)=0.284\pm 0.013$~\cite{HeavyFlavorAveragingGroup:2022wzx}.  Recent measurements from LHCb have been included in the average~\cite{LHCb:2023zxo,LHCb:2023cjr}.  As seen, there is an overall 3.3$\sigma$ deviation from the SM predictions~\cite{HeavyFlavorAveragingGroup:2022wzx}.  Because $B\to M \tau \bar\nu$ is mediated by the tree-level charged weak currents in the SM, the required mechanism to enhance $R(M)$ should have non-universal lepton couplings and be induced at the tree level.   Although $R(J/\Psi)$ and $R(\Lambda_c)$ have the potential to observe the breakdown of lepton universality as well, their statistical errors in the experimental data are still too large to be conclusive~\cite{LHCb:2017vlu,LHCb:2022piu,Fedele:2022iib}.  Hence, we concentrate solely on $R(D)$ and $R(D^*)$ in this work.

Without further introducing a heavy charged gauge boson (e.g., $W'$) or vector leptoquark (LQ) for the $R(D^{(*)})$ anomalies, the new mediating bosons of interest include a charged Higgs boson~\cite{Crivellin:2012ye,Crivellin:2013wna,Crivellin:2015hha,Chen:2017eby,Akeroyd:2017mhr,Chen:2018hqy} and a scalar LQ~\cite{Becirevic:2016yqi,Bhattacharya:2016mcc,Crivellin:2016ejn,Crivellin:2017zlb,Chen:2017hir,Chen:2017usq,Crivellin:2019dwb,Heeck:2022znj}. The flavored $U(1)_{\mu-\tau}$ symmetry strictly limits the Yukawa couplings to different lepton flavors, resulting in a suppressed contribution of the charged Higgs to $R(M)$ by $m_b m_\tau/v^2$ in this model, where $v$ is the combined vacuum expectation value (VEV) of the introduced Higgs doublets. Hence, the introduction of scalar LQ emerges as a more apposite solution. We find that among various LQ representations, the simplest choice to explain the observed excess in $R(M)$ is the $S^{\frac{1}{3}}=(\bar 3, 1, 2/3)$ representation under the $SU(3)_C\times SU(2)_L\times U(1)_Y$ gauge symmetries.  Additionally, based on the flavored $U(1)_{\mu-\tau}$ gauge symmetry, there is a natural suppression in the LQ Yukawa couplings to the light leptons, while the $\tau$-lepton and $\tau$-neutrino respectively couple to up- and down-type quarks to resolve the observed anomaly in $R(M)$. It is worth mentioning that using the exclusive- and hadronic-tag approaches with 362 fb$^{-1}$ of data, the Belle II Collaboration recently has observed the first evidence of $B^+\to K^+ \nu\bar\nu$ decay with a $2.7\sigma$ deviation from the SM prediction, and the measured result is given as~${\cal B}(B^+ \to K^+ \nu \bar\nu) =( 2.3 \pm 0.5^{+0.5}_{-0.4})\times 10^{-5}$~\cite{Belle-II:2023esi}. Applying the LQ $S^{\frac{1}{3}}$ in the model,  the branching ratios for $B\to (K, K^*) \nu_\tau \bar\nu_\tau$ can be significantly enhanced. A detailed phenomenological analysis of the neutrino-pair production in $B$ and $K$ meson decays can be found in Ref.~\cite{Chen:2023wpb}.

In addition to the total cross section of CE$\nu$NS and $R(D^{(*)})$, we propose new observables sensitive to new physics as a function of incident neutrino energy for elastic neutrino-nucleus scattering and as a function of invariant mass-square $q^2$ of $\ell \nu$ for semileptonic charmed $B$ decays.  We find that CE$\nu$NS mediated by the light physical $Z_1$ can deviate significantly from the SM in the low neutrino energy regime.  Additionally, $R(D)$ in the large $q^2$ regime is more sensitive to the leptoquark effects and can significantly differ from the SM.

This paper is organized as follows: In Sec.~\ref{sec:model}, we formulate the model and derive the spectrum of scalar bosons and various new couplings.  The $Z'-Z$ mixing and lepton flavor mixing are also discussed in detail.  With the new interactions, Sec.~\ref{sec:pheno} discusses the new physics effects on various phenomena, including the cross section of CE$\nu$NS, values of $R(D)$ and $R(D^*)$, new Higgs decay channels $h\to HH/Z_1 Z_1/Z_1 Z_2$, LFV processes, lepton $g-2$, and the oblique parameters.  Constraints on the model parameters and detailed numerical analysis are presented in Sec.~\ref{sec:num}.  A summary of our findings is given in Sec.~\ref{sec:summary}.

\section{The Model} \label{sec:model}

We consider in this work a model that extends the SM gauge symmetry by the $U(1)_{\mu-\tau}$ gauge symmetry, under which only the $\mu$ and $\tau$ leptons in the SM are charged.  
 Due to the opposite $U(1)_{\mu-\tau}$ charges within the second and third generations of leptons, it can be easily checked that the loop-induced triangle anomalies mediated by the muon and $\tau$-lepton for $U(1)^3_{\mu-\tau}$, $U(1)^2_{\mu-\tau} U(1)_Y$, $U(1)_{\mu-\tau} U(1)^2_Y$,  $SU(2)^2 U(1)_{\mu-\tau}$, and ${\rm gravity}^2$-$U(1)_{\mu-\tau}$ cancel out automatically without the need of introducing extra chiral fermions. As a result, the gauged $U(1)_{\mu-\tau}$ symmetry model stands free from gauge anomalies.

In addition to the SM Higgs doublet, denoted by $H_2$, whose neutral component has a VEV, $v_2$, to spontaneously break $SU(2)_L \times U(1)_Y$, we introduce an additional Higgs doublet, denoted by $H_1$, which carries not only the $U(1)_{\mu-\tau}$ charge, twice that of $\mu$, but also the weak isospin and $U(1)_Y$ hypercharge.  The new Higgs doublet is assumed to also develop a VEV, $v_1$, in its neutral component to break $U(1)_{\mu-\tau}$ besides $SU(2)_L \times U(1)_Y$, resulting in a massive $Z'$ boson.  Therefore, unlike the conventional two-Higgs-doublet model (2HDM), the model has one charged Higgs and two CP-even Higgs bosons but has no CP-odd Higgs boson, as it has become the longitudinal component of $Z'$.  Finally, we include an $SU(2)_L$-singlet scalar LQ with hypercharge $Y=2/3$ that also has the same $U(1)_{\mu-\tau}$ charge as $\mu$.  The quantum number assignments of the leptons, the Higgs doublets, and the LQ are given in Table~\ref{tab:charge}.  As we will see, such a model can simultaneously accommodate the measured lepton $g-2$, $R(D^{(*)})$, and $W$ mass anomalies while the cross section of the CE$\nu$NS process can be enhanced to deviate from the SM expectation by up to $25 \%$.

\begin{table}[hptb]
\caption{Quantum numbers of the leptons, Higgs doublets, and scalar leptoquark.
}
\begin{tabular}{ccccccc} 
\hline \hline
&  ~~~$e_{L(R)}$~~~  & ~~~$\mu_{L(R)}$~~~ & ~~~ $\tau_{L(R)}$ ~~~&~~~$H_2$~~~& ~~~ $H_1$  ~~~ &~~~$S^{\frac{1}{3}}$~~~\\ \hline 
${L_\mu -L_\tau}$ &   0   &   $q_X$      & $-q_X$    & $0$  &  $2 q_X$  & $q_X$   
\\
$ SU(2)_L$ & 2(1) & 2(1) & 2 (1) & 2& 2 & 1   
\\
$U(1)_Y$ & $-1 (-2)$ & $-1 (-2)$ & $-1 (-2)$ & 1 & 1 & $2/3$  
\\ \hline \hline 
\end{tabular}
\label{tab:charge}
\end{table}

In the following subsections, we analyze the spectra of Higgs and gauge bosons and determine their physical eigenstates. In addition, we also derive the gauge, Yukawa, and trilinear couplings of Higgs bosons, which are used for the phenomenological analysis presented in the paper.

\subsection{Spectra of Higgs bosons and Higgs-related trilinear couplings}

We first write down the scalar potential consistent with the $SU(2)_L\times U(1)_Y \times U(1)_{\mu-\tau}$ gauge symmetry as:
\begin{align}
V(H_1, H_2,S^{\oovt}) = & \mu^2_1 H^\dagger_1 H_1 + \mu^2_{2} H^\dagger_2 H_2 + \frac{\lambda_1}{2} (H^\dagger_1 H_1)^2 + \frac{\lambda_2}{2} (H^\dagger_2  H_2)^2 + \lambda_{3}  H^\dagger_1 H_1 H^\dagger_2 H_2 \nonumber \\
& + \lambda_{4}  H^\dagger_1 H_2 H^\dagger_2 H_1 + \mu^2_{S} S^{-\oovt} S^{\oovt} + S^{-\oovt} S^{\oovt} \left( \lambda^S_{1} H^\dagger_1 H_1  +  \lambda^S_{2} H^\dagger_2 H_2 \right) \,.
\label{eq:scalar_V}
\end{align}
Owing to the $U(1)_{\mu - \tau}$ symmetry, there is no so-called $\mu$ term that couples $H_{1,2}$ quadratically and all terms in Eq. (\ref{eq:scalar_V}) are self-Hermitian due to the $U(1)_{\mu-\tau}$ symmetry, rendering all the coefficients real.  The components of two Higgs doublets can be parametrized as ($i = 1,2$):
\begin{equation}
H_i = 
\begin{pmatrix}
\phi^+   \\
\frac{1}{\sqrt2}(v_i + \phi^0_i + i \eta_i)
\end{pmatrix}
\,.
\end{equation}
Using the tadpole conditions $\partial V/\partial v_{1,2}=0$, we obtain two equalities:
\begin{align}
\begin{split}
 & \mu^2_1 + \frac{\lambda_1}{2} v^2_1 + \frac{\lambda_{34}}{2} v^2_2 = 0 \,, 
 \\
 &  \mu^2_2 + \frac{\lambda_2}{2} v^2_2 + \frac{\lambda_{34}}{2} v^2_1 = 0 \,,
\end{split}
 \label{eq:mini}
\end{align}
 with $\lambda_{34} \equiv \lambda_{3} + \lambda_4$. To achieve spontaneous breakdown of the $SU(2)_L\times U(1)_Y \times U(1)_{\mu-\tau}$ gauge symmetry, we require $\mu^2_{1,2}<0$. For the vacuum stability, where the scalar potential is bounded from below in all field configurations, the quartic couplings have to satisfy the criteria given by~\cite{Klimenko:1984qx,Kannike:2012pe}
\begin{align}
\lambda_{1,2} \ge 0\,,~\lambda_{3} + \sqrt{\lambda_1 \lambda_2} \ge 0
\,, ~ 
\lambda_{34}+\sqrt{\lambda_1 \lambda_2} \ge 0\,.
\end{align}
  
Two neutral Goldstone bosons result from the mixing between the two CP-odd components:
\begin{equation}
\begin{pmatrix}
 G^0_{Z'}  \\
G^0_{Z}      \\    
\end{pmatrix}
= 
\begin{pmatrix}
c_{\beta}  & s_\beta \\
-s_{\beta}     &  c_{\beta} \\    
\end{pmatrix}
\begin{pmatrix}
\eta_1    \\
\eta_2    \\    
\end{pmatrix}
\equiv
U_\beta 
\begin{pmatrix}
\eta_1    \\
\eta_2    \\    
\end{pmatrix}
\,, 
\label{eq:Goldstone}
\end{equation}
where $\beta$ is defined by $t_\beta \equiv \tan\beta=v_2/v_1$, $v=\sqrt{v^2_1 + v^2_2}$, $c_\beta \equiv \cos\beta$, and $s_\beta \equiv \sin\beta$.  To obtain the states of charged Goldstone and charged Higgs bosons, we can use Eq.~(\ref{eq:Goldstone}) by substituting $(G^\pm, H^\pm)$ and $(\phi^+_1, \phi^+_2)$ for $(G^0_{Z'}, G^0_Z)$ and $(\eta_1, \eta_2)$, respectively.  As a result, the mass-squared of the charged Higgs boson is solely dependent on the parameter $\lambda_4$ as follows:
\begin{equation}
m^2_{H^\pm} = - \frac{\lambda_4}{2} v^2\,.
\end{equation}
Since the massive LQ is irrelevant to the EWSB, its mass-squared with the assumption that $\mu^2_{S}>0$ is found to be:
\begin{equation}
m^2_{S}= \mu^2_{S} + \frac{v^2}{2} \left( \lambda^S_{1} c^2_\beta + \lambda^S_2 s^2_\beta \right)
~,
\end{equation}
and can be as heavy as ${\cal O}({\rm TeV})$.

From the scalar potential in Eq.~(\ref{eq:scalar_V}) and the tadpole conditions in Eq.~(\ref{eq:mini}), the mass terms for the $CP$-even scalars can be written as:
\begin{equation}
\frac{1}{2} \begin{pmatrix}
 \phi^0_1\, ,  &  \phi^0_2  \\    
\end{pmatrix}
\begin{pmatrix}
\lambda_1 v^2_1 & v_1 v_2 \lambda_{34} \\
 v_1 v_2 \lambda_{34}    &  \lambda_2 v^2_2 \\    
\end{pmatrix}
\begin{pmatrix}
\phi^0_1    \\
\phi^0_2    \\    
\end{pmatrix}
\,. \label{eq:mass_scalar}
\end{equation}
Eq.~(\ref{eq:mass_scalar}) can be diagonalized by a  $2\times 2$ orthogonal matrix, and   the resulting  eigenstates of neutral Higgses can be parametrized using a mixing angle $\alpha$ as: 
\begin{equation}
\begin{pmatrix}
H \\
h  
\end{pmatrix}
= 
\begin{pmatrix}
c_{\alpha}  & s_\alpha \\
-s_\alpha     &  c_\alpha
\end{pmatrix}
\begin{pmatrix}
\phi^0_1    \\
\phi^0_2   
\end{pmatrix}
\equiv
U_\alpha 
\begin{pmatrix}
\phi^0_1    \\
\phi^0_2  
\end{pmatrix}
\,, \label{eq:H_h}
\end{equation}
where $h$ is the 125-GeV SM-like Higgs boson, $c_\alpha \equiv \cos\alpha$, and $s_\alpha \equiv \sin\alpha$.  In the following, we would focus on the scenario where the new $CP$-even state is lighter than the SM-like Higgs boson, {\it i.e.}, $m_h > m_H$. Using the parameters $\lambda_i$ and $v_i$,  the masses of the $h$ and $H$ states, as well as the mixing angle between them, can be obtained as:
\begin{align}
\begin{split}
m^2_{h, H}  &=  \frac{\lambda_1 v^2_1 + \lambda_2 v^2_2}{2} \pm \frac{1}{2} \sqrt{(\lambda_1 v^2_1 - \lambda_2 v^2_2)^2 + 4 v^2_1 v^2_2 \lambda^2_{34}}\,, 
 \\
\tan2\alpha & = - \frac{2 v_1 v_2 \lambda_{34} }{\lambda_2 v^2_2 - \lambda_1 v^2_1}\,.
\end{split}
\end{align}

The scalar potential in the model involves six parameters, namely, $\mu^2_{1,2}$ and $\lambda_{1-4}$.  One can write them in terms of the physical parameters $\{m_{H^\pm, h, H}, v, \alpha, \beta \}$ as
\begin{subequations}
\begin{align}
\mu^2_1 & =  -\frac{1}{2c_\beta} \left(  -s_\alpha s_{\beta-\alpha} m^2_h + c_\alpha c_{\beta-\alpha} m^2_{H}  \right)\,, \\
\mu^2_2 & =  -\frac{1}{2s_\beta} \left(  c_\alpha s_{\beta-\alpha} m^2_h + s_\alpha c_{\beta-\alpha} m^2_{H}  \right)\,, \\
\lambda_1 & = \frac{1}{v^2 c^2_\beta} \left( m^2_h s^2_\alpha + m^2_H c^2_\alpha \right)\,, \\
\lambda_2 & = \frac{1}{v^2 s^2_\beta} \left( m^2_h c^2_\alpha + m^2_H s^2_\alpha \right)\,, \\
\lambda_3 & = -\frac{s_{2\alpha}}{v^2 s_{2\beta}} \left( m^2_h -m^2_H \right) + \frac{2 m^2_{H^\pm}}{v^2}\,, \\
\lambda_4 & = - \frac{2 m^2_{H^\pm}}{v^2}\,.
\end{align}
\end{subequations}

An important parameter of the scalar potential in the SM is the quartic coupling $\lambda_{\rm SM}$, which not only determines the mass of the SM Higgs boson via $m^2_h=\lambda_{\rm SM} v^2$, but also controls the potential shape. Therefore, to probe the existence of extra scalars, it becomes crucial to precisely determine the Higgs self-coupling through the $hh$ production that involves the Higgs trilinear coupling~\cite{Baur:2002rb}.  In the 2HDM, the SM-like Higgs field is a linear combination of $\phi^0_{1,2}$ and, instead of a factor of $3 m^2_h/v$ for the SM, the Higgs self-coupling also involves the parameters $\beta$ and $\alpha$.  Moreover, when $m_H < m_h/2$, the decay channel $h\to HH$ becomes accessible.  Current measurements of Higgs decays can impose stringent constraints on the related parameters.  To take these constraints into account, we present all the Higgs trilinear couplings as follows:
\begin{align}
-{\cal L} \supset 
& - \frac{s_{2\alpha} s_{\beta-\alpha} }{v s_{2\beta}} \left(m^2_h + 2 m^2_H \right) \frac{h H^2}{2}+ \frac{s_{2\alpha} c_{\beta-\alpha}}{v s_{2\beta}} \left(2 m^2_h + m^2_H \right) \frac{h^2 H}{2} 
\nonumber \\
& + \frac{3 m^2_{h}}{v} \left( s_{\beta-\alpha} + \frac{2}{s_{2\beta} } c_{\beta+\alpha} c^2_{\beta-\alpha}\right) \frac{h^3}{3!} + \frac{3m^2_H}{v} \left( c_{\beta-\alpha} + \frac{2}{s_{2\beta} } s_{\beta+\alpha}s^2_{\beta-\alpha}\right) \frac{H^3}{3!} 
\nonumber \\
& + v \left(\lambda^S_+ c_{\beta-\alpha} + \lambda^S_{-} c_{\beta + \alpha} \right) H S^{-\oovt} S^{\oovt} + v \left(\lambda^S_+ s_{\beta-\alpha} -\lambda^S_{-} s_{\beta + \alpha} \right) h S^{-\oovt} S^{\oovt}
\,. \label{eq:tri_hHH}
\end{align}
Taking the limits of $\alpha\to 0$ and $s_\beta \to1$, it can be seen that only the self-couplings of $h$ and $H$ remain.  We note that the scalar couplings to the LQ are also included, which can be used to analyze the loop-induced Higgs boson decays.

\subsection{ $Z'-Z$ mixing  and gauge couplings of scalars}

The masses of the gauge bosons and the gauge couplings of scalars are determined by the kinetic terms of $H_{1,2}$, with the  covariant derivatives given as:
\begin{align}
\begin{split}
D_\mu H_1 & = \left(\partial_\mu+ i \frac{g}{2} \vec\tau \cdot \vec{W}_\mu + i \frac{g'}{2} B_\mu + g_{Z'} X Z'_\mu \right)H_1\,, 
\\
D_\mu H_2 & = \left( \partial_\mu+ i \frac{g}{2} \vec\tau \cdot \vec{W}_\mu + i \frac{g'}{2} B_\mu \right)H_2\,, 
\\
D_\mu S^{\oovt} & = \left( \partial_\mu  + i Q_S g' B_\mu + i g_{Z'} q_X Z'_\mu \right) S^{\oovt}\,,
\end{split}
\end{align}
where $g$, $g'$, and $g_{Z'}$ denote the gauge couplings of $SU(2)_L$, $U(1)_Y$, and $U(1)_{\mu-\tau}$, respectively, $X=2 q_X$ is the $U(1)_{\mu-\tau}$ charge of $H_1$, and $Q_S=1/3$ is the electric charge of LQ.  As in the conventional 2HDM, the tree-level $W$ boson mass can be obtained as $m_W = g v/2$.  However, since $H_1$ carries the charges of both electroweak and $U(1)_{\mu-\tau}$ symmetries, its VEV breaks not only $SU(2)_L\times U(1)_Y$ but also $U(1)_{\mu-\tau}$ at the same time.  As a result, the $Z$ and $Z'$ states are not physical and generally mix with each other.  More explicitly, the mass-squared terms for $Z$ and $Z'$ are given by:
\begin{equation}
\frac{1}{2} \begin{pmatrix}
 Z' \, ,  &  Z  
\end{pmatrix}
\begin{pmatrix}
m^2_{Z'} & m^2_{Z'Z} \\
 m^2_{Z'Z}    & m^2_Z 
\end{pmatrix}
\begin{pmatrix}
Z'    \\
Z   
\end{pmatrix}
\,, \label{eq:mass_ZZp}
\end{equation}
where $m^2_{Z'}$, $m^2_{Z}$, and $m^2_{Z'Z}$ are defined as: 
\begin{align}
\begin{split}
m^2_{Z'} & = g^2_{Z'} X^2 v^2_1= \frac{(g_{Z'} X v)^2}{1+t^2_\beta}\,, 
\\
m^2_{Z} & = \frac{g^2 + g'^2}{4} v^2 = \frac{g^2 v^2}{4}\left( 1+t^2_W \right)\,, 
\\
m^2_{Z'Z}&= - \frac{g g_{Z'} X}{2c_W} v^2_1=  - \frac{g g_{Z'} X v^2}{2c_W (1+t^2_\beta)} \,.
\end{split}
\end{align}
  The states of the photon and $Z$ boson fields are written, as in the SM, as:
   \begin{align}
   \begin{split}
      A_\mu & = c_W B_\mu +s_W W^3_\mu\,,
   \\
   Z_\mu & = -s_W B_\mu + c_W W^3_\mu\,,   
   \end{split}
\end{align}
where $c_W \equiv \cos\theta_W$, $s_W \equiv \sin\theta_W$, and $\theta_W$ is the weak mixing angle.  The mass-squared matrix in Eq.~(\ref{eq:mass_ZZp}) can be diagonalized using a $2\times 2$ orthogonal matrix, parametrized by a mixing angle $\theta_Z$, in a fashion analogous to Eq.~(\ref{eq:H_h}).  Assuming that $m_{Z'}\ll m_Z$ and taking $Z_1$ and $Z_2$ as the physical states of the neutral gauge bosons, their mass-squares and mixing angle can be approximately obtained as follows:
\begin{align}
  \begin{split}
  m^2_{Z_1} & \simeq  m^2_{Z'} - \frac{m^4_{Z'Z}}{m^2_Z }=m^2_{Z'} \frac{t^2_\beta }{1+t^2_\beta}\,, 
  \\
  m^2_{Z_2} & \simeq m^2_Z + \frac{m^4_{Z'Z}}{m^2_Z }\,, 
  \\
  s_{\theta_{Z}} & \simeq -{\rm sign}(\theta_Z) \frac{m^2_{Z'Z}}{m^2_{Z}}={\rm sign}(\theta_Z) \frac{2 c_W}{g t_\beta} \frac{m_{Z_1} }{v}\,,
  \end{split}
  \label{eq:Zp_Z_mixing} 
\end{align}
where sign$(\theta_Z)=\pm 1$ represents the sign of the mixing angle.  Apparently, the mixing angle is suppressed by $m_{Z_1}/(vt_\beta)$ as $t_\beta$ gets large.  If the mass of $m_{Z_1}$ is of ${\cal O}(10)$~MeV, $s_{\theta_Z}$ is at most of ${\cal O}(10^{-5})$.

To study the loop-induced processes or variables ({\it e.g.}, lepton $g-2$) mediated by the $Z_1$ boson, we also need the gauge couplings of scalars and LQ as follows:
 \begin{align}
 {\cal L} 
 \supset &
 \,  i \frac{g}{2 c_W} \left[ (\partial^\mu H^-) H^+-H^- \partial^\mu H^+ \right] \left( s_{2W} A_\mu + c_{2W} Z_{\mu} \right) \nonumber \\ 
& +\left[ -i \frac{g s_{\beta-\alpha}}{2} W^+_\mu \left( H\partial^\mu H^{-} - H^- \partial^\mu H  \right) \right. \nonumber \\
& \qquad \left. + i \frac{g c_{\beta-\alpha}}{2}W^+_\mu  \left(  h\partial^\mu H^{-} - H^- \partial^\mu h \right)+ 
\mbox{H.c.} \right] \nonumber \\
& + g m_W W^-_\mu W^{+\mu} \left( s_{\beta-\alpha} h+ c_{\beta-\alpha} H\right)  \nonumber \\
& -i Q_S e \left( A_\mu - t_W Z_\mu \right) \left( S^{-\oovt} \partial^\mu S^{\oovt} - S^{\oovt} \partial^\mu S^{-\oovt} \right) \nonumber \\
&+ \frac{2m^2_Z}{v} \left( c_{\beta-\alpha} H + s_{\beta-\alpha} h \right)   \frac{Z_\mu Z^\mu}{2} + \frac{2 m^2_{Z'}}{v_1} \left( c_\alpha H -s_\alpha h \right)\frac{Z'_\mu Z^{\prime\mu} }{2}  \nonumber \\
& - \frac{g m_{Z'}}{c_W} \left( c_\alpha H -s_\alpha h \right) Z'_\mu Z^{\mu}
\,. \label{eq:h_V}
 \end{align}

\subsection{Yukawa couplings of fermions}

The Yukawa sector plays a crucial role in flavor physics as it governs the mass generation of the SM fermions and the couplings of scalars to fermions in the model. The Lagrangian of the Yukawa sector under $SU(2)_L\times U(1)_Y\times U(1)_{\mu-\tau}$ gauge symmetry can be written based on the quantum number assignments in Table~\ref{tab:charge} as:
\begin{align}
-{\cal L}_{Y} 
= &
\ \overline{Q_L} H_2  {\bf Y}^d d_R + \overline{Q_L} \tilde H_{2} {\bf Y}^u u_R + \bar L_\ell H_2 {\bf Y}^\ell \ell_R + y_{\mu \tau} \bar L_\mu H_1 \tau_R
\nonumber \\
&+ \overline{Q^c_L} i \tau_2  {\bf y}^q_L L_\tau S^{\oovt} + \overline{u^c_R} {\bf y}^u_R \tau_R S^{\oovt} + \text{H.c.}
~,
\label{eq:LY}
\end{align}
where, except for the $\bar L_\mu H_1 \tau_R$ term, the flavor indices are all suppressed, $Q^T_L=(u, d)^T_L$ and $L^T=(\nu_\ell ,\ell)^T_L$ represent the quark and lepton doublets, respectively, $\ell_R$ denotes the right-handed charged lepton, and $F^c=C\gamma^0 F^*$ for a fermion $F$ with $C$ being the charge conjugation operator. The $U(1)_{\mu-\tau}$ gauge symmetry restricts  the $3\times 3$ Yukawa matrix ${\bf Y}^\ell $ to be a diagonal matrix, i.e., ${\bf Y}^\ell = \mbox{diag}(y^e, y^\mu, y^\tau)$.  We note that because $H_1$ and $H_2$ simultaneously couple to the charged leptons, the term $\bar L_\mu H_1 \tau_R$ will induce flavor-changing neutral-currents (FCNCs) at tree level.  After diagonalizing the quark mass matrices and using the physical states of scalars, the Yukawa couplings of quarks to $h(H)$ and $H^\pm$ are found to be the same as those in type-I 2HDM~\cite{Branco:2011iw}.  Although the charged Higgs boson could in principle enhance the $b\to c \tau \nu$ transition~\cite{Crivellin:2012ye,Crivellin:2013wna,Crivellin:2015hha,Chen:2017eby,Akeroyd:2017mhr,Chen:2018hqy}, the involved Yukawa couplings in this model are suppressed by $m_{b,c} / (\tan\beta\sqrt{v^2_1+v^2_2})$ and are irrelevant for our later discussions.  The explicit expressions of the couplings can be found in Ref.~\cite{Branco:2011iw}.

While the diagonal ${\bf Y}^\ell$ matrix contributes to the charged lepton masses, the $\bar L_\mu H_1 \tau_R$ term induces flavor mixing between the $\mu$ and $\tau$ leptons.  Thus, the electron mass is simply $m_e = y^e v_2 /\sqrt{2}$, and the mass matrix for the $\mu$ and $\tau$ leptons is expressed as:
\begin{equation}
( \bar \mu_L\,,~ \bar \tau_L)  {\bf M}_{\ell}  
\begin{pmatrix}
\mu_R    \\
\tau_R  
\end{pmatrix}
=
( \bar \mu_L\,,~ \bar \tau_L) 
\begin{pmatrix}
\hat{m}_{\mu} &  \hat m_{\mu \tau} \\
0    & \hat m_{\tau}
\end{pmatrix}
\begin{pmatrix}
\mu_R    \\
\tau_R  
\end{pmatrix}
\,,
\end{equation}
where $\hat{m}_{\mu(\tau)}= y^{\mu(\tau)} v_2 /\sqrt{2}$ and $\hat m_{\mu \tau} = y_{\mu \tau} v_1/\sqrt{2}$.  The matrix ${\bf M}_\ell$ can be diagonalized through a bi-unitary transformation: ${\bf m}_{\ell} = V^\ell_L {\bf M}_\ell V^{\ell \dagger}_R$.  Accordingly, the Yukawa couplings of the Higgs bosons to the leptons are found to be:
\begin{align}
-{\cal L}_Y \supset & ~\overline{\ell_L} {\bf m}_{\ell} \ell_R + \overline{\ell_L}   \frac{{\bf m}_{\ell}}{v}  \ell_R\left( \frac{c_\alpha}{s_\beta} h+\frac{s_\alpha}{s_\beta} H\right) + \overline{\ell_L} \frac{{\bf X_\ell}}{v} \ell_R \left( - \frac{2 c_{\beta-\alpha}}{s_{2\beta}} h + \frac{2 s_{\beta-\alpha}}{s_{2\beta}}H\right) \nonumber \\
& +  \overline{\nu_{ L}} \left( \frac{\sqrt{2} {\bf m}_\ell}{v t_\beta} - \frac{2\sqrt{2} {\bf X}_\ell }{s_{2\beta} v}  \right)\ell_R H^+  + \text{H.c.} \,, \label{eq:h_Yu}
\end{align}
where ${\bf X}_\ell$ is defined as:
\begin{equation}
{\bf X}_\ell = V^\ell_L 
\begin{pmatrix}
0 &  \hat m_{\mu \tau} \\
0    & 0
\end{pmatrix}
V^{\ell\dag}_R\,.
\end{equation}
It is worth mentioning that ${\bf X}_\ell$ induces the tree-level FCNCs mediated by the Higgs bosons in the lepton sector.  To see the decoupling and large $\tan\beta$ limits, it is useful to rewrite $c_\alpha/s_\beta$ and $s_\alpha/s_\beta$ as:
\begin{align}
\begin{split}
\frac{c_\alpha}{s_\beta} & = s_{\beta-\alpha} + t^{-1}_\beta c_{\beta-\alpha}\,,
\\
\frac{s_\alpha}{s_\beta} & = c_{\beta-\alpha} - t^{-1}_\beta s_{\beta-\alpha}\,. 
\end{split}
\end{align}

When the lepton Yukawa couplings are real, we can obtain the $2\times 2$ flavor mixing matrices $V^\ell_{R,L}$ using the identities:
 \begin{align}
 \begin{split}
 {\bf m}^\dag_\ell {\bf m}_\ell & = V^\ell_R {\bf M}^\dag_\ell {\bf M}_\ell V^{\ell \dag}_{R}\,,
 \\
  {\bf m}_\ell {\bf m}^\dag_\ell & = V^\ell_L {\bf M}_\ell {\bf M}^\dag_\ell V^{\ell \dag}_{L}\,.
 \end{split}
 \end{align}
By parametrizing $V^\ell_{R, L}$ in the same form as $U_\alpha$ in Eq.~(\ref{eq:H_h}), we can obtain the mixing angles $\theta_{R, L}$ as:
  \begin{align}
  \begin{split}
    \tan2\theta_R & = -\frac{2\hat{m}_\mu \hat{m}_{\mu\tau}}{\hat{m}^2_\tau + \hat{m}^2_{\mu\tau} - \hat{m}^2_\mu}\,, 
    \\
    \tan2\theta_L & = -\frac{2\hat{m}_\tau \hat{m}_{\mu\tau}}{\hat{m}^2_\tau - \hat{m}^2_{\mu\tau} - \hat{m}^2_\mu}\,.
  \end{split}
\end{align}
In the limit when $\hat{m}_\mu \hat{m}_{\mu\tau}/\hat{m}^2_\tau$ is negligible, these mixing angles can be obtained to a good approximation as:
  \begin{equation}
  \theta_R\approx 0, ~ s_{\theta_L} \simeq - \hat{m}_{\mu\tau}/\hat{m}_\tau \,.  \label{eq:theta_RL}
  \end{equation}
As a free parameter with the mass dimension that appears only in the $\mu-\tau$ element of ${\bf X}_\ell$, $\hat{m}_{\mu\tau}$ can be parametrized in terms of a free parameter $\chi_{\mu\tau}$ as $\hat{m}_{\mu\tau}= \chi_{\mu\tau} \sqrt{m_\mu m_\tau}$, where $m_{\mu,\tau}$ are the physical masses of $\mu$ and $\tau$ leptons.  Using the approximate mixing angles in Eq.~(\ref{eq:theta_RL}), we obtain
  \begin{align}
  \begin{split}
    \hat{m}^2_\mu & \simeq  m^2_\mu \left( 1- \chi^2_{\mu \tau} \frac{m_\mu}{m_\tau}\right) \approx m^2_\mu\,, 
    \\
   \hat{m}^2_\tau & \simeq  m^2_\tau \left( 1- \chi^2_{\mu \tau} \frac{m_\mu}{m_\tau}\right)\,, 
   \\
   {\bf X}_\ell & \simeq  \left( \begin{array}{cc}
0 & \chi_{\mu \tau} \sqrt{m_\mu m_\tau}\\
0    & \chi^2_{\mu \tau} m_\mu \\    
\end{array}
\right)\,.
  \end{split}
\end{align}

We now discuss the LQ couplings to quarks and leptons.  Since the Yukawa couplings ${\bf y}^q_L$ and ${\bf y}^u_R$ are free parameters, the up-type quark flavor mixings can be absorbed into these parameters.  As such, the up-type quark fields appearing in the LQ couplings in Eq.~(\ref{eq:LY}) can be treated as the physical states.  However, the same ${\bf y}^q_L$ also appears in the couplings to the down-type quarks.  Therefore, in addition to $V^d_L$, the LQ couplings to the down-type quarks must include $V^u_L$.  With $V^\ell_R\simeq \mathbbm{1}$ and $V_{\rm CKM}=V^u_L V^{d\dag}_L$,  we can express the Yukawa couplings of the LQ as:
 \begin{align}
- {\cal L}_Y \supset   \left( \overline{u^C_L} {\bf y}^q_L V^{\ell\dag}_{L\tau \ell}  P_L \ell + \overline{u^C_R} {\bf y}^u_R P_R  \tau \right) S^{\oovt} - \overline{d^C_L} V^T_{\rm CKM} {\bf y}^q_L P_L \nu_{\tau} S^{\oovt} + \text {H.c.} \label{eq:LQ_C}
 \end{align}

\subsection{Gauge couplings of fermions}

Next, we consider the gauge couplings of the fermions.  Since the $U(1)_{\mu-\tau}$ gauge symmetry does not affect the weak charged currents, they remain the same as those in the SM.  Although quarks do not carry the $U(1)_{\mu-\tau}$ charge and thus do not directly couple to the $Z'$ gauge boson, their couplings to the $Z'$ boson can be induced through the mixing with the SM $Z$ boson.  Intriguingly, the distinct $U(1)_{\mu-\tau}$ charges carried by the muon and tau-lepton lead to a lepton FCNC in the interaction $\bar \mu_L \gamma^\mu \tau_L Z'_\mu$.  Due to the $Z'-Z$ mixing, they then result in $Z$-mediated lepton FCNCs although such effects are suppressed by $s_{\theta_L} s_{\theta_Z}$.  Using the results shown in Eq.~(\ref{eq:Zp_Z_mixing}) and   Eq.~(\ref{eq:theta_RL}) for the lepton-flavor and $Z'-Z$ mixings, respectively, we obtain the neutral gauge couplings to fermions as follows: 
 \begin{align}
 {\cal L}^N_{ ffV} 
 = & 
 - \sum_f Q_f e \bar f \gamma^\mu f A_\mu - \frac{g}{2c_W} \sum_f \bar f \gamma^\mu \left( C^f_{iV} - C^f_{iA} \gamma_5\right) f Z_{i\mu} 
 \nonumber \\
 & + \left[ g_{Z'} q_X s_{2\theta_L} \overline{\mu_L} \gamma^\mu \tau_L (c_{\theta_Z} Z_{1\mu} - s_{\theta_Z} Z_{2\mu}) + \text{H.c.} \right]
 \,, \label{eq:L_N}
 \end{align}  
where the coefficients $C^f_{iV, iA}$ are explicitly given by:
\begin{align}
\begin{split}
C^{f}_{1V} & = c^f_V s_{\theta_Z} + \frac{c_W m_{Z_1} c_{\theta_Z} }{g v} X^f_V \sqrt{2+t^2_\beta+t^{-2}_\beta}
\,, 
\\
C^{f}_{1A} & = c^f_A s_{\theta_Z} + \frac{c_W m_{Z_1} c_{\theta_Z} }{g v} X^f_A \sqrt{2+t^2_\beta+t^{-2}_\beta}
\,, 
\\
C^{f}_{2V} & = c^f_V c_{\theta_Z} -  \frac{c_W m_{Z_1} s_{\theta_Z} }{g v} X^f_V \sqrt{2+t^2_\beta+t^{-2}_\beta}
\,, 
\\
C^{f}_{2A} & = c^f_A c_{\theta_Z} - \frac{c_W m_{Z_1} s_{\theta_Z} }{g v} X^f_A \sqrt{2+t^2_\beta+t^{-2}_\beta}
 \,, 
\end{split}
\end{align}
with $c^f_V = T^3_f -2s^2_W Q_f$, $c^f_A= T^3_f$ given in terms of the weak isospin $T^3_f$ and the electric charge $Q_f$ of the fermion $f$, and $X^{f}_V=(0,1/2,-1/2, 0, 1, -1,0,0)$ and $X^f_A= (0,1/2,-1/2, 0, 0, 0, 0,0)$ for $f=(\nu_e,\nu_\mu, \nu_\tau, e, \mu, \tau, u, d)$.  Because only vector currents are involved in the $Z'$ couplings to the charged leptons, $X^\ell_A =0$.  However, $X^{\nu_\mu, \nu_\tau}_A$ are non-vanishing because neutrinos are left-handed particles in the model.

\section{Phenomenology} \label{sec:pheno}

In this section, we derive the formalisms for the processes studied in this work. These include the cross section for CE$\nu$NS via the $Z'-Z$ mixing, the $R(D)$ and $R(D^*)$ from LQ interactions, new  Higgs decay modes, lepton $g-2$, and the effects on the oblique parameters and the $W$ mass.

\subsection{CE$\nu$NS through the $Z'-Z$ mixing }

In the model, elastic electron- and muon-neutrino (including anti-neutrino) scatterings off a nucleus arise from gauge interactions with the neutral gauge bosons $Z_{1}$ and $Z_2$. Using the gauge couplings given in Eq.~(\ref{eq:L_N}), we can write the effective Hamiltonian for neutrino scattering at the quark level as:
\begin{align}
{\cal H}_{\nu_\ell q \to \nu_\ell q}
&= 
\sqrt{2} G_F  \left( \frac{m_Z c_{\theta_Z}}{ m_{Z_2}} \right)^2 
\left[ 1 + \Delta^\ell (q^2)  \right] 
\left[ \bar\nu_\ell \gamma^\mu P_L \nu_\ell \right] 
\left[\bar q \gamma_\mu \left( c^q_V -c^q_A \gamma_5 \right) q \right]
\,,  \label{eq:Heff}
\\
\Delta^\ell (q^2) 
&= 
{\rm sign}(\theta_Z) \frac{m^2_{Z_1}}{c^2_{\theta_Z} (q^2 + m^2_{Z_1})}  \frac{m^2_{Z_2}}{ m^2_Z t^2_\beta }
\left( 1 + \delta^\ell_\mu \frac{1 + t^2_\beta }{2} \right) 
\nonumber \\ 
&  \simeq  
{\rm sign}(\theta_Z)  \frac{m^2_{Z_1} }{q^2 + m^2_{Z_1}} 
\left( \frac{1}{t^2_\beta} +\frac{ \delta^\ell_\mu}{2} \right)
\,, \label{eq:Dell}
\end{align}
where the Kronecker delta $\delta^\ell_\mu$ indicates that only the muon-neutrino or anti-muon-neutrino contributes.  The second line in Eq.~(\ref{eq:Dell}) results from the limits of $c_{\theta_Z}\simeq 1$, $m_{Z_2}\simeq m_{Z}$ and large $t_\beta$. We will demonstrate that due to the $h\to HH$ and $h\to Z_1 Z_1$ constraints, a large  $t_\beta$ is required for the model.  As a result, the electron-neutrino scattering becomes insignificant and negligible.  Since the structure of the four-fermion interaction in Eq.~(\ref{eq:Heff}) is the same as that in the SM, the new physics contribution can be obtained simply by replacing $C_{\rm SM}$ with $C_{\rm SM} \left[ 1+\Delta^\ell (q^2) \right]$.  In contrast to the effects induced by the kinetic mixing in the conventional $U(1)_{\mu-\tau}$ model, the $g_{Z'}$ dependence has been absorbed into $m_{Z_1}$. Thus, the new physics effect depends only on $m_{Z_1}$ in the large-$t_\beta$ scheme.  Because the LQ mass is of ${\cal O}(1)$~TeV, its contribution is negligible.  As such, we skip the discussion related to the LQ effects.

The cross section for the elastic neutrino-nucleus scattering can be written as~\cite{Abdullah:2022zue}:
\begin{align}
\frac{d\sigma}{dE_r} & = \frac{G^2_F m_A}{\pi} \left( 1- \frac{m_A E_r}{2E^2_\nu} - \frac{E_r}{E_\nu}\right) 
\left| Q^\ell_w (q^2) \right|^2
\,,  \label{eq:vector}\\
Q^\ell_w(q^2) & = Z g^\ell_p F_p(q^2) + N g^\ell_n F_n (q^2)\,, 
\end{align}
 where $m_A$ is the nucleus mass, $Z(N)$ is the proton (neutron) number of the target nucleus, $E_\nu$ is the incident neutrino energy, $E_r$ is the nuclear recoil energy, and $q^2\simeq 2 m_A E_r$. The couplings to the proton $g_p$ and the neutron $g_n$ are respectively given by:
 \begin{align}
 g^\ell_p 
 &=
 ( 2 c^u_V+ c^d_V) \left[ 1+\Delta^\ell (q^2) \right]
 \,, \nonumber\\
 g^\ell_n
 &=
 (c^u_V + 2 c^d_V) \left[ 1+\Delta^\ell (q^2) \right]
 \,.
 \end{align}
Since the contribution from the weak axial-vector currents is much smaller than that from the vector currents, we have ignored their effects in Eq.~(\ref{eq:vector}).
 To include the nuclear  effects, we adopt the  Klein-Nystrand approach~\cite{Klein:1999qj} for $F_{p/n}(q^2)$, expressed as~\cite{Abdullah:2022zue}
 \begin{align}
 F_{KN}(q^2)= \frac{3 j_1(q R_A)}{q R_A} \frac{1}{1+ q^2  a^2_K}\,, 
 \end{align}
where $R_A=1.2 A^{1/3}$ with $A$ being the mass number, $j_1$ is the spherical Bessel function of order one, and $a_K$ denotes the range of a short-range Yukawa potential. For a numerical estimate, we take $a_K=0.7$~fm.  The neutrinos detected in the COHERENT experiment are produced by the stopped $\pi^+$ decay via $\pi^+\to \nu_\mu + \mu^+$ and by the subsequent $\mu^+$ decay through $\mu^+ \to e^+ + \nu_e + \bar\nu_\mu$.   
In this study, we assume that the shapes of neutrino fluxes are the same as their energy spectra, expressed as~\cite{COHERENT:2020iec,Barbeau:2021exu,Bertuzzo:2021opb}:  
\begin{align}
\frac{d\phi_\mu(E_\nu) }{dE_\nu} 
& = 
{\cal N} \delta\left(E_\nu-  \frac{m^2_\pi -m^2_\mu}{2m_\pi} \right)
\,, \nonumber \\
\frac{d\phi_{\bar\mu}(E_\nu) }{dE_\nu} 
& =
{\cal N} \frac{64}{m_\mu}  \frac{E^2_\nu}{m^2_\mu} \left( \frac{3}{4} - \frac{E_\nu}{m_\mu} \right)
\,, \nonumber \\
\frac{d\phi_{e}(E_\nu) }{dE_\nu} & = {\cal N} \frac{192}{m_\mu}  \frac{E^2_\nu}{m^2_\mu} \left( \frac{1}{2} - \frac{E_\nu}{m_\mu} \right) \label{eq:nu_flux}
\,,
\end{align}
with ${\cal N}$ being a normalization factor. Hence, the total cross section  averaged over the neutrino fluxes can be obtained as:
 \begin{equation}
 \langle \sigma \rangle_{\phi}= \sum_{\ell=e,\mu,\bar{\mu}} \int^{E^{\rm max}_\nu}_{E^{\rm min}_\nu} dE'_\nu  \int^{E^{\rm max}_r}_{E^{\rm min}_r} dE_r \frac{d\sigma(\nu_\ell A \to \nu_\ell A)}{dE_r}  \frac{d\phi_\ell (E'_\nu)}{dE'_\nu}\,, \label{eq:cross_avg}
 \end{equation}
where $E^{\rm max,\nu_\mu}_r = 2 E^{2}_{\nu_\mu}/(m_A + 2 E_{\nu_\mu})$, $E_{\nu_\mu}=(m^2_\pi -m^2_\mu)/2 m_\pi$, $E^{\rm max,\nu_{e, \bar\mu}}_r=2 E^{\prime 2}_{\nu}/(m_A + 2 E'_{\nu})$, $E^{\rm max}_\nu=m_\mu/2$, $E^{\rm min}_r$ denotes the nuclear threshold recoil energy, and $E^{\rm min}_\nu$ is the minimum incident neutrino energy required to reach $E^{\rm min}_r$.

\subsection{$R(D)$ and $R(D^*)$   } \label{sec:FDDv}

The model has two different mechanisms contributing to the  $b\to c \ell \nu$ process: one involves the charged Higgs boson, and the other is from the LQ.  However, the effects of the charged Higgs are not significant as its couplings to quarks and leptons are suppressed by $m_{b,c,\ell}/ (v t_\beta)$.  We thus focus exclusively on the LQ contributions.  Based on the Yukawa couplings of LQ in Eq.~(\ref{eq:LQ_C}), the effective Hamiltonian for $b\to c \ell \nu$ mediated by the $W$ gauge boson and $S^{\frac{1}{3}}$ can be obtained as~\cite{Chen:2017hir}:
 \begin{align}
 {\cal H}_{b\to c \ell \nu} 
 = & 
 \frac{4G_F V_{cb}}{\sqrt{2}} \left[(\delta^{\ell'}_{\ell} + C^\ell_V \delta^{\ell'}_\tau)\bar c \gamma^\mu P_L b~ \bar \ell \gamma_\mu P_L \nu_{\ell'}+ C^\ell_S  \bar c P_L b~ \bar\ell P_L \nu_\tau  
 +  C^\ell_T \bar c \sigma_{\mu \nu} P_L  b~ \bar\ell \sigma^{\mu \nu} P_L \nu_\tau \right]
 \,, \label{eq:Hbc}
 \end{align}
where the effective Wilson coefficients at the $m_S$ scale are given as:
  \begin{align}
  \begin{split}
    C^{\ell}_V & = \frac{\sqrt{2}}{4 G_F V_{cb} }  \frac{y^q_{L3} y^q_{L2} V^\ell_{L\ell \tau} }{ 2 m^2_S}  
  \,, 
  \\
 C^{\ell}_S & =  - \frac{\sqrt{2} }{4 G_F V_{cb} }  \frac{ y^q_{L3} y^u_{R2} V^\ell_{L\ell \tau}}{2m^2_S}  
 \,, 
 \\
 C^{\ell}_T & = \frac{\sqrt{2} }{4G_F V_{cb} } \frac{ y^q_{L3} y^u_{R2} V^\ell_{L\ell \tau}}{8m^2_S} \,.
  \end{split}
\end{align}
We note that since the electron does not mix with the $\mu$ and $\tau$ leptons, the $b\to c e \nu$ process only arises from the SM contribution. In addition, because the LQ contribution to $b\to c \ell \nu$ only involves the tau-neutrino, the induced $b\to c \mu \nu_\tau$ decay does not interfere with the SM contribution.  The effective couplings $C_S^{\ell}$ and $C_T^{\ell}$ at the $m_b$  scale can be obtained from the LQ mass scale via the renormalization group (RG) equations.  Following the results in Ref.~\cite{Dorsner:2013tla},  we obtain $C_S^\ell(m_b)\approx 1.57\, C^\ell_S (m_S)$ and $C_T^\ell(m_b) =0.86 \,C_T^\ell(m_S)$.

To calculate the BRs for the $\bar B \to (D, D^*) \ell \nu$ decays, one requires the hadronic effects for the $B\to D^{(*)}$ transitions.  The parametrization of form factors for different weak currents can be found in Appendix~\ref{app:FF}.  By utilizing these form factors,  the differential decay rate for the $\bar B \to D \ell \bar \nu$ process as a function of the invariant mass $q^2$ of $\ell \nu$ can be expressed as:
\begin{align}
\frac{d\Gamma^{\ell}_D}{dq^2} 
=& 
\frac{G^2_F |V_{cb}|^2   \sqrt{\lambda_D}}{256 \pi^3 m^3_B} \left( 1- \frac{m^2_{\ell}}{q^2}\right)^2 \left[ \frac{2}{3} \left( 2+ \frac{m^2_{\ell}}{q^2}\right) \left( \delta^\ell_\mu |X^{e}_{+}|^2 + |X^{\ell}_+|^2  \right)  \right.
\nonumber \\
& 
+ \frac{2 m^2_{\ell}}{q^2}  \left( \delta^\ell_\mu \left| X^{e}_0\right|^2 + \left|X^{\ell}_0 
+ \frac{\sqrt{q^2}}{m_{\ell}} X^{\ell}_S \right|^2 \right) 
\nonumber \\
& \left.
+ 16  \left(\frac{2}{3}\left( 1+ \frac{2 m^2_{\ell}}{q^2}\right) |X^{\ell}_T|^2-\frac{m_{\ell}}{\sqrt{q^2}} X^{\ell}_T X^{\ell}_0 \right) \right]
\,, \label{eq:diffD}
\end{align}
where $X^{\ell}_{+,0,S,T}$ and $\lambda_M$ are defined as:
 \begin{align}
 \begin{split}
 X^{e, \mu, \tau}_+ & = \sqrt{\lambda_D} F_{+} \left(1, ~ C^\mu_V, ~1 + C^{\tau}_V  \right) \,, 
 \\  
 X^{e,\mu,\tau}_0 &= (m^2_B -m^2_D)  F_0  \left(1, ~ C^\mu_V, ~1+C^{\tau}_V  \right)\,, 
 \\
 X^{\ell}_S & =  (m_B + m_D )  C^{\ell}_S \sqrt{q^2} F_S\,,  
 \\ 
 X^{\ell}_T &= - \frac{\sqrt{q^2 \lambda_D}}{m_B + m_D}  C^{\ell}_T F_T ~, 
 \\
 \lambda_M & = m^4_B + m^4_M + q^4 -2(m^2_B m^2_M + m^2_M q^2 + q^2 m^2_B) \,.
 \end{split}
 \label{eq:Hfunc}
 \end{align}
The $q^2$ dependence of the form factors has been suppressed.

The $\bar B \to D^* \ell \bar\nu_{\ell}$ decay involves $D^*$ polarizations, and the transition form factors are more complicated. Using the parametrization in Eq.~(\ref{eq:ffV}), the differential decay rate after summing all $D^*$ helicities is given by:
 \begin{equation}
  \frac{d\Gamma^{\ell}_{D^*}}{dq^2} 
  =  
  \sum_{h=L,+,-} \frac{d\Gamma^{\ell h}_{D^*}}{dq^2} 
  = 
  \frac{G^2_F |V_{cb}|^2   \sqrt{\lambda_{D^*}}}{256 \pi^3 m^3_B} \left( 1- \frac{m^2_{\ell}}{q^2}\right)^2 
  \sum_{h=L,+,-} V^{\ell h}_{D^*}(q^2)\,,  \label{eq:diffD*}
 \end{equation}
where $\lambda_{D^*}$ can be found in Eq.~(\ref{eq:Hfunc}) and 
\begin{align}
V^{\ell L}_{D^*}(q^2) 
=& 
\frac{2}{3} \left(2+ \frac{m^2_{\ell}}{q^2} \right) \left( \delta^\ell_\mu |h^e_0|^2 + |h^\ell_0|^2  \right)+ \frac{2}{3} \left(1+ 2\frac{m^2_{\ell}}{q^2} \right)|h^{0\ell}_T|^2 
\nonumber \\
& 
+ \frac{2 m^2_{\ell}}{q^2} \lambda_{D^*}  \left( \delta^\ell_\mu \left| X^e_V A_0\right|^2 + \left| X^\ell_V A_0+ \frac{C^{\ell}_S\, q^2\, F_P}{m_{\ell} (m_B + m_{D^*})}  \right|^2  \right)- \frac{16 m_{\ell}}{\sqrt{q}} h^\ell_{0} h^{0\ell}_{T} 
\,, \nonumber \\
V^{\ell \pm}_{D^*}(q^2) 
=& 
\frac{2 q^2 }{3} \left( 2+ \frac{m^2_{\ell}}{q^2}\right) \left( \delta^\ell_\mu |h^e_{\pm}|^2 + |h^\ell_{\pm }|^2 \right) \nonumber \\
& 
+ \frac{32 q^2 }{3} \left( 1 + \frac{2 m^2_{\ell}}{q^2}\right) |h^{\pm\ell}_{T}|^2 - 16 m_{\ell} \sqrt{q^2} h^\ell_{\pm} h^{\pm\ell}_{ T}  
\,.
\end{align}
The quantities $h^\ell_0$, $h^{0\ell}_T$, $h^\ell_\pm$, and $h^{\pm\ell}_T$ are defined by:
\begin{align}
h^{e,\mu,\tau}_0 
&= 
\frac{X^{e,\mu,\tau}_V}{2m_{D^*}} \left( (m^2_B -m^2_{D^*} -q^2) (m_B + m_{D^*}) A_1 - \frac{\lambda_{D^*}}{m_B + m_{D^*}} A_2\right)  
\,, \nonumber \\
h^{0\ell}_T
&= 
\frac{C^{\ell}_T \sqrt{q^2} }{2m_{D^*}} \left( (m^2_B + 3 m^2_{D^*} -q^2) T_2 - \frac{\lambda_{D^*}}{m^2_B - m^2_{D^*}} T_3\right) 
\,, \nonumber \\
h^{e, \mu , \tau}_{\pm} 
&=
X^{e,\mu,\tau}_V \left[(m_B+m_{D^*})A_1 \mp \frac{\sqrt{\lambda_{D^*}} }{m_B+m_{D^*}}V \right]
\,, \nonumber \\
h^{\pm\ell}_{ T} 
& = 
\frac{C^{\ell}_T}{\sqrt{q^2}} \left[  \left( m^2_B - m^2_{D^*} \right)T_2  \pm \sqrt{\lambda_{D^*}} T_1 \right] 
\,,
\end{align}
with $X^{e,\mu,\tau}_V =(1, C^{\mu}_V, 1 + C^{\tau}_{V})$, respectively.  Based on Eqs.~(\ref{eq:diffD}) and (\ref{eq:diffD*}), $R(M)$ ($M=D,D^*$) can be calculated by:
\begin{equation}
R_M 
= 
\frac{\displaystyle \int^{q^2_{\rm max}}_{m^2_\tau }dq^2 ~\frac{d\Gamma^{\tau}_M}{dq^2}}
{\displaystyle \int^{q^2_{\rm max}}_{m^2_{\ell}} dq^2 ~\frac{ d\Gamma^{\ell'}_M}{dq^2}}  
~,
\label{eq:R_M}
\end{equation} 
with $q^2_{\rm max}=(m_B -m_M)^2$ and $\Gamma^{\ell'}_M = (\Gamma^e_M + \Gamma^\mu_M)/2$.

\subsection{New Higgs decays}

Eq.~(\ref{eq:Zp_Z_mixing}) shows that utilizing an additional Higgs doublet to spontaneously break the $U(1)_{\mu-\tau}$ gauge symmetry leads to a strong correlation among $m_{Z_1}$, $g_{Z'}$, and $t_\beta$.  As a result, several processes involving the same set of parameters exhibit distinct behaviors.  In the following, we discuss these interesting processes.

With focus on the scenario with $m_H< m_h/2$ and $m_{Z_1} < 200$~MeV, the new Higgs decay channels $h\to HH$ and $h\to (Z_1 Z_1, Z_1 Z_2)$ become kinematically accessible.  Using the Higgs trilinear and gauge couplings given in Eqs.~(\ref{eq:tri_hHH}) and (\ref{eq:h_V}), the partial decay rates for these channels are obtained as:
\begin{align}
\Gamma(h\to HH) 
& = 
\frac{m_h}{32 \pi} \left( \frac{\xi m^2_h}{v^2}\right) \left( 1+\frac{2m^2_H}{m^2_h}\right)^{2} \sqrt{1-\frac{4 m^2_H}{m^2_h}}
\,,\nonumber \\
& \qquad \mbox{with }
\xi 
= 
s^2_{\beta-\alpha} \left( c_{\beta-\alpha} + t_\beta s_{\beta-\alpha}\right)^2 \left( c_{\beta-\alpha} - t^{-1}_\beta s_{\beta-\alpha} \right)^2
\label{eq:hHH}
\,, \\
\Gamma(h\to Z_1 Z_1) 
& \simeq 
\frac{m_h}{32 \pi} \frac{m^2_h }{ v^2}  \left(s_{\beta-\alpha} - \frac{t^2_\beta-1}{t_\beta } c_{\beta-\alpha}\right)^2 
\,, \nonumber \\
\Gamma(h\to Z_1 Z_2) 
& \simeq \frac{m_h}{16 \pi} \left( \frac{g m_h  }{2 c_W m_{Z_2} }  c_{\beta-\alpha} \right)^2 \left( 1 - \frac{m^2_{Z_2}}{m^2_h}\right)^3
\,. \label{eq:hZ1Z2}
\end{align}
In the decoupling limit when $s_{\beta-\alpha}\to 1$, as required by the current Higgs signal strength measurements, the processes $h\to (HH, Z_1 Z_1)$ can in principle have large decay rates.  Hence, the observed Higgs width $\Gamma_h$ strongly constrains the values of $t_\beta$ and $c_{\beta-\alpha}$. Therefore, from Eq.~(\ref{eq:hHH}),  the condition of  $c_{\beta - \alpha}\sim s_{\beta - \alpha}/t_\beta \ll 1$ has to be satisfied, i.e., a large $t_\beta$ scheme is demanded by data in the model. Interestingly,  when we use a large $t_\beta$ value, the same condition can be used to suppress the partial decay width of $h\to Z_1 Z_1$. Moreover, since $h\to Z_1 Z_2$ does not depend on the  $t_\beta$ parameter, we can use the limit of $\Gamma(h\to Z_1 Z_2)$ as an independent constraint on $c_{\beta-\alpha}$.  Although our analysis does not focus on the search for collider signals,  the percent-level BR for $h\to Z_1 Z_2$ with invisible $Z_1$ decay could be an interesting channel for detecting the new physics.  We note that $c_{\beta-\alpha}\sim 0.1$ is still permissible when considering the constraints from the current measurements of Higgs decays. We will see later that the BRs of new Higgs decay modes can reach the percent level with $c_{\beta-\alpha}\sim 0.05$.

In addition to the flavor-conserving Higgs Yukawa couplings, which are suppressed by $m_\ell /v$ according to  Eq.~(\ref{eq:h_Yu}), there is a tree-level LFV Higgs coupling, i.e., $h\bar \mu_L \tau_R$, where the strength of this LFV coupling is primarily determined by $\chi_{\mu \tau} c_{\beta-\alpha} \sqrt{m_\mu m_\tau} / (s_{2\beta} v)$. The partial decay rate for $h\to \mu \tau$ can thus be written as:
\begin{align}
\Gamma(h\to \mu \tau)
&= 
\frac{m_h}{16 \pi }  \left|  c_{\beta-\alpha} \zeta_{\mu \tau}\right|^2
\,,~ 
\mbox{with }
\zeta_{\mu \tau} 
= 
\frac{\sqrt{m_\mu m_\tau}}{v} \chi_{\mu \tau}  \sqrt{2+t^2_\beta + t^{-2}_\beta}
\,.
\end{align}
 When $c_{\beta-\alpha}$ and $t_\beta$ are determined from the processes $h\to HH/Z_1 Z_2$, the $h\to \mu \tau$ decay rate then depends only on $\chi_{\mu \tau}$.

From Eq.~(\ref{eq:L_N}), it can be seen that the tree-level lepton FCNC arises not only from the Higgs couplings but also from the $Z_i$ couplings.  For a light $Z_1$ gauge boson, the $\tau \to \mu Z_1$ decay can be induced at the tree level and the BR can be obtained as:
\begin{equation}
BR(\tau \to \mu Z_1 )  \simeq \frac{m_\tau  \left(g_{Z'} q_X s_{2\theta_L}  c_{\theta_Z} \right)^2 }{32 \pi \Gamma_\tau} \left(1+ \frac{m^2_\tau }{m^2_{Z_1}}\right) \,,  \label{eq:taumuZ1}
\end{equation}
where we have dropped the $m_{\mu, Z_1}/m_\tau$ factors because $m_{\mu, Z_1}\ll m_\tau$.  The $1/m^2_{Z_1}$ factor in the parentheses from the contribution of the longitudinal component of $Z_1$ will largely enhance the BR as $m_{Z_1}$ is taken at the sub-GeV level.  Since the BR of this decay is mainly determined by $g_{Z'}$, $m_{Z_1}$, and $s_{\theta_L}$, we can use $\tau\to \mu Z_1$ to constrain the $\theta_L$ parameter when $g_{Z'}$ and $m_{Z_1}$ are fixed by other processes.

\subsection{Lepton $(g-2)$'s}

Our model makes additional contributions to the lepton $(g-2)$'s through the mediations of $Z_1$, $H$, and LQ at the one-loop level.  One can neglect the contribution from LQ as it is suppressed by $m^2_\mu/m^2_S$.  Based on the gauge couplings given in Eq.~\eqref{eq:L_N}, although the LFV coupling $\mu \tau Z_1$ can contribute to the muon and tau $(g-2)$'s, its effect is negligible as the coupling is proportional to $g_{Z'} s_{\theta_L}$, where $g_{Z'}$ is of ${\cal O}(10^{-4})$ and $s_{\theta_L}$ is highly constrained by the $\tau \to \mu Z_1$ decay, as argued at the end of last subsection.  On the contrary, the contribution from the light $H$ is through the LFV coupling $\mu \tau H$.  From Eq.~(\ref{eq:h_Yu}), it can be seen that although this coupling is suppressed by a factor of $m_\ell/v$, the factor $1/c_\beta$ can enhance the lepton $(g-2)$'s in the regime of large $t_\beta$ and small $m_H$.

The explicit expressions of the $Z_1$ and $H$ contributions to the lepton $(g-2)$'s are respectively given by:
\begin{align}
\Delta a^{Z_1}_\ell 
= & 
\frac{g^2}{32 \pi^2 c^2_W}   \left( C^{\ell^2}_{1V} +C^{\ell^2}_{1A} \right)  \int^1_0 dx \frac{2 m^2_\ell x^2 (1-x)}{ m^2_{Z_1} (1-x)+m^2_\ell x^2 }
\,, \nonumber\\
& 
-\frac{g^2}{32 \pi^2 c^2_W}    C^{\ell^2}_{1A}   \int^1_0 dx \frac{8 m^2_\ell x (1-x)}{ m^2_{Z_1} (1-x)+m^2_\ell x^2 }
\,, \label{eq:aZ1_mu}
\\
\Delta a^{H}_{\ell'} 
=& 
\frac{m^2_{\ell'}}{8 \pi^2 m^2_H} \left|s_{\beta-\alpha}  \zeta_{\mu\tau}\right|^2
\,, \label{eq:aH}
\end{align}
with $\ell=(e, \mu, \tau)$ and $\ell'=(\mu, \tau)$. Although the couplings of $Z_2$, excluding the SM part, can  contribute to the lepton $(g-2)$'s, the suppression factors of $(s_{\theta_Z}, g_{Z'}) m^2_\ell/m^2_{Z_2}$ make the effects negligible.  We, therefore, disregard the new physics contribution from $Z_2$.

\subsection{Oblique parameters and the $W$ mass}

An important set of precision measurements for constraining new physics comprises the oblique parameters denoted by $S$, $T$, and $U$.  These quantities are related to the loop-induced vacuum polarizations of vector gauge bosons, and their detailed definitions can be found in Refs.~\cite{Peskin:1990zt,Peskin:1991sw}.  In our model, in addition to the SM-like Higgs doublet $H_2$, the oblique parameters receive effects from the extra $SU(2)$ Higgs doublet $H_1$ and the new gauge coupling to $Z'$.  Since we will focus on $g_{Z'} \sim {\cal O}(10^{-4})$, we ignore the $Z'$ contribution and take $m_{Z_2}\simeq m_Z$ in the analysis.  However, a distinctive difference is that the pseudoscalar in the conventional 2HDM becomes the longitudinal component of $Z'$.  Thus, the main contributions running in the loops to the oblique parameters are from $H^\pm$, $h$, and $H$.

To calculate the $S$, $T$, and $U$ parameters in the model, we use the results obtained in Ref.~\cite{Grimus:2008nb}, where the resulting oblique parameters are suitable for the multi-Higgs-doublet models, and even for the models with new singlet charged scalars. Except for the absence of pseudoscalar contributions, the effects from  $H^\pm$, $h$, and $H$ are similar to the conventional 2HDM.  The detailed expressions for the $S$, $T$, and $U$ parameters as functions of the scalar masses and couplings are given in Appendix~\ref{app:STU}.

Using the obtained oblique parameters, the $W$ mass under the influence of new radiative corrections can be expressed as~\cite{Peskin:1991sw,Maksymyk:1993zm,Burgess:1993vc}:
\begin{align}
m_W  
& \equiv 
m^{\rm SM}_W \delta_O 
=
m^{\rm SM}_W \left[ 1 +  \frac{\alpha_{\rm em}}{c^2_W-s^2_W} \left( c^2_W T - \frac{S}{2} + \frac{c^2_W-s^2_W}{4s^2_W} U\right)\right]^{1/2} \,,
\label{eq:mW}
\end{align}
where $m^{\rm SM}_W$ denotes the $W$ mass in the SM, and its relationship with $m_Z$ is defined to be the same as that in the SM, i.e., $m^{\rm SM}_W =m_Z c_W$.  It is worth mentioning that the tree-level $Z'-Z$ mixing can affect the oblique parameters and modify the relation between $m^{\rm SM}_W$ and $m_Z$~\cite{Burgess:1993vc}.  However, since the mixing angle $\theta_Z$ in the model is of ${\cal O}(10^{-5})$ in our study, the effects can be safely ignored.

\section{Numerical Analysis and Discussions} \label{sec:num}

Before conducting a numerical analysis of the physical processes studied in this work, we should first find the viable ranges of new physics parameters in the $U(1)_{\mu-\tau}$-extended model.  For example, as alluded to before, the most influential parameter for the CE$\nu$NS is $m_{Z_1}$ and its cross section can be potentially enhanced by a larger value of $m_{Z_1}$.  The magnitude of $m_{Z_1}$, on the other hand, is proportional to $g_{Z'}$ whose value can be constrained by, e.g., the observed muon $g-2$.  In the following, we start by setting bounds on the parameter space and then make predictions for the CE$\nu$NS cross section, $R(D^{(*)})$, and the oblique parameters and $W$ boson mass.  We will also study the decays of the $Z_1$ and $H$ bosons in the model.

\subsection{Constraints of parameters} \label{sec:limits}

\begin{figure}[phtb]
\begin{center}
\includegraphics[scale=0.45]{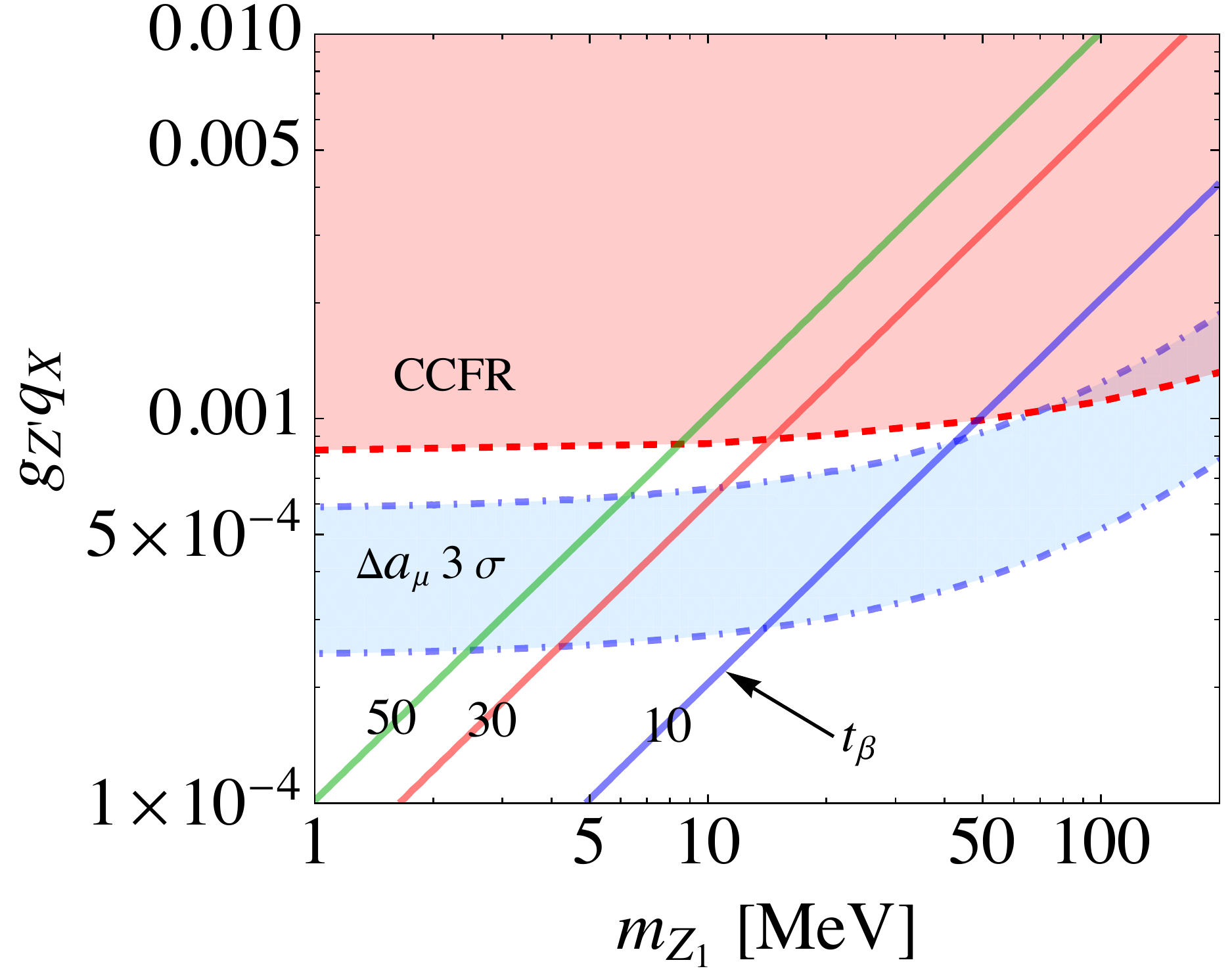}
\caption{Parameter space preferred by the muon $g-2$ (shaded region bounded by blue dot-dashed curves) and ruled out by the CCFR experiment (shaded region above the red dashed curve).  The solid lines represent the contours for $t_\beta$. }
\label{fig:gZp_mZ1}
\end{center}
\end{figure}

The free parameters considered in this study are $m_{H}$, $m_{Z_1}$, $m_S$, $g_{Z'}$, $\chi_{\mu\tau}$, $c_{\beta-\alpha}$, and $t_\beta$, where $\chi_{\mu\tau}$ parametrizes the $\mu-\tau$ mixing effect through $s_{\theta_L}\simeq \chi_{\mu\tau} \sqrt{m_\mu/m_\tau}$ and the $Z'-Z$ mixing is determined by $m_{Z_1}$ and $t_\beta$.  Based on the constraints from the neutrino trident process~\cite{Altmannshofer:2014pba}, measured by CCFR~\cite{CCFR:1991lpl}, and the $4\mu$ final states in the BaBar experiment~\cite{BaBar:2016sci}, we can conservatively take the bounds of $g_{Z'} q_X\lesssim 1.3\times 10^{-3}$ and  $m_{Z_1}<200$~MeV.  According to Eq.~(\ref{eq:aZ1_mu}), the $Z_1$ boson makes an important contribution to the muon $g-2$.  Therefore, we show in Fig.~\ref{fig:gZp_mZ1} the CCFR bound~\cite{Altmannshofer:2014pba} and the $\pm 3\sigma$ contours (blue dot-dashed curves) of the measured muon $g-2$ in the $m_{Z_1}$-$g_{Z'} q_X$ plane, where the shaded region above the red dashed curve is ruled out by the CCFR experiment.  Although $\Delta a^{Z_1}_\mu$ depends on $t_\beta$ via the $Z'-Z$ mixing, its effect is negligibly small because $s_\theta\sim {\cal O}(10^{-5})$ in the considered range of $m_{Z_1}$.  As a result, the electron $g-2$ mediated by $Z_1$ and induced through $Z'-Z$ mixing is estimated to be $\Delta a^{Z_1}_e \approx -1.4 \times 10^{-16}$, completely negligible.  We will show later that due to the small lepton flavor mixing, as constrained by other processes, the effect mediated by $H$ for the lepton $g-2$ is also highly suppressed.  In the model, $m_{Z_1}$ and $g_{Z'} q_X$ are not independent parameters and are related by $m_{Z_1}= 2 v t_\beta g_{Z'} q_X / ( 1+t^2_\beta)$.  In Fig.~\ref{fig:gZp_mZ1}, we also show contours of $t_\beta$ using solid lines.  The large $t_\beta$ scheme, as required to restrict the $h\to HH$ and $h\to Z_1 Z_1$ rates, to be discussed in more detail below, further narrows down the preferred $m_{Z_1}$ range.

The SM prediction for the Higgs boson width is $\Gamma^{\rm SM}_h\approx 4.1$~MeV~\cite{LHCHiggsCrossSectionWorkingGroup:2016ypw}, while the current measurement gives $\Gamma^{\rm exp}_{h}= 3.2^{+2.8}_{-2.2}$~MeV~\cite{PDG2022}.  As an illustrated example, we assume that each new Higgs decay channel in the model contributes less than $5\%$ of $\Gamma^{\rm SM}_h$, i.e., $\Gamma^{\rm NP}_h \le 0.20$~MeV.  This assumption is consistent with the current upper limit on the Higgs invisible decays, $BR(h\to \text{invisible}) < 0.19$~\cite{PDG2022}.  To fit the observed Higgs signal strengths, the Higgs couplings to the fermions and the $W^\pm$ and $Z$ gauge bosons should have $s_{\beta-\alpha}\approx 1$.

\begin{figure}[phtb]
\begin{center}
\includegraphics[scale=0.45]{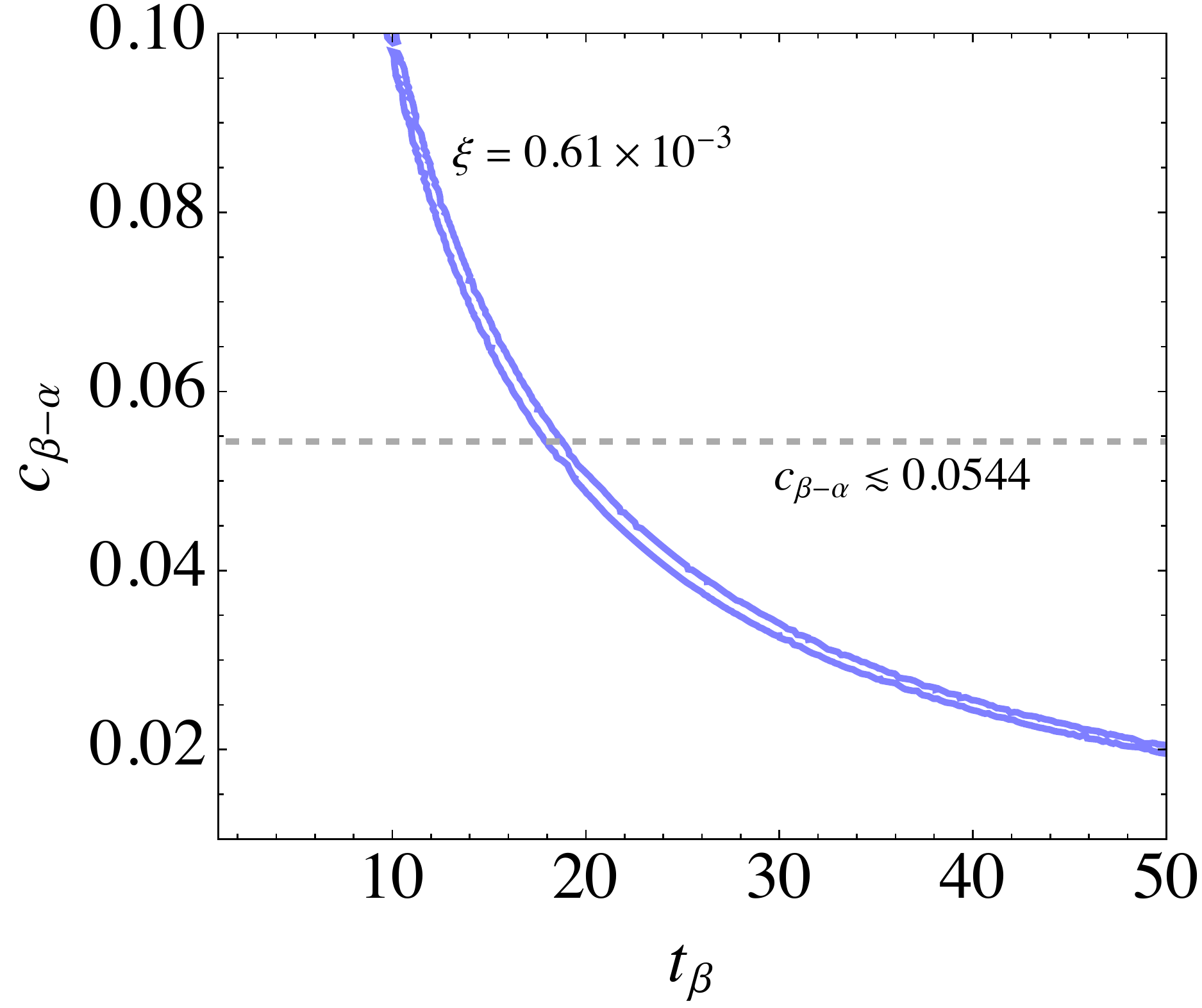}
 \caption{Contours of $\xi$ (blue solid curves) in the $t_\beta$-$c_{\beta-\alpha}$ plane, assuming $\xi=0.61\times 10^{-3}$ and $\Gamma(h\to HH)=0.20$~MeV.  The dashed line denotes the upper bound on $c_{\beta-\alpha}$ from $\Gamma(h\to Z_1Z_2) \le 0.20$~MeV.}
\label{fig:xi}
\end{center}
\end{figure}

We now use $\Gamma(h\to HH)$ to bound $c_{\beta-\alpha}$ and $t_\beta$.  Since the $h\to HH$ process depends on $m_H$, we show the upper bound on $\xi$, defined in Eq.~(\ref{eq:hHH}), for some benchmarks of $m_H$:
\begin{equation}
\xi \lesssim \frac{10^{-3} \Gamma(h\to HH) }{\text{0.20 MeV}}  \times \left\{  \begin{array}{cc}
0.59 & \text{$m_H=30$ GeV} \,, \\
0.61  & \text{$m_H=50$ GeV} \,,  \\    
1.07 & \text{$m_H=60$ GeV} \,.
\end{array}
\right.
\end{equation} 
To illustrate the dependence of $\xi$ on $c_{\beta-\alpha}$ and $t_\beta$, we show in Fig.~\ref{fig:xi} the contour plot of $\xi$ in the $t_\beta$-$c_{\beta - \alpha}$ plane, where we have fixed $\xi=0.61\times 10^{-3}$ and $\Gamma(h\to HH)=0.20$~MeV.  It is found that there are two slightly separated contours, which are insensitive to the chosen value of $\xi$ and indicate that $c_{\beta-\alpha}$ decreases as $t_\beta$ increases.  With the choice of $t_\beta=25$, $m_H=50$~GeV, and $\xi=0.61\times 10^{-3}$, we obtain $c_{\beta-\alpha}\approx 4.095\%$ and $s_{\beta-\alpha} \approx 99.92\%$.  The values in turn determine that $\Gamma(h\to Z_1 Z_1)\approx 0.17$~MeV  and $\Gamma(h\to Z_1 Z_2)\approx 0.11$~MeV.

In the large $t_\beta$ scheme, $\Gamma(h\to Z_1 Z_2)$ only depends on $c_{\beta-\alpha}$.  With $m_h=125$~GeV and $m_{Z_2}=91.187$~GeV, the limit of $c_{\beta-\alpha}$ can be determined as:
 \begin{equation}
 c_{\beta-\alpha} \lesssim 0.0544 \left(\frac{\Gamma(h\to Z_1 Z_2)}{\text{0.20 MeV}} \right)^{1/2}\,.
 \end{equation}
The assumption of $\Gamma(h\to Z_1Z_2) \le 0.20$ MeV then translates into the dashed line in Fig.~\ref{fig:xi}.  According to the result in Eq.~(\ref{eq:Zp_Z_mixing}), we can estimate the $Z'-Z$ mixing angle to have:
 \begin{equation}
 |s_{\theta_Z}| \approx 1.08 \times 10^{-5} \left(\frac{20}{t_\beta} \right) \left(\frac{m_{Z_1}}{\text{20 MeV}} \right)\,. 
 \end{equation}
Clearly, $s_{\theta_Z}$ can be  larger than the loop-induced kinetic mixing between $Z'$ and $\gamma$, characterized by the mixing parameter~\cite{Holdom:1985ag,Chen:2017cic}:
  \begin{equation}
  \epsilon = \frac{g_{Z'} e}{6 \pi^2} \ln\frac{m_\tau}{m_\mu} \approx 8.68\times 10^{-6} \left(\frac{g_{Z'}}{6 \times 10^{-4}} \right)\,.
  \end{equation}
Consequently, we concentrate on the contributions from the $Z'-Z$ mixing in this study.

The $\chi_{\mu \tau}$ parameter contributes to $h\to \mu \tau$, $\tau\to \mu Z_1$, and $\Delta a^H_\ell$.  Since the $\tau\to \mu Z_1$ process is strongly enhanced by the factor of $m^2_\tau/m^2_{Z_1}$, its measurement will put a strict constraint on $\chi_{\mu \tau}$.  
To bound the $\chi_{\mu\tau}$ parameter using available data, we can use the upper limit of the process $\tau\to \mu + \text{light boson}$ as an estimate, where the current data give $BR(\tau \to \mu + \text{light boson}) < 5 \times 10^{-3}$~\cite{PDG2022, Belle-II:2022heu}.  With $c_{\theta_L}\approx c_{\theta_{Z}}\approx 1$ and the result in Eq.~(\ref{eq:taumuZ1}),  we obtain an upper bound on  $\chi_{\mu\tau}$ as:
 \begin{equation}
 \chi_{\mu \tau} < 1.82 \times 10^{-5} \left(\frac{25}{t_\beta} \right) \,. 
 \end{equation}
The resulting $BR(h\to \mu \tau)$ and $\Delta a^{H}_\mu$ are then less than ${\cal O}(10^{-11})$ and ${\cal O}(10^{-16})$, respectively.

The primary purpose of introducing the scalar LQ, $S^{\frac{1}{3}}$, in the model is to address the $R(D^{(*)})$ anomalies.  Along with the mass of LQ, the related parameters are $y^{q}_{L3, L2}$, $y^u_{R2}$, and $V^\ell_{L\ell \tau}$.  Due to the $\tau \to \mu Z_1$ constraint, the lepton flavor mixing matrix can be approximated as $V^\ell_{L}\approx \mathbbm{1}$, allowing us to ignore its contribution to the muon mode.  Consequently, the LQ only couples to the third-generation leptons.  According to Eq.~(\ref{eq:LQ_C}), unlike the independent couplings to the different up-type quarks, the LQ couplings to the different down-type quarks are related by the CKM matrix and can be written as 
 \begin{align}
 (V^T_{\rm CKM} {\bf y}^q_L)_{d} 
 & \approx 
 \frac{10}{3} \lambda^4 y^q_{L3}  -\lambda y^q_{L2} + y^q_{L1} 
 \,, \nonumber \\
 (V^T_{\rm CKM} {\bf y}^q_L)_{s}  
 & \approx 
 -\frac{4}{5} \lambda^2 y^q_{L3} + y^q_{L2} + \lambda y^q_{L1}
 \,, \nonumber \\
 (V^T_{\rm CKM} {\bf y}^q_L)_{b}
 & \approx 
 y^q_{L3}
 \,, \label{eq:VydL}
 \end{align}
where $\lambda\approx 0.2257$ is a Wolfenstein parameter and $V_{ub}\ll V_{cb}\ll V_{tb}\approx V_{cs}\approx V_{ud}\approx 1$ has been applied. 
To suppress the LQ couplings to the first- and second-generation quarks so as to satisfy constraints from low-energy physics, such as $P-\bar P$ mixing and $q_i\to q_j \bar f' f'$, where $P$ and $f'$ are respectively possible neutral mesons and leptons, we require the Yukawa couplings to have the hierarchy:
 \begin{align}
 y^{q}_{L3} \sim {\cal O}(1)
 \,, ~ 
 y^{q}_{L2} \sim {\cal O}(\lambda^2)
 \,, ~ 
 y^{q}_{L1} \sim {\cal O}(\lambda^3)
 ~. 
 \end{align}
 If cancellations are allowed in the terms of $ (V^T_{\rm CKM} {\bf y}^q_L)_{d,s}$, small LQ couplings to the first two generations of down-type quarks can be easily achieved in the model.  Although $D-\bar D$ mixing can constrain $y^u_{R2} y^u_{R1}$, we can take a small $y^u_{R1}$ to avoid this constraint on $|y^u_{R2}|$, for which we need $y^u_{R2}\sim {\cal O}(0.5)$ to enhance $R(D^{(*)})$.

In this model, the LQ couplings to the third-generation quarks are dominant.  Both CMS~\cite{CMS:2020wzx} and ATLAS~\cite{ATLAS:2021oiz} have searched for the scalar LQ with an electric charge of $e/3$ using the $t\tau$ and $b\nu$ production channels.  ATLAS has placed a stronger upper bound on the LQ mass when $BR(S^{-1/3}\to t \tau)=1/2$, obtaining $m_{S} \ge 1.22$~TeV.  If we set $y^u_{R3} = 0$, the ATLAS measurement can be directly applied to our model, and $t\tau$ and $b\nu_\tau$ thus become the dominant decays of the LQ.  To be more conservative, we use $m_{S}=1.5$~TeV in our numerical calculations.

\subsection{Phenomenological analysis}

Here, we present the numerical results of the observables discussed in Sec.~\ref{sec:pheno} and highlight their features while taking into account the constrained parameter space obtained in Sec.~\ref{sec:limits}.

\subsubsection{Cross sections of CE$\nu$NS on Ar and CsI targets}

Since the targets of the measured CE$\nu$NS in the COHERENT experiment are Ar and CsI, we focus on both targets in the following numerical analysis. Because CsI is a compound of cesium and iodide,  the fraction of each nucleus contributing to the cross section is defined by $f_i = A_i/(A_{\rm Cs} + A_{\rm Ar})$~\cite{AristizabalSierra:2019zmy}.  Based on COHERENT's best-fit results for $\langle \sigma\rangle_e$ and $\langle \sigma \rangle_{\mu+\bar\mu}$~\cite{COHERENT:2021xmm}, where the resulting $\langle \sigma \rangle_{\mu+\bar\mu}$ is noticeably smaller than the SM prediction, we choose to present the numerical results with sign$(\theta_Z)= -1$.

To calculate the cross section of CE$\nu$NS for Ar and CsI, the quantities involved in Eq.~(\ref{eq:cross_avg}) are taken as follows: The weak mixing angle is $s^2_W=0.23112$,  the number of the protons and neutrons in $^{40}$Ar, $^{127}$I, and $^{133}$Cs are set to be $(Z, N)_{\rm Ar}=(18, 22)$, $(Z, N)_{\rm I}=(53,74)$, and $(Z, N)_{\rm Cs}=(55,75)$, respectively, and the masses of the nuclei are $m_{\rm Ar}=37.20$~GeV, $m_{\rm I}=118.24$~GeV, and $m_{\rm Cs}=123.86$ GeV. The energy of the prompt $\nu_\mu$ is determined from the $\pi^+$ decay at rest. With $m_\mu=105.65$~MeV and $m_\pi=139.57$~MeV, we obtain $E_{\nu_\mu}\simeq 29.80$~MeV. By neglecting the electron mass, the maximum energy of $\nu_e$ and $\bar\nu_\mu$ from the $\mu^+$ decay is $E^{\rm max}_{\nu_e, \bar\nu_\mu}= m_\mu/2 \simeq 52.8$~MeV. 

As mentioned in the Introduction, the difficulty in measuring the CE$\nu$NS is due to the small nuclear recoil energy (RE). We can estimate the maximum RE of the nuclear targets, argon, iodine, and cesium, by incident $\nu_\mu$ with the energy of  $29.80$~MeV as $E^{\rm max, \nu_\mu}_r=(47.66,\, 15.01,\, 14.33)$~keV, respectively. The maximum RE of (Ar, I, Cs) from $\bar\nu_\mu$ or $\nu_e$ with the maximum incident energy of $52.8$~MeV is given by $E^{\rm max, \bar\nu_\mu(\nu_e)}_r=(149.46,\, 47.11,\, 44.98)$~keV. The nuclear threshold RE in the COHERENT experiment for (Ar, CsI) is $(20, 6.5)$~keV~\cite{Scholberg:2018vwg}. Using $E_\nu \approx \sqrt{m_T E_r/2}$, the minimum neutrino energy of producing the threshold RE for Ar and CsI can be estimated to be $E^{\rm min}_\nu \sim 19$ MeV. If we apply this $E^{\rm min}_\nu$ to Eq.~(\ref{eq:cross_avg}), it is found that the total cross section of CE$\nu$NS will be reduced by $\sim 2.4\%$, which is the same as the uncertainty from the nuclear form factor. Due to the fact that $E_{\nu_\mu}\simeq 29.80$~MeV, the kinematic cut of $E^{\rm min}_{\nu} \sim 19$ MeV does not influence the $\nu_\mu$ scattering. Additionally, according to neutrino fluxes shown in Eq.~(\ref{eq:nu_flux}), $E_{\nu} \lesssim 19$~MeV locates at the front tail of the $\nu_e$ and $\bar\nu_\mu$ fluxes, where the contributions from this region are much smaller than those from 19~MeV to 52.80~MeV. Since our purpose is to demonstrate the sensitivity to the new physics effects, for simplicity, we do not consider the kinematic cut based on the experimental conditions. The detailed event analysis based on the experimental setup can be found in Ref.~\cite{AtzoriCorona:2022moj}.

\begin{figure}[phtb]
\begin{center}
\includegraphics[scale=0.4]{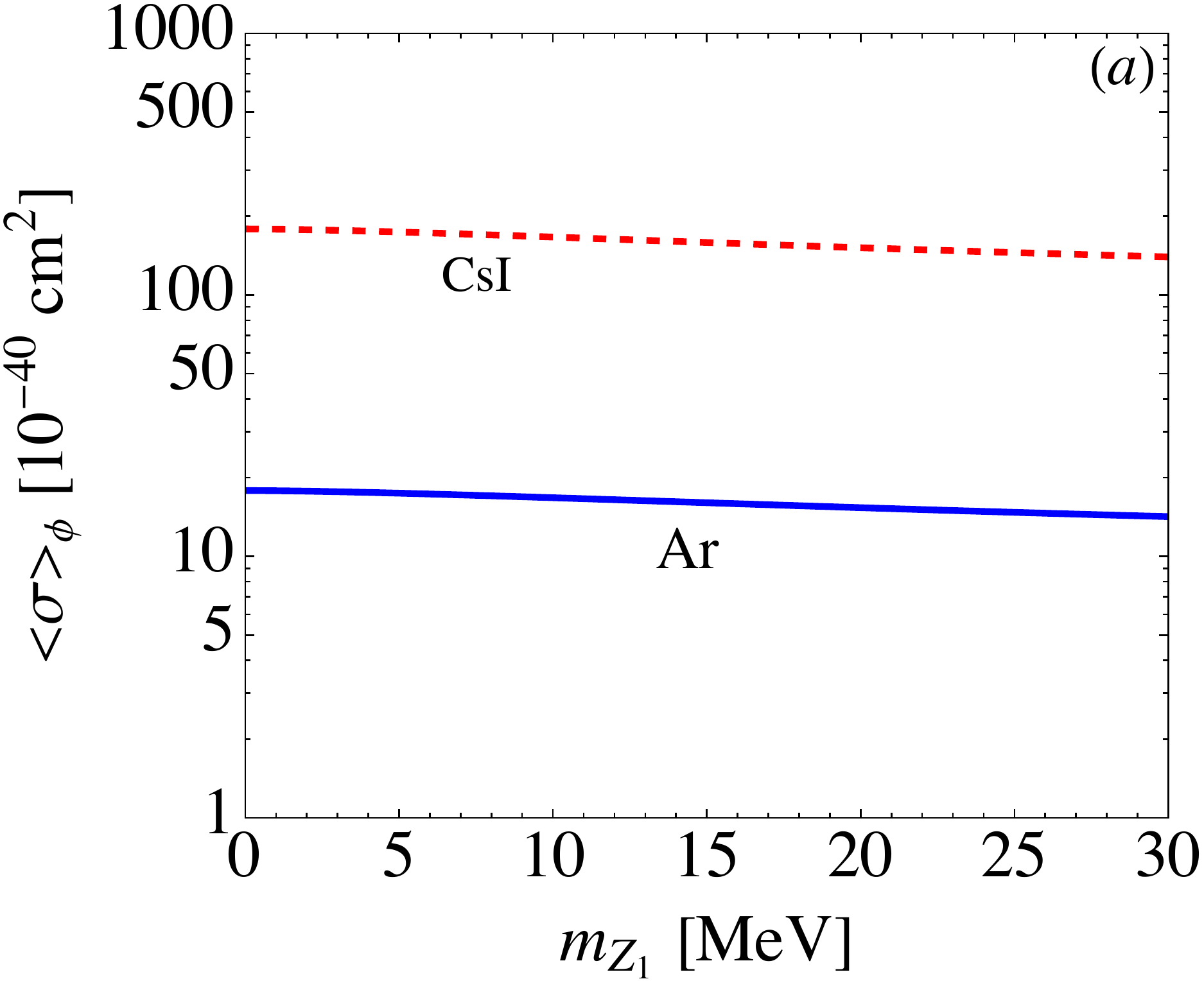}
\hspace{5mm}
\includegraphics[scale=0.41]{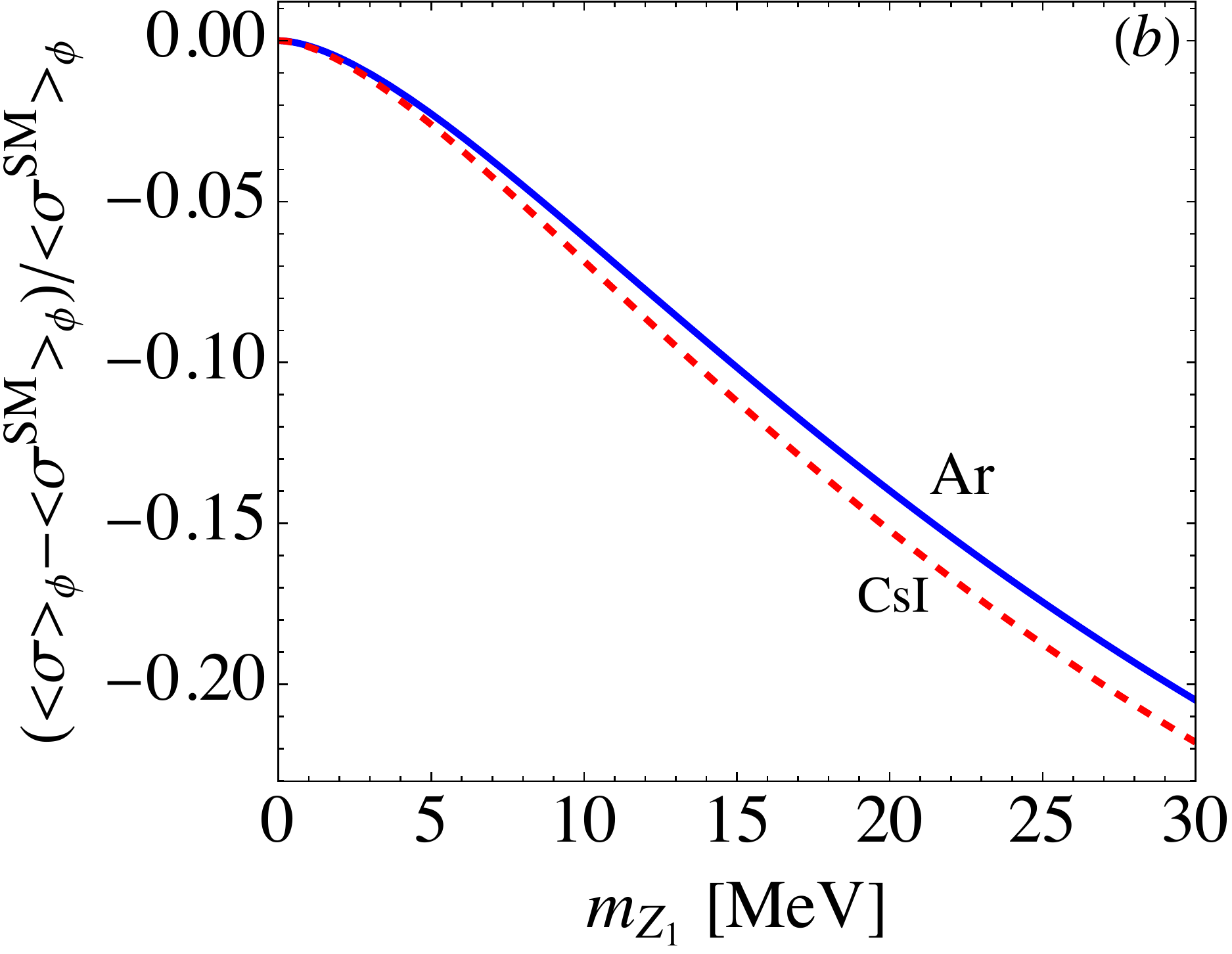}
\caption{(a) Cross section averaged by neutrino fluxes for Ar and CsI targets as a function of $m_{Z_1}$, where the points for $m_{Z_1}=0$ correspond to the SM results.  (b) Fractional deviation on the total cross section $\langle \sigma \rangle_\phi$ from its SM value as a function of $m_{Z_1}$.  In both plots, the solid and dashed curves represent the results for Ar and CsI targets, respectively.   }
\label{fig:sigma_tot}
\end{center}
\end{figure}

Using Eq.~(\ref{eq:cross_avg}), we show the total cross section of CE$\nu$NS for Ar (solid) and CsI (dashed) as a function of $m_{Z_1}$ in Fig.~\ref{fig:sigma_tot}(a).  We estimate the SM results for Ar and CsI to be $18.2 \times 10^{-40}$ cm$^2$ and $183.12\times 10^{-40}$ cm$^2$, respectively.  Since the cross section is plotted in the logarithmic scale, the sensitivity in $m_{Z_1}$ is not obvious.  To illustrate the new physics effects, we show the deviation from the SM result, defined by ($\langle \sigma^{\rm NP+SM} \rangle_\phi- \langle \sigma^{\rm SM} \rangle_\phi)/\langle \sigma^{\rm SM} \rangle_\phi$, in Fig.~\ref{fig:sigma_tot}(b).  It can be seen that the influence of new physics can exceed $10\%$ when $m_{Z_1}\gtrsim 12$~MeV, with a slightly larger influence on CsI than on Ar.

\begin{figure}[phtb]
\begin{center}
\includegraphics[scale=0.40]{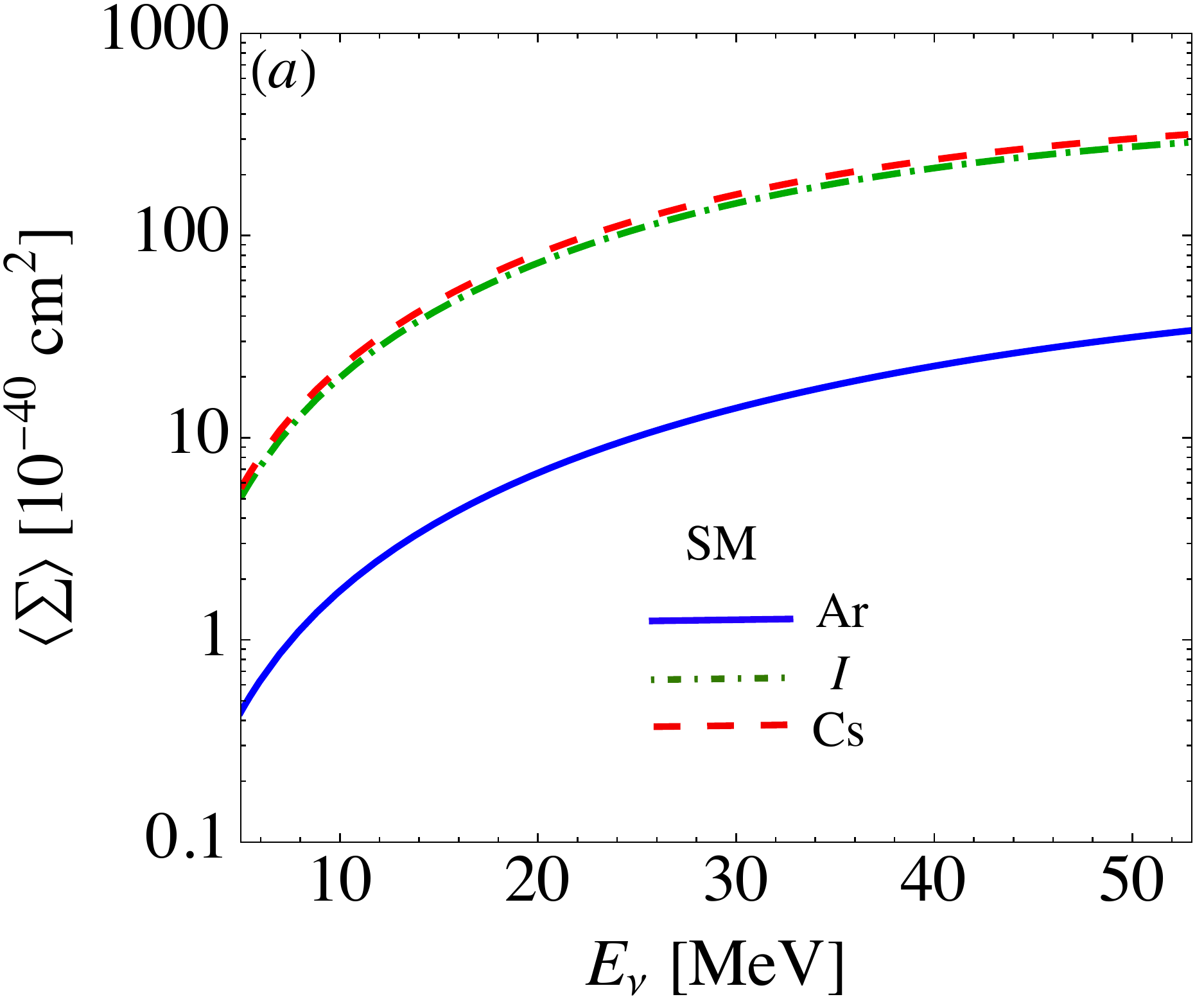}
\hspace{8mm}
\includegraphics[scale=0.37]{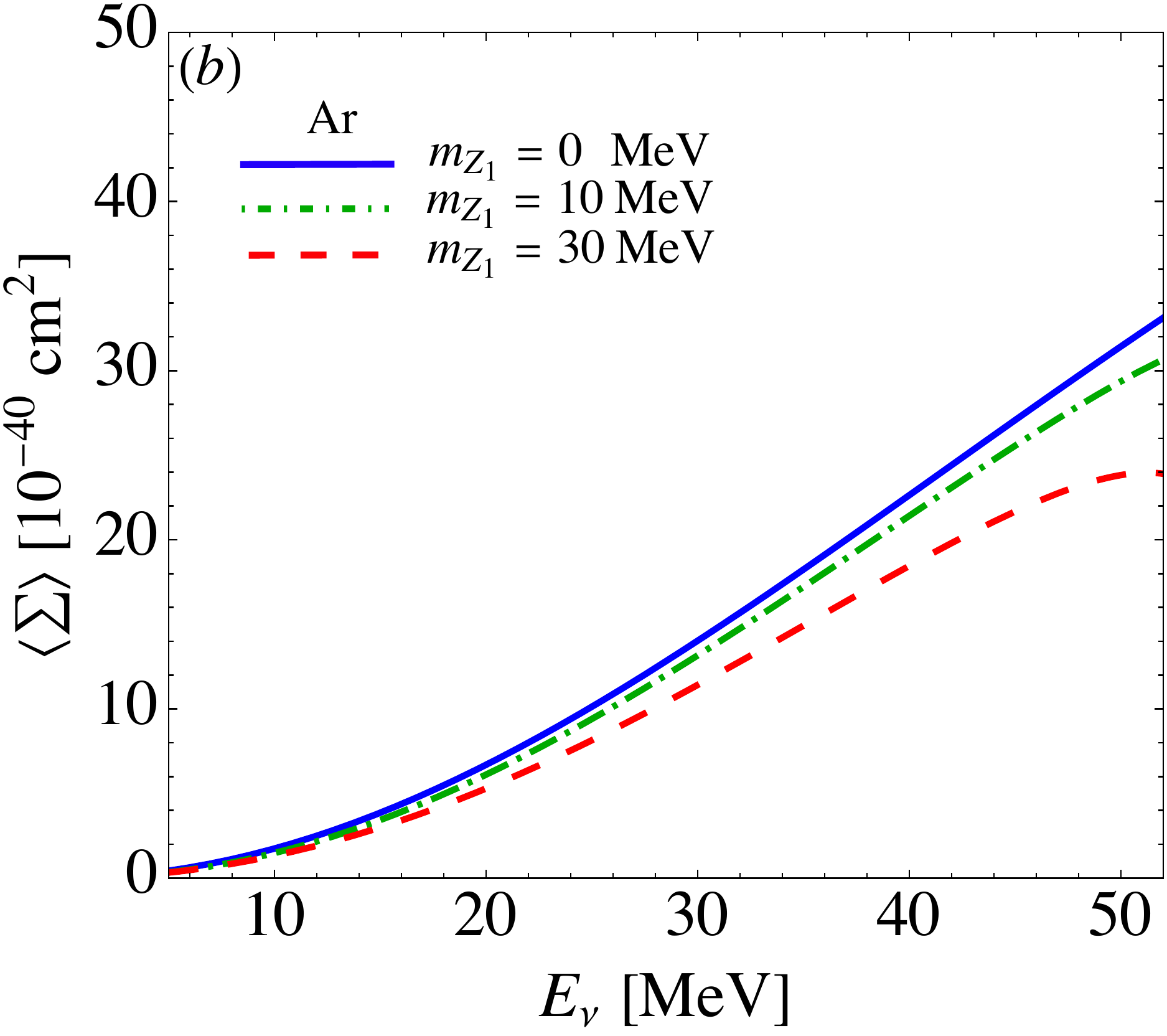} \vspace{2mm} 
\\
\hspace{2mm}\includegraphics[scale=0.38]{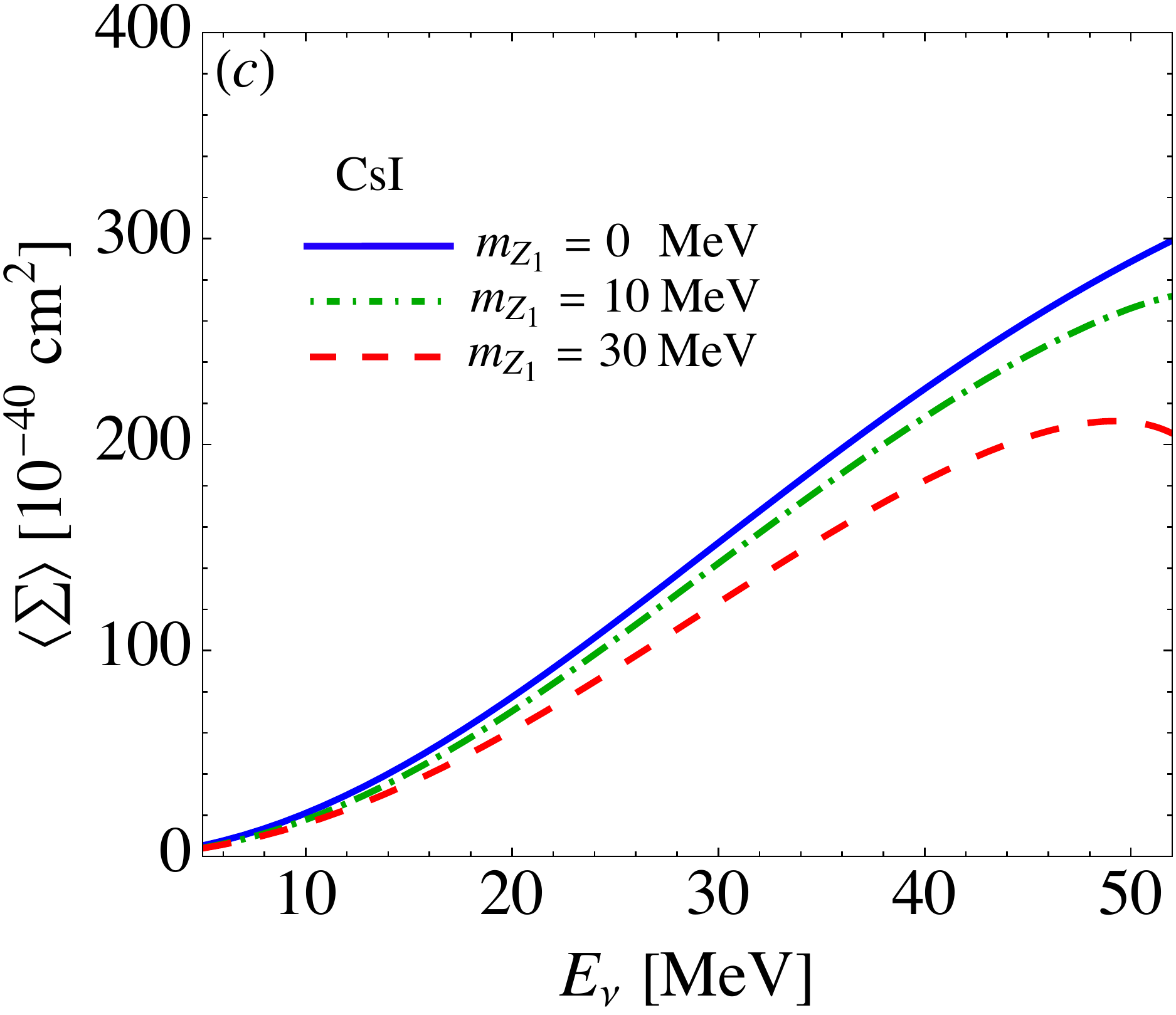}
\hspace{3mm}
\includegraphics[scale=0.40]{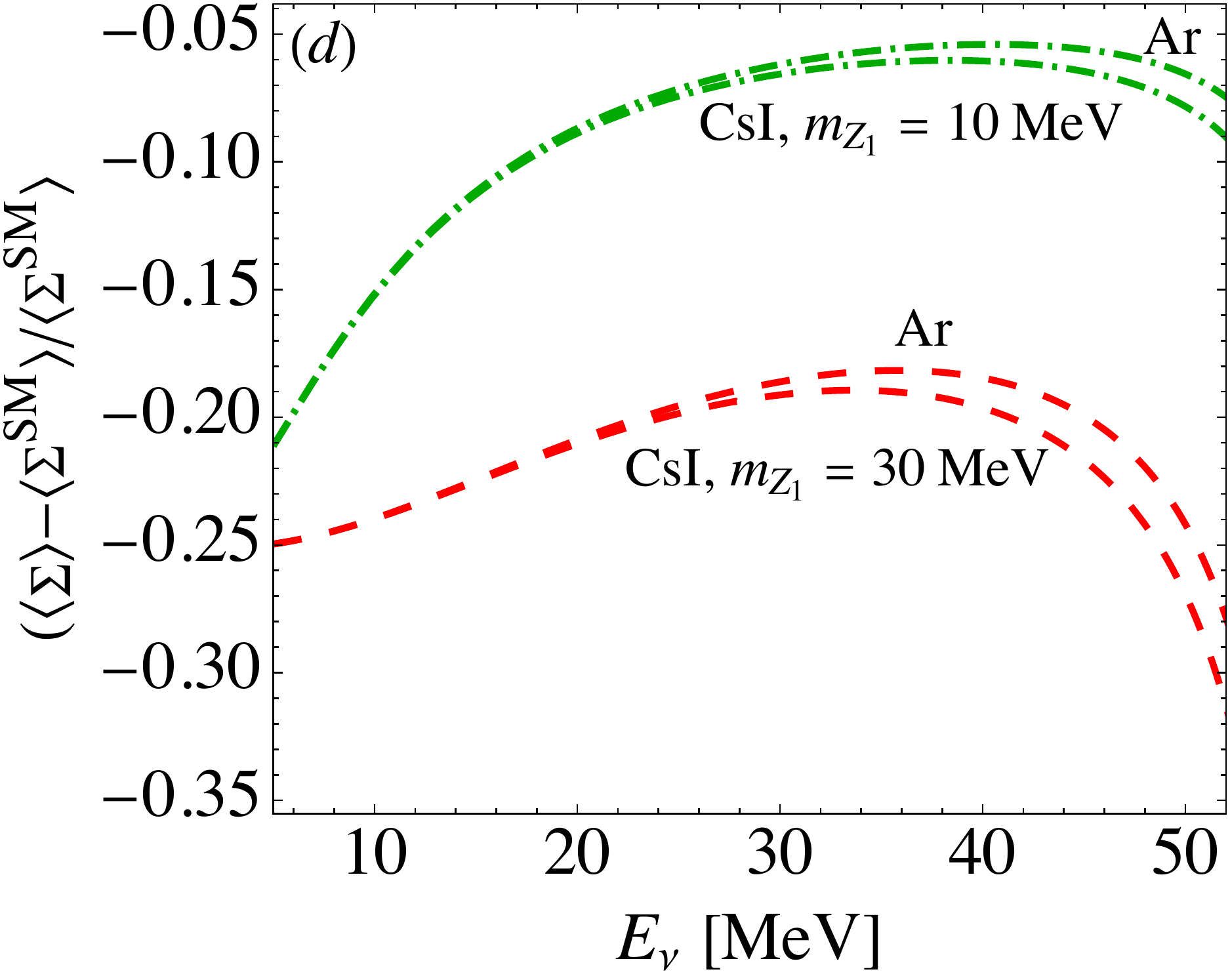}
 \caption{(a) $\langle \Sigma\rangle$ as a function of $E_\nu$ for Ar (solid), I (dot-dashed), and Cs (dashed) in the SM.  $\langle \Sigma\rangle$ for (b) Ar and (c) CsI with $m_{Z_1} = 0$~MeV (solid), $10$~MeV (dot-dashed), and $30$~MeV (dashed).  (d) Sensitivity of $\langle \Sigma\rangle$ on the $Z'$ boson for Ar and CsI with $m_{Z_1}=10$~MeV (dot-dashed) and $30$~MeV (dashed). }
\label{fig:diffEnu}
\end{center}
\end{figure}

In addition to the total cross section of CE$\nu$NS, the cross section at specific incident neutrino energy $E_\nu$ serves as another useful physical observable for probing the new physics effects.  For clarity, we define the averaged total cross section as a function of $E_\nu$ as follows:
  \begin{align}
  \langle \Sigma \rangle 
  & = \frac{1}{ \Phi(E_\nu)} 
  \sum_{\ell=e,\mu,\bar{\mu}} 
  \int^{E^{\rm max}_{r}}_{E^{\rm min}_r} dE_r
  \frac{d\sigma(\nu_\ell A \to \nu_\ell A)}{d E_r} \frac{d\phi_\ell(E_{\nu})}{dE_{\nu}}\,, \nonumber \\
  \Phi(E_\nu) & = \sum_{\ell=e,\mu,\bar\mu} \frac{d\phi_\ell (E_\nu)}{dE_{\nu}}\,.
  \end{align}
In Fig.~\ref{fig:diffEnu}(a),  we show $\langle \Sigma \rangle$ as a function of $E_\nu$ in the SM for the targets of Ar, I, and Cs by the solid, dot-dashed, and dashed curves, respectively.  To demonstrate the sensitivity of  $\langle \Sigma \rangle$  to the new physics effects, we present the results for Ar and CsI in Figs.~\ref{fig:diffEnu}(b) and (c), respectively, where the solid, dot-dashed, and dashed curves denote cases with $m_{Z_1}=(0, 10, 30)$~MeV.  It can be seen that the deviation from the SM increases with $m_{Z_1}$.  To illustrate the sensitivity of $\langle \Sigma \rangle$ on the $Z'$ mass, we exhibit $\delta \langle \Sigma\rangle=(\langle \Sigma^{\rm NP+SM} \rangle-\langle \Sigma^{\rm SM} \rangle)/\langle \Sigma^{\rm SM} \rangle$ in Fig.~\ref{fig:diffEnu}(d) for Ar and CsI, where the dot-dashed and dashed curves are for $m_{Z_1} = 10$ and $30$~MeV.  From the results, we find that the  sensitivity level $|\delta\langle \Sigma \rangle|$ first decreases with $E_\nu$ and then turns to increase with $E_\nu$ at some higher $E_\nu$, e.g., at $E_\nu\sim (41, 36)$~MeV for $m_{Z_1}=(10, 30)$~MeV.  Hence, the deviation from the SM result can reach $\sim 11\%$ ($22\%$) at $E_\nu=15$~MeV and $\sim 7\%$ ($25\%$) at $E_\nu=50$~MeV for $m_{Z_1}= 10~ (30)$~MeV.

\begin{figure}[phtb]
\begin{center}
\includegraphics[scale=0.4]{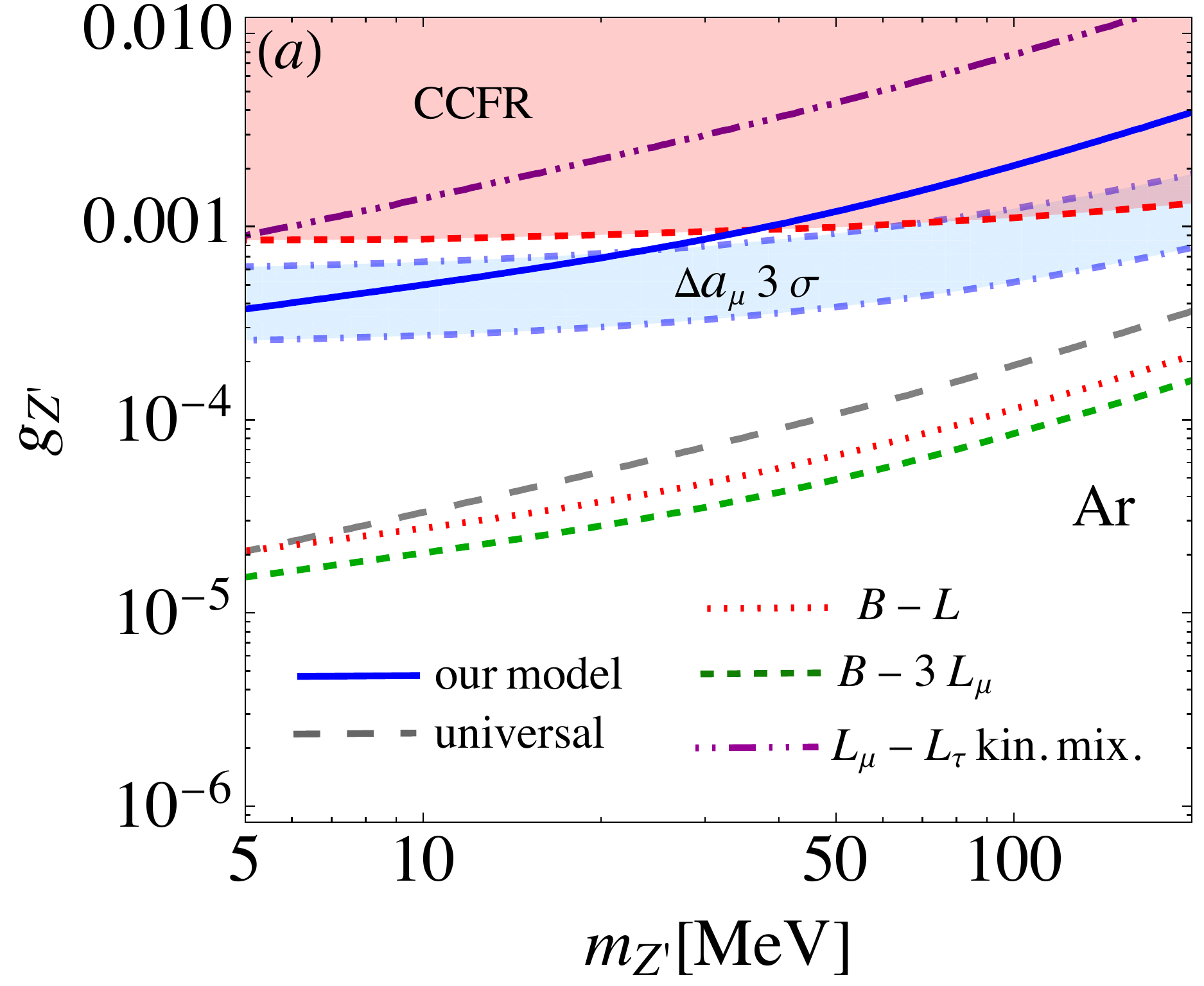}
\hspace{5mm}
\includegraphics[scale=0.4]{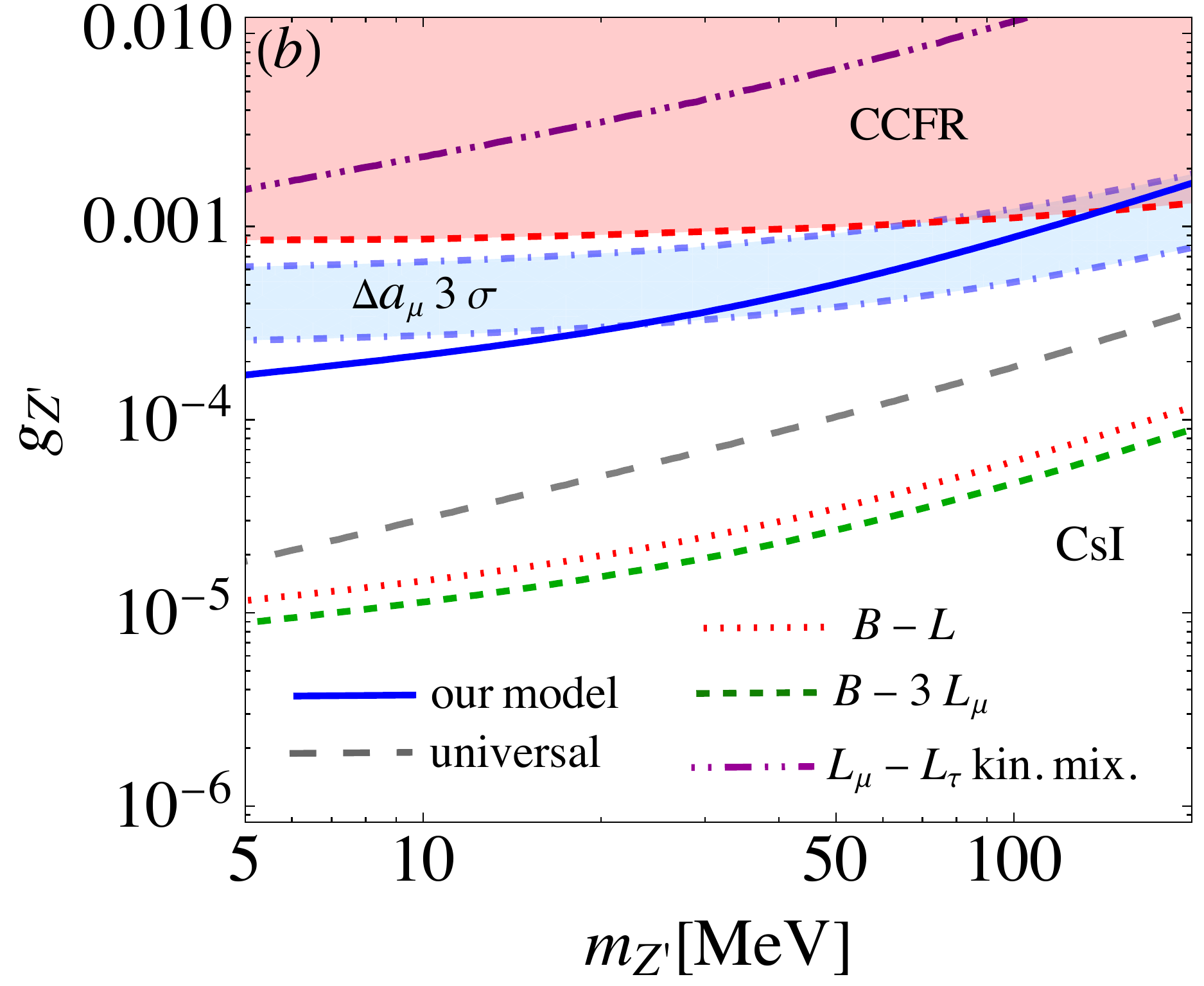}
 \caption{Selected $U(1)$ gauged models contributing to the flux-averaged cross section $\langle \sigma\rangle_\phi$ as functions of $g_{Z'}$ and $m_{Z'}$ for the (a) Ar and (b) CsI targets, where we have taken the upper bounds of $\langle \sigma \rangle_{\phi} = (29, 200) \times 10^{-40}$ cm$^2$ for Ar and CsI, respectively.}
\label{fig:Models}
\end{center}
\end{figure}

As stated in the Introduction, a light $Z'$ gauge boson can be realized by a variety of local $U(1)$ gauge symmetries. The gauged $U(1)$ symmetries can be classified as $U(1)_{X_q- \sum_\ell c_\ell X_\ell}$, where $X_{q, \ell}$ denote the $U(1)$ charges of quark and lepton, respectively. Since the experiments from the searches of visible dark photons place strict constraints on $g_{Z'}$ and $m_{Z'}$, not all $U(1)$ models are of interest in the study. To illustrate the contributions from different gauged $U(1)$ symmetries to CE$\nu$NS, we consider the potential models, including universal, $B-L$, $B-L_e-2L_\mu$, and $L_\mu-L_\tau$ with kinetic mixing, from the model listed in Ref.~\cite{AtzoriCorona:2022moj}, where the charge assignments of the selected $U(1)$'s are given in Table~\ref{tab:Uonep}.  Using the central value of data along with $1\sigma$ errors as the upper bound for CE$\nu$NS, the flux-averaged cross section $\langle \sigma\rangle_\phi$ for the selected $U(1)$ models as a function of $g_{Z'}$ and $m_{Z'}$ is shown in Fig.~\ref{fig:Models}, where the solid, long dashed, dotted, dashed, and dot-dashed curves represent the results from our model, universal, $B-L$, $B-L_e-2L_\mu$, and $L_\mu -L_\tau$ with kinetic mixing, respectively. It can be seen that in the mass region of $10 \leq m_{Z'} \leq 100$~MeV, our model can fit better the constraint from CCFR and the observed muon $g-2$.

 \begin{table}[htbp]
   \caption{ Charge assignments of the selected new $U(1)$ gauged models~\cite{AtzoriCorona:2022moj}. }
  \label{tab:Uonep}
   \begin{tabular}{c|ccccc} \hline \hline
   Model &~  universal~ & ~$B-L$ ~&~ $B-3L_\mu$ ~& ~$B-L_e-2L_\mu$ ~& ~$L_{\mu-\tau}$~   \\ \hline 
   $X_u$ & 1 & 1/3 &  1/3   & 1/3    &  0\\ \hline 
   $X_d$ & 1 & 1/3 &  1/3    & 1/3    & 0\\ \hline 
   $X_e$  & 1& $-1$ & 0      & $-1$  & 0 \\ \hline 
   $X_\mu$ & 1 & $-1$ & $-3$ & $-2$ &  $1$\\ \hline 
   $X_\tau$ & 1& $-1$ & 0       & 0  & $-1$\\ \hline \hline 
     \end{tabular}
\end{table}

\subsubsection{$R(D)$ and $D(D^*)$ mediated by LQ}

The calculations of $R(D)$ and $R(D^*)$ depend on the form factors of the $B\to (D, D^*)$ transitions. In this study, we use the form factors given in Ref.~\cite{Bernlochner:2017jka}, obtained using the heavy quark effective theory (HQET). With the input values of $m_{B^+}=5.28~{\rm GeV}$, $m_{D^0}=1.864~ {\rm GeV}$, $m_{D^{0*}}=2.007~ {\rm GeV}$, $\tau_{B^-}=2.450\times 10^{12}~ {\rm GeV}^{-1}$, and $V_{ub}=0.0395$, the BRs for $B^+\to (D^0, D^{0*}) \ell \nu$ are found to be consistent with current experimental data, as shown in Table~\ref{tab:BDlnu}.  Using the formulas presented in Sec.~\ref{sec:FDDv}, we obtain for the SM that:
 \begin{equation}
 R^{\rm SM}(D)\approx 0.297\,, ~R^{\rm SM}(D^*)\approx 0.258\,.
 \end{equation}
The values  are  within $1\sigma$ errors of those obtained in Ref.~\cite{Bernlochner:2017jka} and are consistent with the results given in Refs.~\cite{MILC:2015uhg,Na:2015kha,Bigi:2016mdz,Bernlochner:2017jka,Jaiswal:2017rve,BaBar:2019vpl,Bordone:2019vic,Martinelli:2021onb}.

\begin{table}[thp]
 \caption{Branching ratios of the $B^-\to D^{0(*)} \ell \nu$ decays in the SM and their experimental measurements. }
\begin{center}
\begin{tabular}{c|cccc} \hline \hline
  Mode & ~~$B^-\to D^0 \ell \nu$~~ & ~~$B^-\to D \tau \nu$~~  &  ~~$B^-\to D^{0*} \ell \nu$~~ &~~$B^-\to D^{0*} \tau \nu$   \\ \hline
 SM & $2.32 \%$ & $6.89 \times 10^{-3}$ & $5.84\%$ &$1.50\%$ \\ \hline
 Exp~\cite{PDG2022} & $(2.30\pm 0.09)\%$ &  $(7.7\pm 2.5)\times 10^{-3}$  & $(5.58\pm 0.22)\%$ &  $(1.88\pm 0.20)\%$ \\ \hline \hline
 
\end{tabular}
\end{center}
\label{tab:BDlnu}
\end{table}%

\begin{figure}[phtb]
\begin{center}
\includegraphics[scale=0.4]{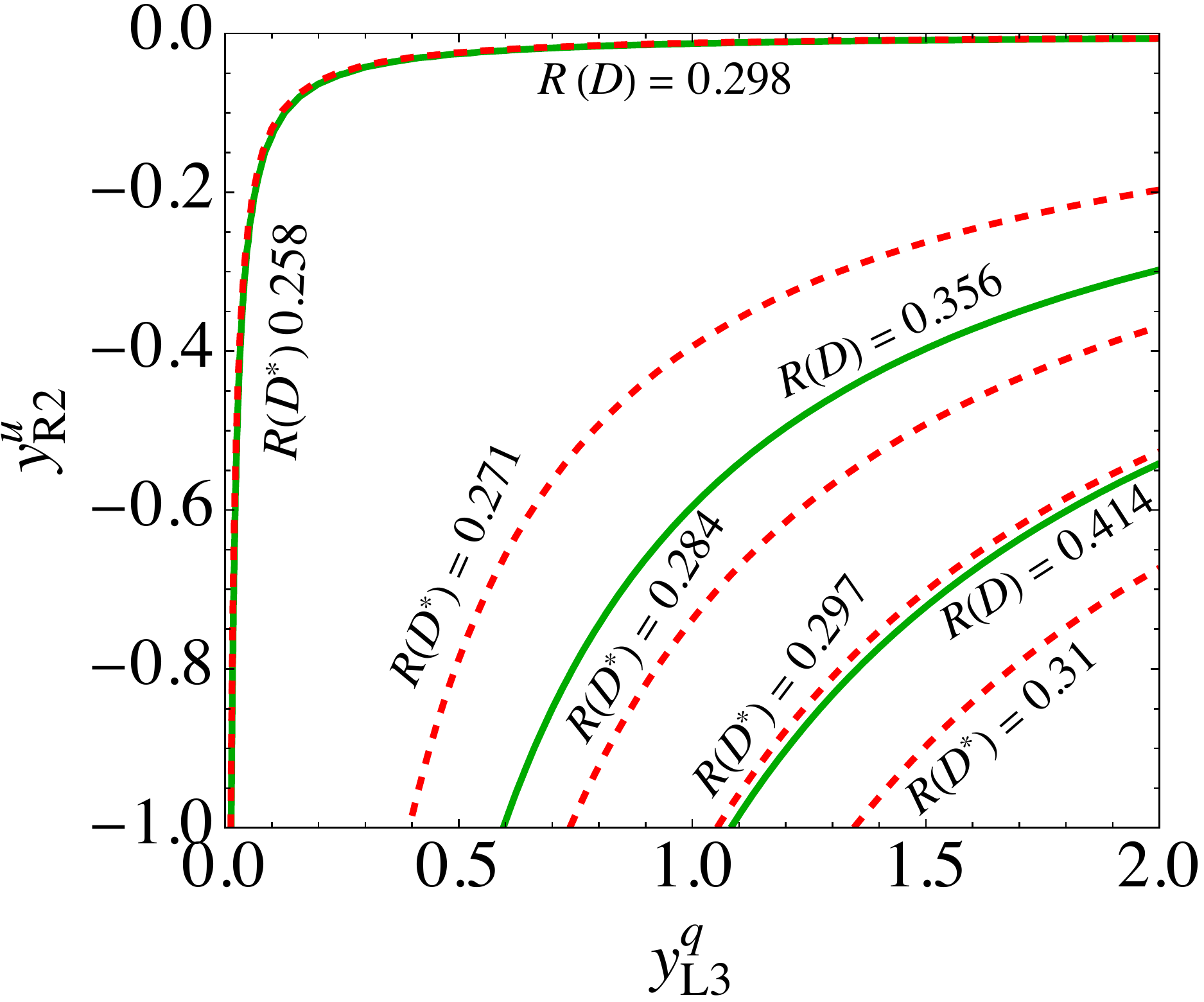}
\hspace{5mm}
\includegraphics[scale=0.4]{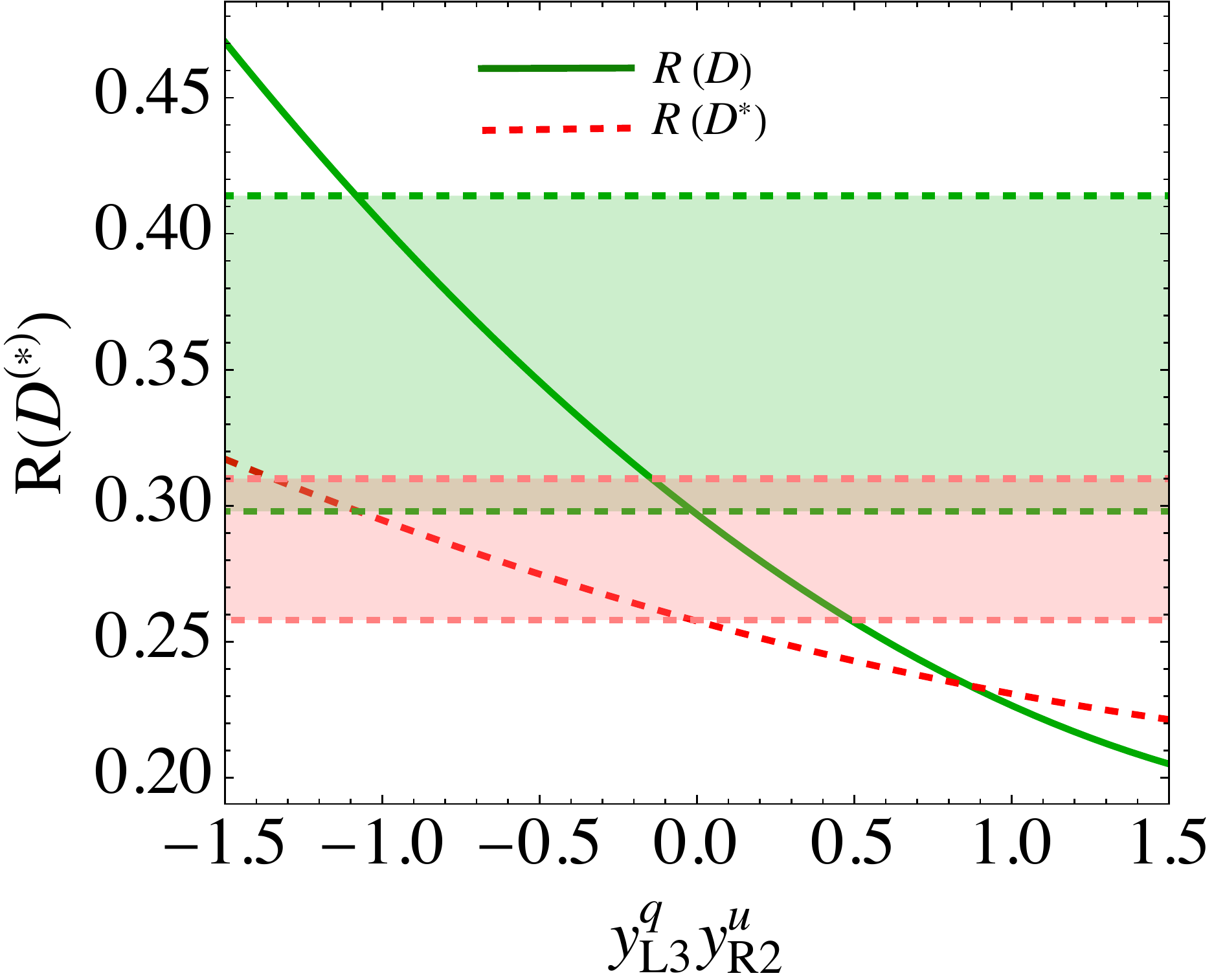}
 \caption{Left: Contours of $R(D)$ and $R(D^*)$ in the $y^q_{L3}$-$y^u_{R2}$ plane.  The solid (darker-green)  and dashed (red)  lines cover the $2\sigma$ range of the world-averaged $R(D)$ and  $R(D^*)$, respectively.  Right: Dependence of $R(D^{(*)})$ on $y^q_{L3} y^u_{R2}$. The light-green [pink] shaded region represents the $2\sigma$ range of the world-averaged $R(D)$ [$R(D^*)$].}
\label{fig:RDRDv}
\end{center}
\end{figure}

The parameters involved in the $b\to c \tau \nu$ transition mediated by the LQ appear in the combinations of $y^q_{L3}y^q_{L2}/m^2_{S}$ and  $y^q_{L3}y^u_{R2}/m^2_{S}$.  For the numerical analysis, we fix $m_S=1.5$~TeV.  From Eq.~(\ref{eq:VydL}), we see that $y^q_{L2}\sim {\cal O}(\lambda^2) \ll y^q_{L3}$, indicating that the dominant effect on $R(D)$ and $R(D^*)$ comes from the combination $y^q_{L3}y^u_{R2}$. To simplify the analysis, we take the assumption that $y^q_{L2}=0$, in which case $R(D^{(*)})$ is found to deviate from that with $y^q_{L2}=0.04$ by only $\sim 2\%$.  We present the contours of $R(D)$ and $R(D^*)$ in the $y^q_{L3}$-$y^u_{R2}$ plane in the left plot of Fig.~\ref{fig:RDRDv}, with the shaded areas (light-green and grey, respectively) covering the $2\sigma$ ranges of their world averages.  It is seen that the low boundaries of $R(D)$ and $R(D^{*})$ match exactly, while the upper boundary for $R(D) = 0.414$ is close to the contour of $R(D^*)=0.297$.  This illustrates that an accurate measurement of $R(D)$ can indirectly constrain the value of $R(D^*)$, and vice versa.  The right plot of Fig.~\ref{fig:RDRDv} shows the dependence of $R(D^{(*)})$ on the product $y^q_{L3} y^u_{R2}$.  To explain the $R(D)$ and $R(D^*)$ anomalies, we need $-1<y^q_{L3} y^u_{R2} < 0$ for $m_{LQ}=1.5$~TeV.  It is observed that $R(D)$ is more sensitive to the $S^{\frac{1}{3}}$ contribution.

\begin{figure}[phtb]
\begin{center}
\includegraphics[scale=0.4]{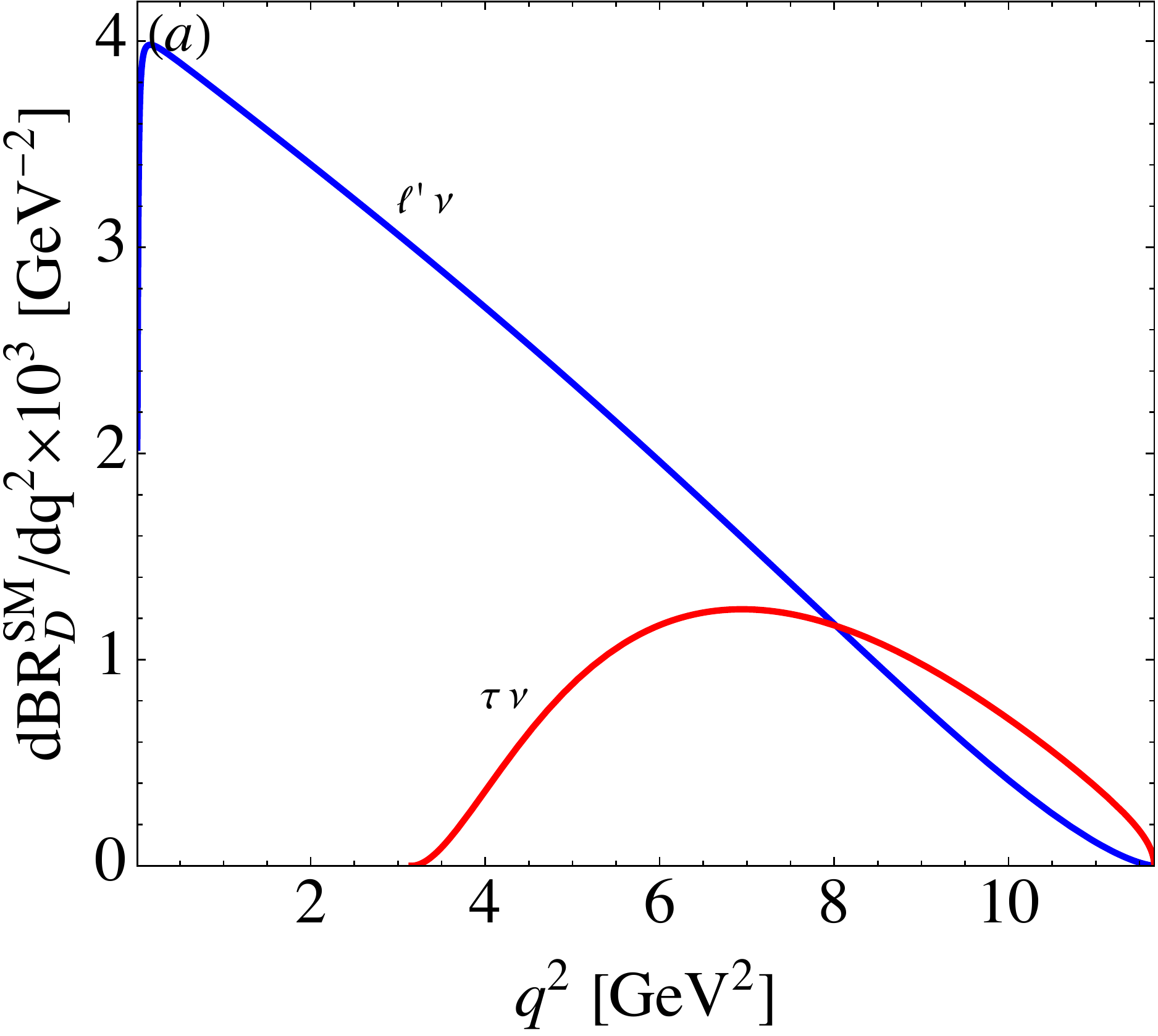}
\hspace{5mm}
\includegraphics[scale=0.4]{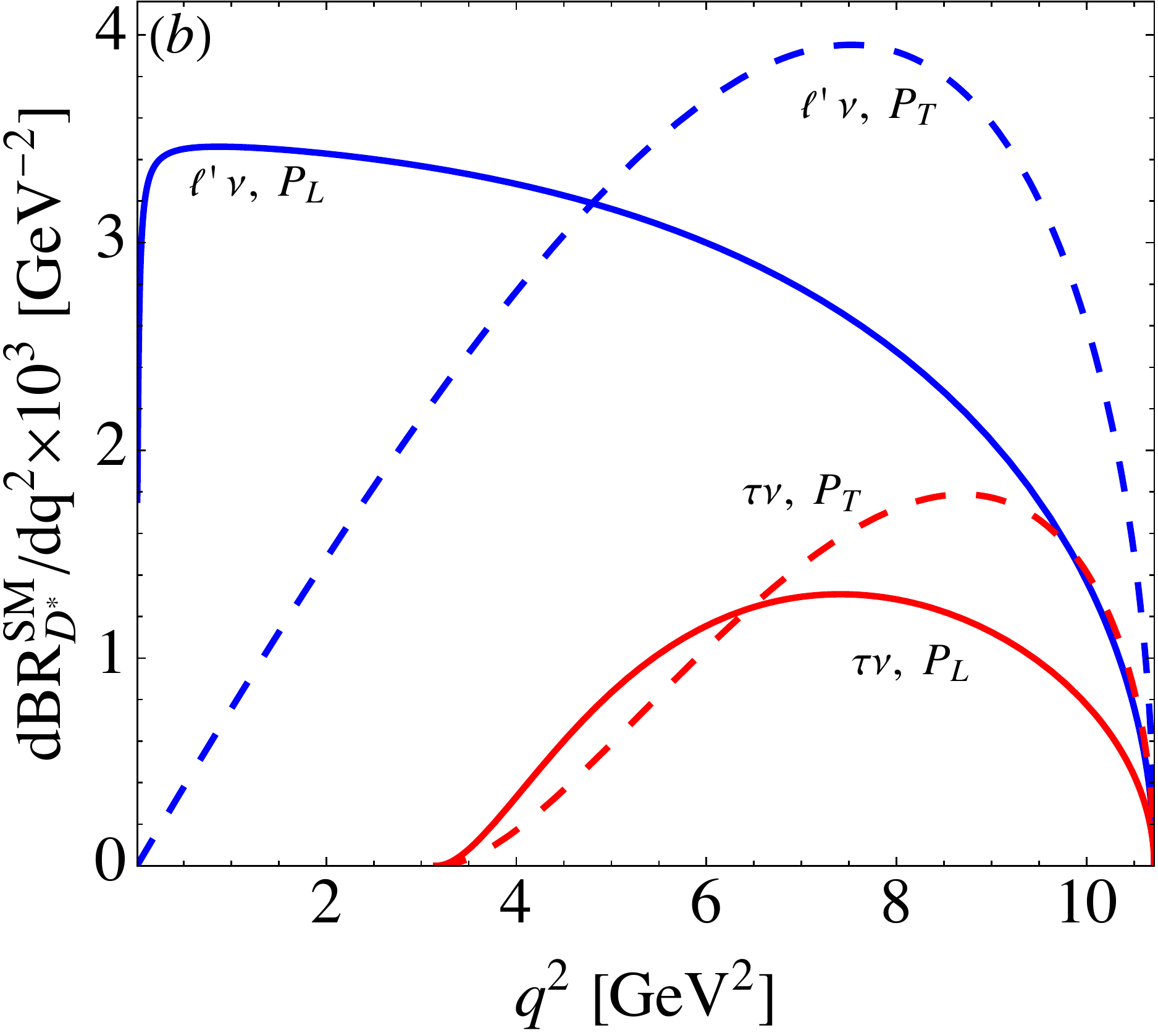} \vspace{2mm}
\\
\hspace{-1mm}\includegraphics[scale=0.4]{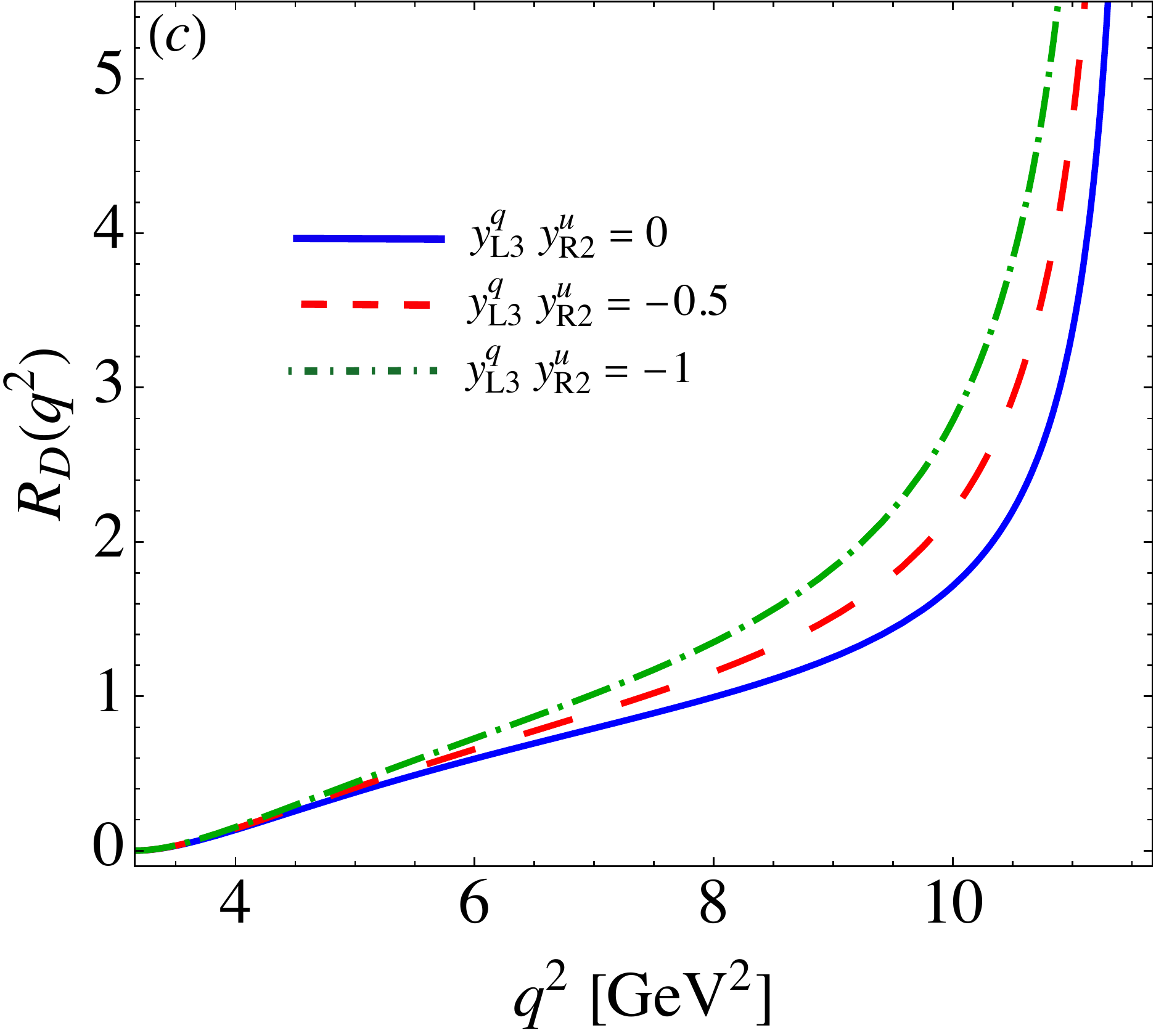}
\hspace{2mm}
\includegraphics[scale=0.415]{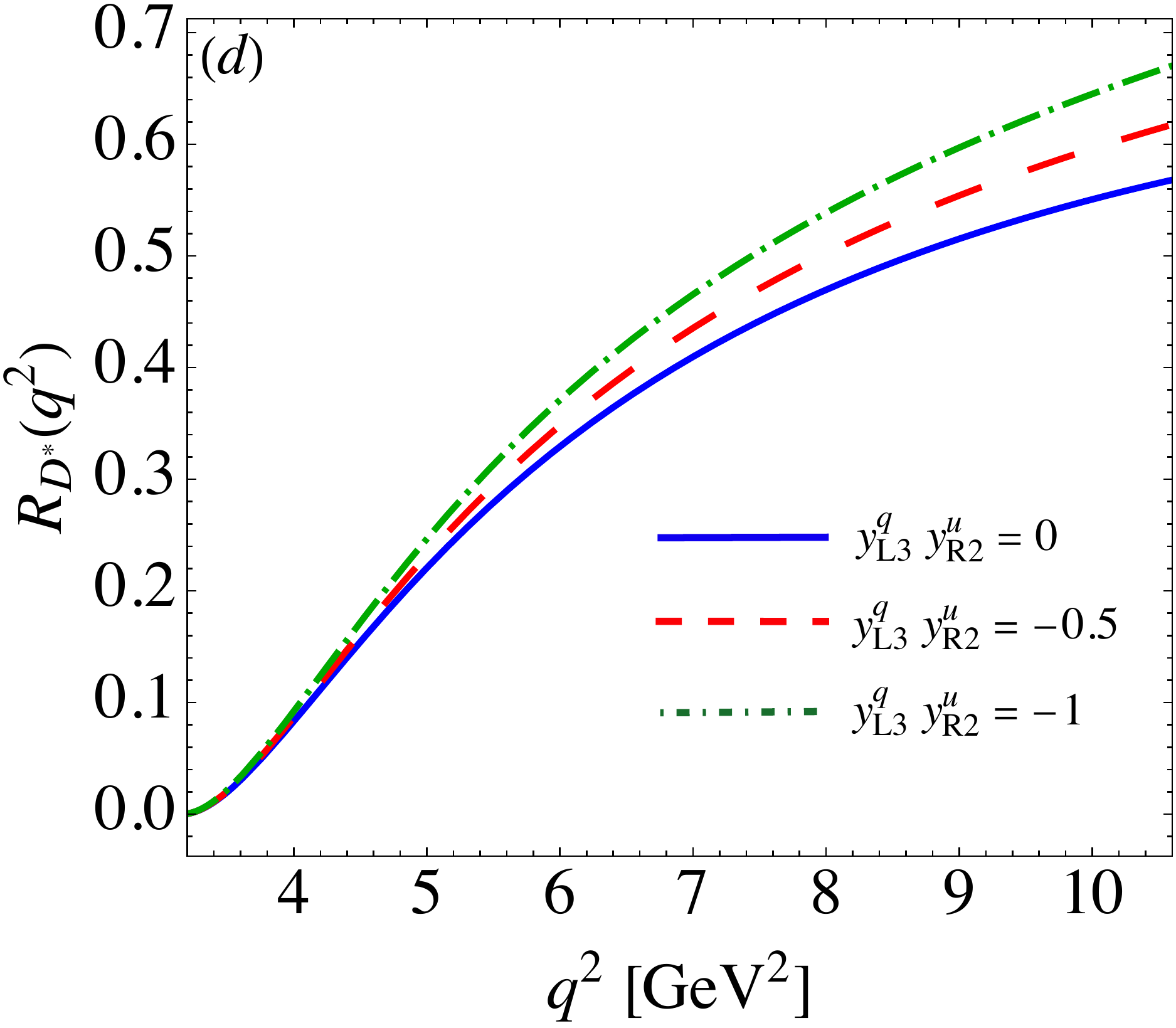}
 \caption{Differential BR as a function of $q^2$ in the SM for (a) $B^-\to D^0 \ell'' \nu$ and (b) $B^-\to D^* \ell'' \nu$, where  the longitudinal and transverse polarizations of $D^*$ are illudtrated separately. The ratios $R_D(q^2)$ (c) and $R_{D^*}(q^2)$ (d) as functions of $q^2$, where the solid, dashed, and dot-dashed curves are plotted for $y^q_{L3} y^u_{R2}=0$, $-0.5$, and $-1$, respectively. }
\label{fig:RMq2}
\end{center}
\end{figure}

In addition to the ratio of the BR for $\tau \nu$ to that for $\ell \nu$, other physical observables may be sensitive to the new physics, such as the forward-backward asymmetry of the charged lepton, $\tau$ polarization~\cite{Chen:2017eby,Chen:2018hqy}, and $q^2$-dependent differential decay rates.  The BR is sensitive to the CKM matrix elements and the form factors of the $B\to (D, D^*)$ transitions.  To eliminate these factors, we propose the ratio of the $q^2$-dependent differential decay rates, defined to be:
\begin{align}
R_{M}(q^2) = \frac{d\Gamma^\tau_M/dq^2}{d\Gamma^{\ell'}_M/dq^2} H(q^2-m^2_\tau)\,, \label{eq:RMq}
\end{align}
where $H(x)$ is the Heaviside step function, and $d\Gamma^{\ell'}_M/dq^2$ is the average of the electron and muon modes.  Because the threshold invariant mass-squared of $\tau \nu$ in the $B\to M \tau \nu$ decay is $q^2=m^2_\tau$, we thus require that the denominator $d\Gamma^{\ell'}_M/dq^2$ also starts from the same invariant mass-squared.  To appreciate the benefit of considering the observable defined in Eq.~(\ref{eq:RMq}), we first show the $q^2$-dependent BRs for $B^-\to (D^0, D^{0*}) \ell'' \nu$ ($\ell''=\ell', \tau$) in the SM in Figs.~\ref{fig:RMq2}(a) and (b), respectively. Plot (a) shows that when $q^2\gtrsim 8$ GeV$^2$, the decay $B^-\to D^0 \tau\nu$ becomes larger than the light lepton mode, and it is expected that $R_D(q^2)>1$ in this region. $D^*$ is a vector meson and has longitudinal ($P_L$) and transverse ($P_T$) components. To exhibit their contributions, we separately show $P_L$ and $P_T$ in Fig.~\ref{fig:RMq2} (b). The results indicate that $P_T$ becomes larger than $P_L$ at somewhat large $q^2$ regions in both light lepton and $\tau$ modes. In contrast to the $B^-\to D^{0}\ell'' \nu$ decay, $d\Gamma^{\ell'}_{D^*}/dq^2$  is always larger than $d\Gamma^{\tau}_{D^*}/dq^2$  in the allowed kinematic region, thus, it is expected that $R_{D^*}(q^2) < 1$.

\begin{figure}[phtb]
\begin{center}
\includegraphics[scale=0.4]{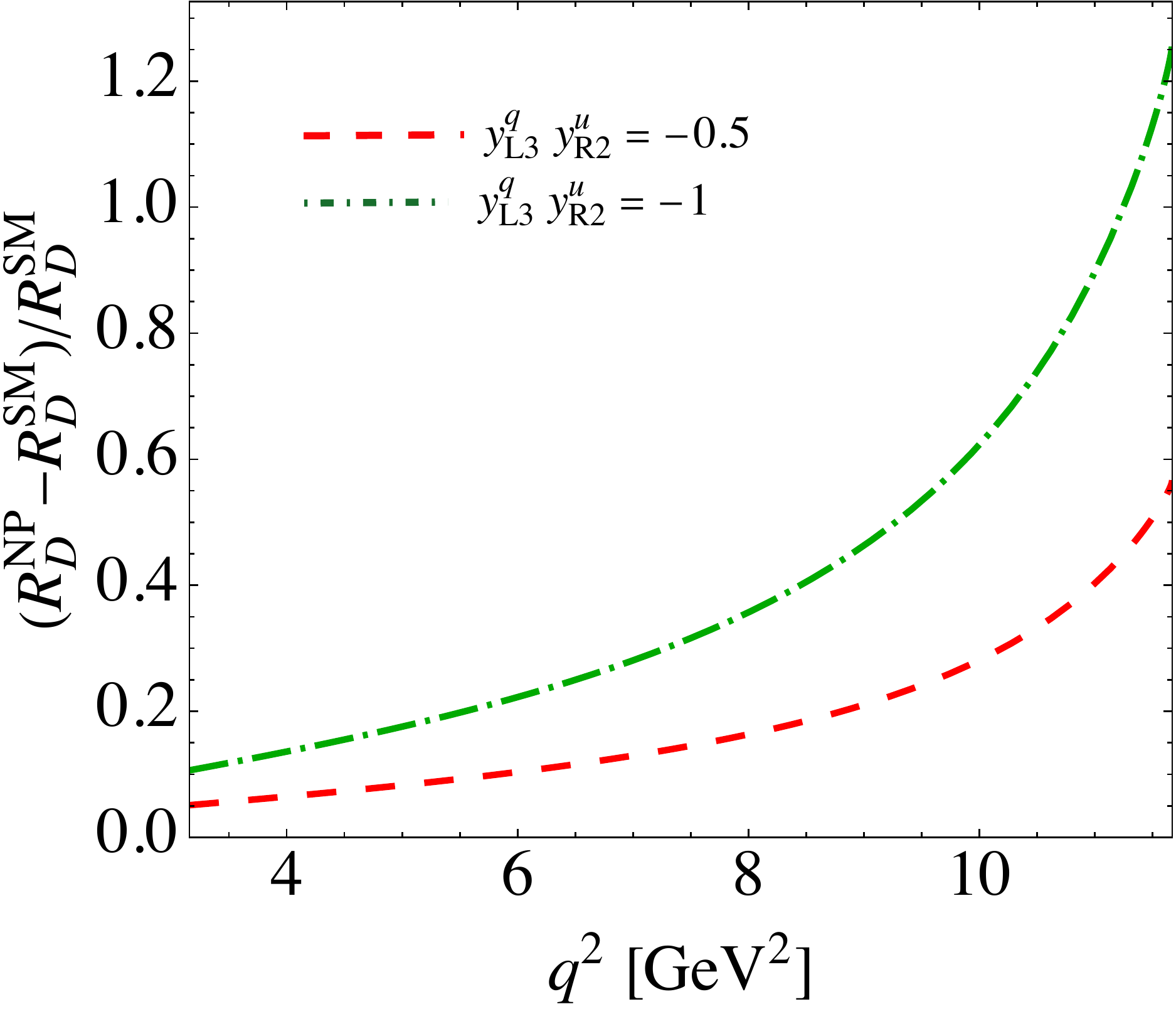}
\hspace{3mm}
\includegraphics[scale=0.41]{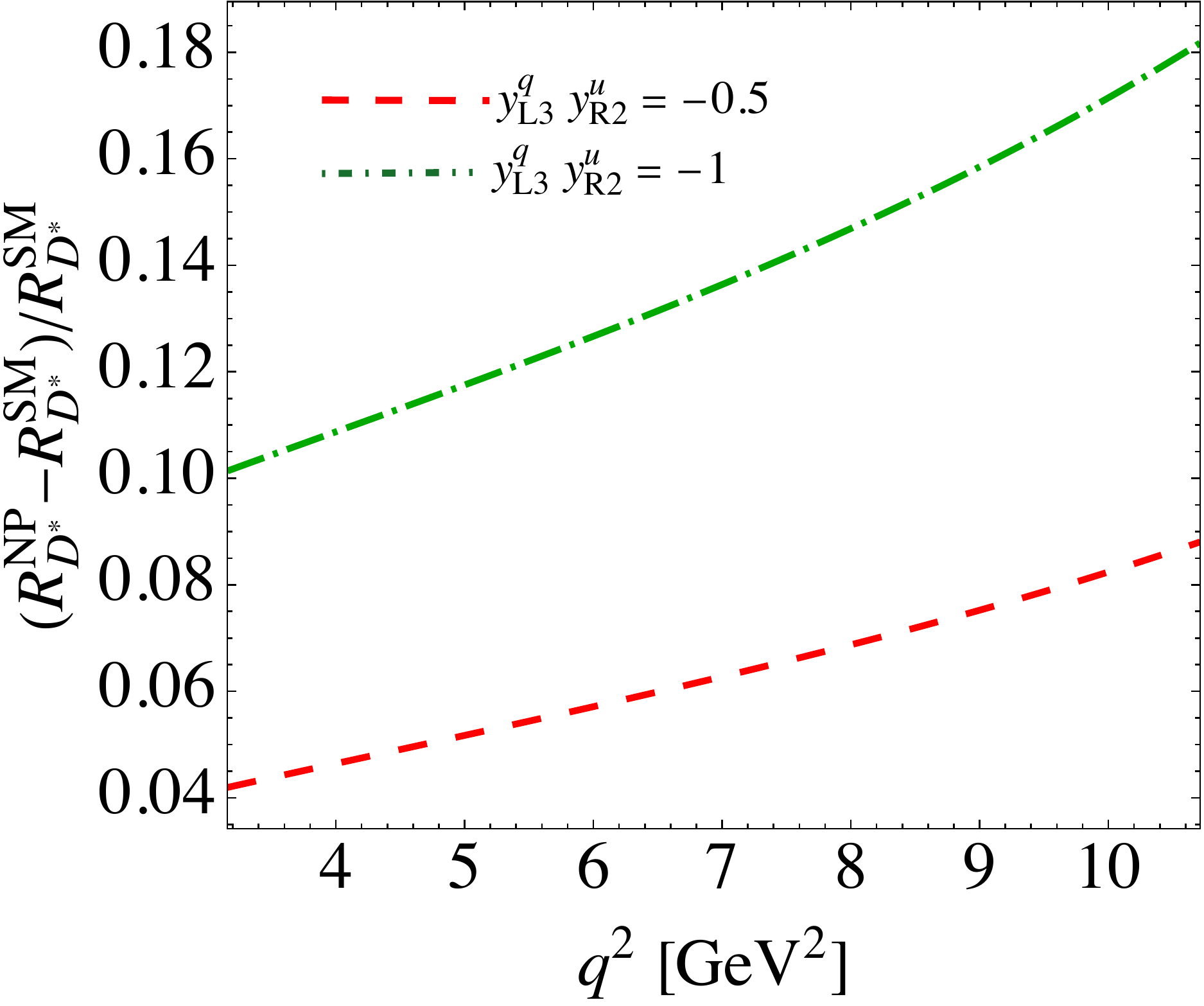}
 \caption{Deviation of $R_{M}(q^2)$ from the SM result for $B^-\to D^0\ell \nu$ (left) and $B^-D^{0*} \ell \nu$ (right).  }
\label{fig:rRMq2}
\end{center}
\end{figure}

The $q^2$-dependence of $R_D(q^2)$ and $R_{D^*}(q^2)$ in the SM is shown in Figs.~\ref{fig:RMq2}(c) and (d), respectively, using the solid curves.  It is confirmed that $R_{D}(q^2)\gtrsim 1$ at $q^2\gtrsim 8$~GeV, while $R_{D^*}(q^2) < 1$ in the physical kinematic region.  Additionally, we find that $R_{M}(q^2)$ increases monotonically with $q^2$.  This means that the decreasing rate of $d\Gamma^{\ell'}_M/dq^2$ in $q^2$ is faster than that of $d\Gamma^\tau_{M}/dq^2$.  To see how sensitive $R_M(q^2)$ is to new physics effects, we show the results using benchmarks of $y^q_{L3} y^u_{R2}=-0.5$ (dashed) and $y^q_{L3} y^u_{R2}=-1$ (dot-dashed) for $R_D(q^2)$ and $R_{D^*}(q^2)$ in the corresponding plots.  We also consider the quantity $(R^{\rm NP}_M(q^2)-R^{\rm SM}(q^2) )/R^{\rm SM}_M(q^2)$ to exhibit the deviation caused by the new physics effects in $R_M(q^2)$ from the SM prediction, and the results are shown in Fig.~\ref{fig:rRMq2}.  The variations of these curves show that $R_D(q^2)$ is more sensitive to new physics than $R_{D^*}(q^2)$ in the model.

\subsubsection{The oblique parameters and $W$-mass}

\begin{figure}[phtb]
\begin{center}
\includegraphics[scale=0.4]{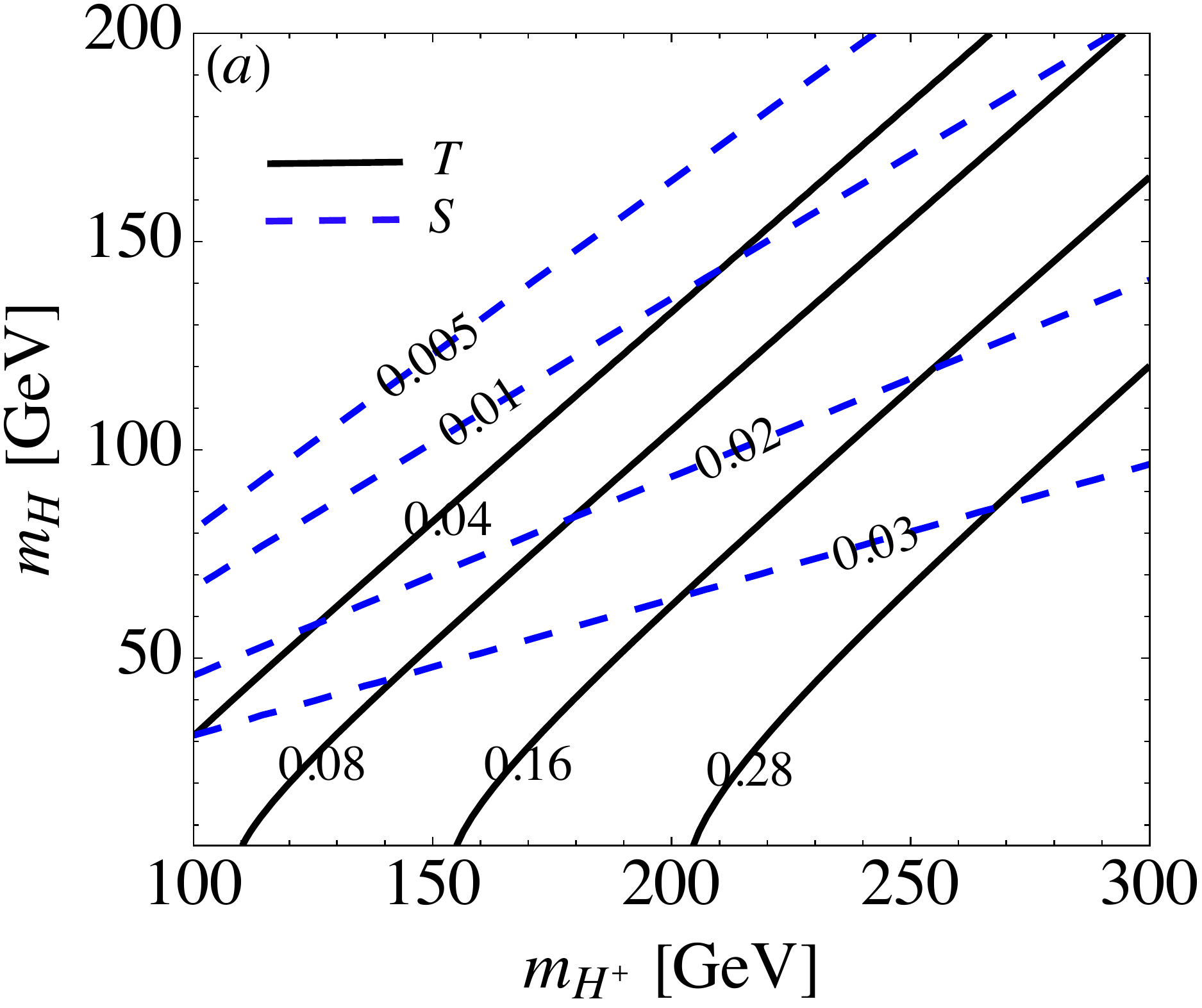}
\hspace{5mm}
\includegraphics[scale=0.4]{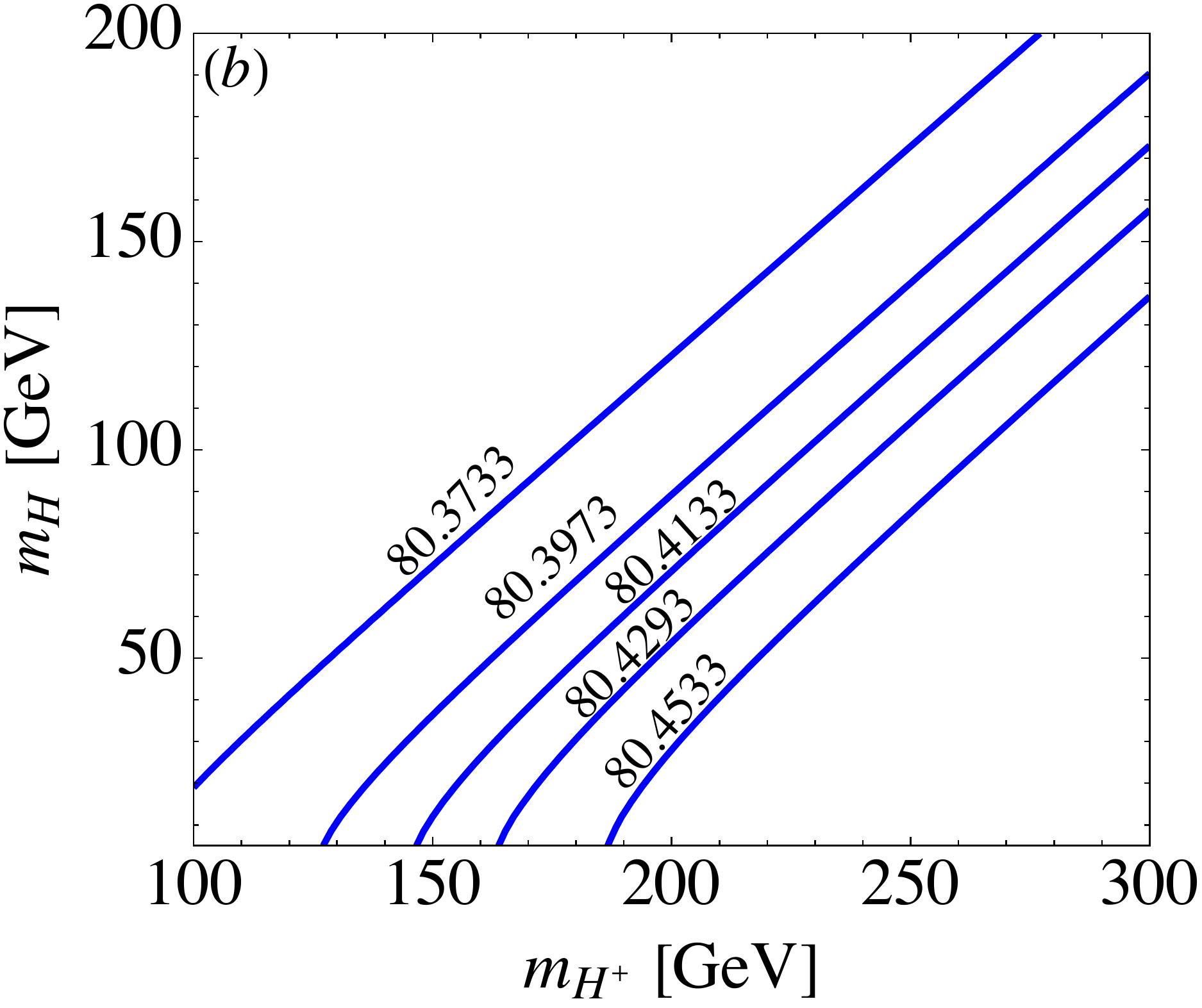}
 \caption{ (a) Contours of the oblique parameters, $S$ and $T$, in the $m_H$-$m_{H^+}$ plane.  (b) Contours of $m_W$ in the $m_H$-$m_{H^+}$ plane. }
\label{fig:Oblique}
\end{center}
\end{figure}

By combining the CDF II measurement of $m_W$ with others, the oblique parameters are determined to be~\cite{deBlas:2022hdk}:
 \begin{align}
 U&\equiv0\,,~ S=0.10\pm 0.073\,,~ T=0.202\pm0.056\,, 
 \nonumber \\ 
 U&=0.134 \pm 0.087\,,~ S=0.05\pm 0.096\,,~ T=0.040\pm 0.120\,.
 \end{align} 
We can use these results to constrain the free parameters in the model.  Based on Eqs.~(\ref{eq:T}), (\ref{eq:S}), and (\ref{eq:U}),  the oblique parameters have a quadratic dependence on $c_{\beta-\alpha}$. However, $c_{\beta-\alpha}\lesssim {\cal O}(0.04)$ as previously discussed, meaning that its effects on the $S$, $T$, and $U$ parameters are negligible.  Therefore, these parameters can be approximated for the model as follows:
  \begin{align}
   T & \simeq    \frac{1}{16 \alpha_{\rm em}  \pi^2 v^2}  s^2_{\beta-\alpha} F(m^2_{H^+}, m_H) 
   \,,  \nonumber \\
   S &\simeq \frac{1}{24 \pi} \left[ (c^2_W -s^2_W)^2 G(m^2_{H^+}, m^2_{H^+}, m^2_Z) + \ln\frac{m^2_{H^+} }{m^2_H} \right]
   \,, \nonumber \\
   U & \simeq  \frac{1}{24\pi} \left[  s^2_{\beta-\alpha} G(m^2_{H^+}, m^2_H, m^2_W) - (2 s^2_W -1)^2 G(m^2_{H^+}, m^2_{H^+}, m^2_Z)  \right]\,.
  \end{align}
In this simplified form, the oblique parameters depend only on the ratio $m_{H^+} / m_H$.  The contours for $T$ (solid) and $S$ (dashed) in the plane of $m_{H^+}$ and $m_H$ for the model are drawn in Fig.~\ref{fig:Oblique}(a), where $s_{\beta-\alpha}\approx 1$ is taken in the estimates.  Due to the fact that $U \ll T$, we do not show the results of $U$ in the plot. The values of $S$ and $U$ in the model can only be up to the percent level and can be neglected in the numerical estimates for further phenomenological analyses.  Thus, using the obtained $T$ parameter, the loop-corrected $W$ mass in the model is shown in Fig.~\ref{fig:Oblique}(b), where the contours correspond to the central value, $\pm 2\sigma$ and $\pm 5\sigma$ of the world average of $m_W=80.4133\pm 0.0080$~\cite{deBlas:2022hdk}.  We observe that $m_W$ increases with $m_{H^+}$ for a given $m_H$, while a lower $m_H$ is needed to increase $m_W$ when $m_{H^+}$ is fixed.  For instance, $m_W\approx 80.43$~GeV can be achieved for $m_{H}\approx 50$~GeV and $m_{H^+}\approx 150$~GeV.

\subsubsection{$Z_1$ and $H$ decays}

Finally, let's discuss possible decays of the light $Z_1$ and $H$.  Because the mass of the light gauge boson is limited in the region of $m_{Z_1}\in (10, 100)$~MeV, it can only decay dominantly into on-shell light leptons through two-body decays.  The $Z_1$ partial decay rate for possible final leptons is given by:
\begin{equation}
\Gamma(Z_1 \to f \bar{f} ) \simeq \frac{g^2 m_{Z_1}}{96 c^2_W} \left( |C^f_{R}|^2 + |C^f_{L}|^2\right)\,, 
\end{equation}
where $C^f_{R(L)}=C^f_{1V}\mp C^f_{1A}$, $f$ denotes the possible light leptons (such as the three active neutrinos and the electron), and $m^2_f/m^2_{Z_1} \approx 0$ is applied. The effective  couplings of $C^f_{R, L}$ for each involved $f$ are given as follows: 
 \begin{align}
 C^{\nu_\ell}_R &=0 ~,~ C^{\nu_e}_L=s_{\theta_Z} ~,
 \nonumber \\
 C^{\nu_{\mu, \tau} }_L& = s_{\theta_Z} \pm \frac{ c_W m_{Z_1} c_{\theta_Z}}{g v}  \sqrt{2+t^2_\beta + t^{-2}_\beta} 
 \,, \\
 C^e_R & = 2 s^2_W s_{\theta_Z}\,, ~ C^e_L = (-1+2s^2_W) s_{\theta_Z}\,. 
 \nonumber
 \end{align}
Although $Z'$ does not couple to the first-generation leptons, the physical $Z_1$ can decay to them via $Z'-Z$ mixing.

\begin{figure}[phtb]
\begin{center}
\includegraphics[scale=0.4]{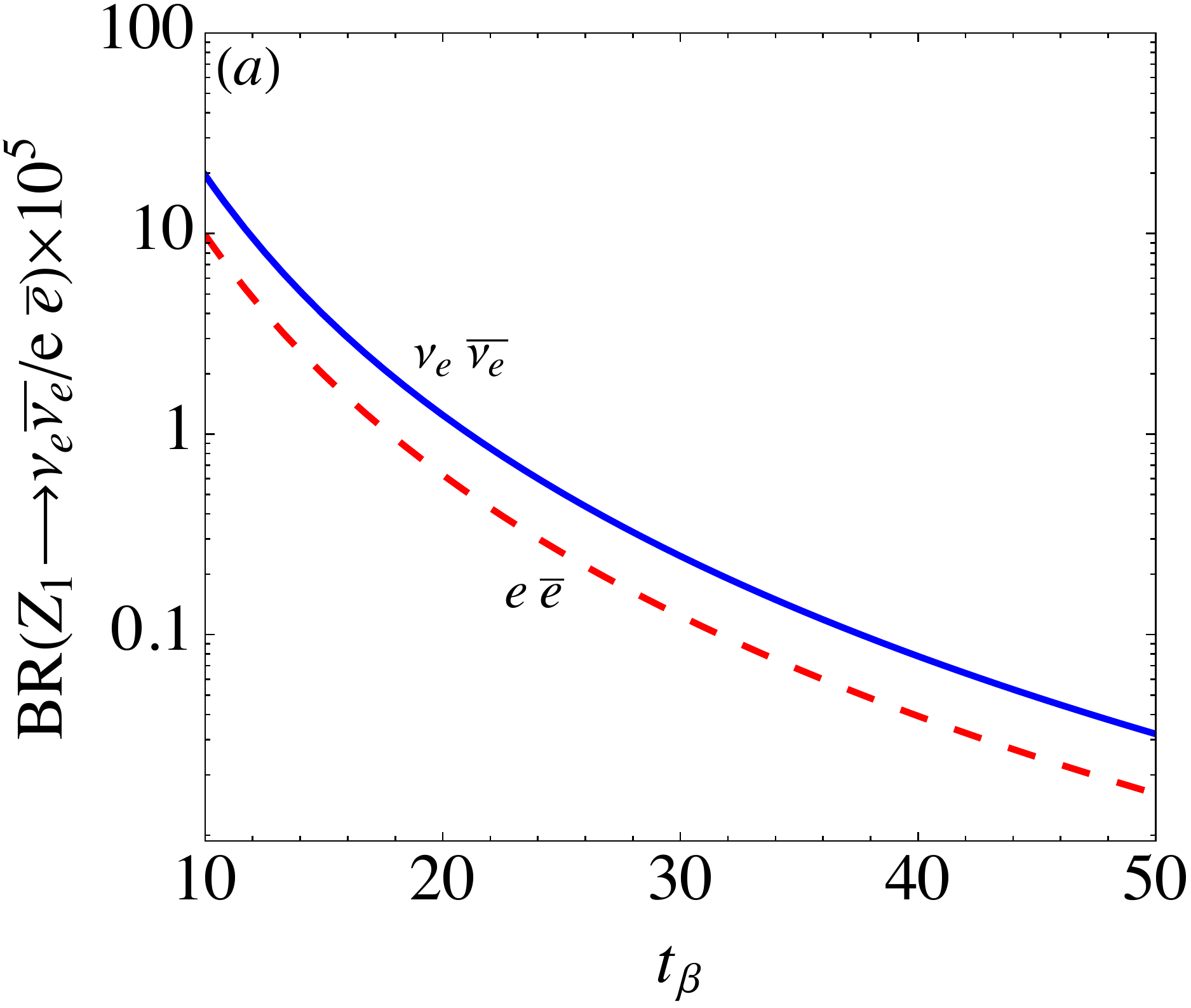}
\hspace{5mm}
\includegraphics[scale=0.40]{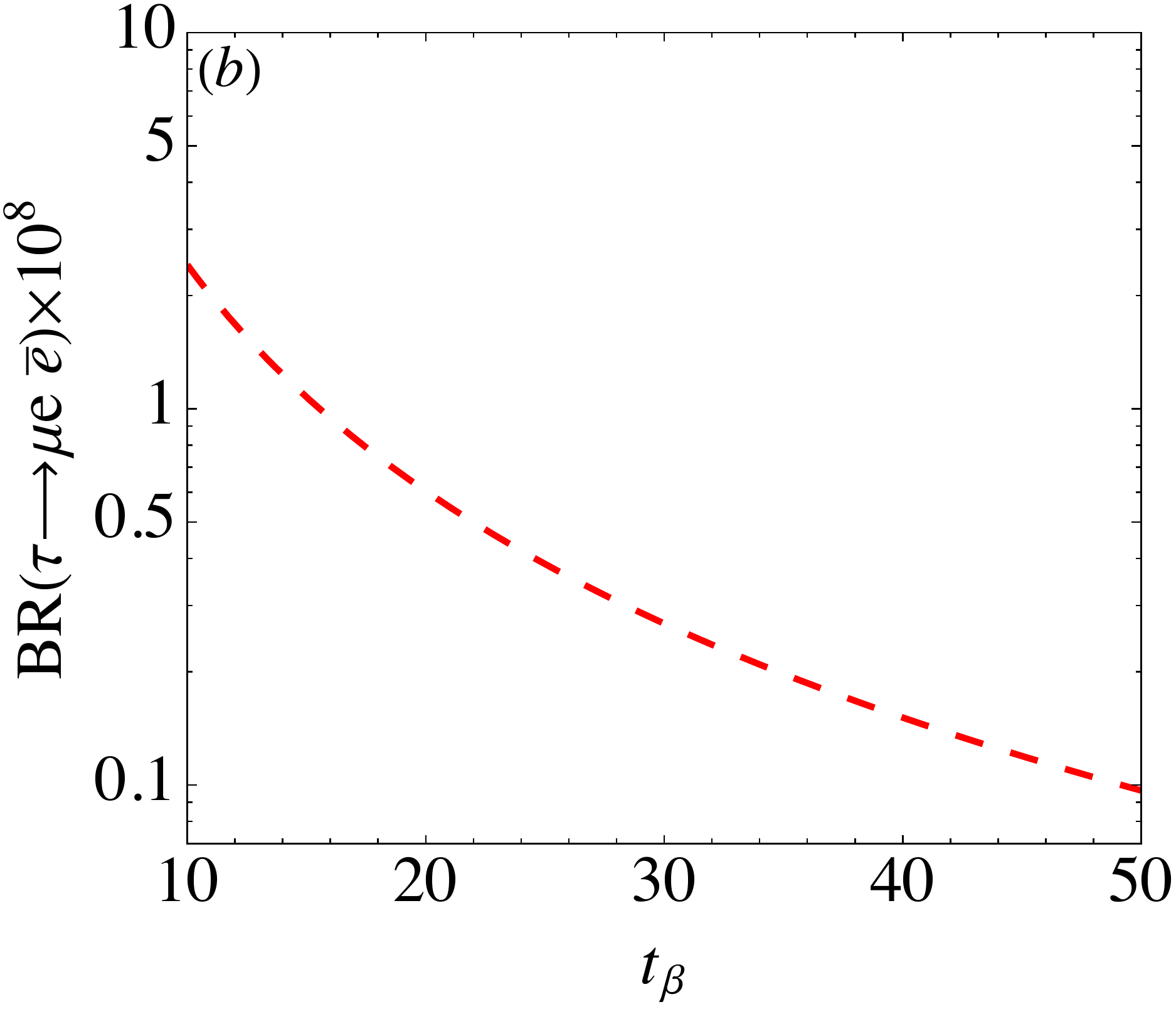}
\caption{BRs for (a) $Z_1\to (\nu_e \bar\nu_e, e \bar e)$  and (b) $\tau\to \mu Z_1 \to \mu e^- e^+$ as functions of $t_\beta$. }
\label{fig:BRZ1}
\end{center}
\end{figure}

If $s_{\theta_Z}$ were not significantly smaller than $g_{Z'}$, the decay rates for $Z_1\to (\bar \nu_e \nu_e,\, e^- e^+)$ could be sizable compared to the $Z_1 \to \bar \nu_{\ell'} \nu_{\ell'}$ decays.  However, due to the large $t_\beta$ enhancement in the $Z_1$ gauge coupling to $\nu_{\mu,\tau}$, the dominant decay channels are $Z_1\to \nu_\mu \bar\nu_\mu/ \nu_\tau \bar\nu_\tau$, with estimated BRs of  approximately $ 50.5\%$ and $49.5\%$, respectively. The BRs for $\nu_e \bar\nu_e$ and $e^- e^+$ as functions of $t_\beta$ are presented in Fig.~\ref{fig:BRZ1}(a).  It is found that the BRs are more sensitive to $t_\beta$ and less sensitive to $m_{Z_1}$.  Because $Z_1$ can be produced in the $\tau\to \mu Z_1$ decay, which depends on the lepton flavor mixing $\theta_L$, a significant $BR(Z_1\to e^- e^+)$ thus implies a large BR for the LFV process $\tau\to \mu Z_1 \to \mu e^- e^+$, where the current upper limit is $BR(\tau\to \mu e^- e^+) < 1.8 \times 10^{-8}$~\cite{PDG2022}.  Our estimate of $BR(\tau\to \mu e^- e^+)$ is shown in Fig.~\ref{fig:BRZ1}(b), where $\chi_{\mu\tau}=10^{-5}$ is used.  Since $\tau\to \mu Z_1$ is also not sensitive to $m_{Z_1}$, the dependence of $m_{Z_1}$ in $BR(\tau\to \mu e^- e^+)$ is not manifest.  Assuming the integrated luminosity of $50$~ab$^{-1}$, Belle~II will be capable of probing the LFV process BRs down to the level of $10^{-10}-10^{-9}$~\cite{Banerjee:2022vdd}.  The BR of ${\cal O}(10^{-9})$ for $\tau\to \mu e^- e^+$ predicted in this model can thus be probed at Belle~II.

As discussed earlier, when $m_H < m_h/2$, $H$ can be produced through the $h\to HH$ decay.  The partial decay width of this process can provide a strict limit on the $t_\beta$ and $c_{\beta-\alpha}$ parameters.  In the following, we concentrate on this scenario, even though $H$ generally can be heavier.

For two-body decays, $H$ should decay into a pair of fermions, as long as the phase space permits.  From Eq.~(\ref{eq:h_Yu}), its Yukawa couplings to fermions are suppressed by $m_f/v$ and $(c_{\beta-\alpha} -  s_{\beta-\alpha}/t_\beta)$, with no other factors that can enhance the partial decay width.  As a result, $\Gamma(H\to f \bar{f})$ is small and negligible.  However, even though suppressed by $m^2_{Z_1}/v$ from the gauge coupling as shown in
 \begin{equation}
 {\cal L}_{HZ_1 Z_1} \simeq  \frac{2 (t^2_\beta -1 )}{t_\beta }  \frac{m^2_{Z_1} }{v}  s_{\beta-\alpha} \frac{H Z_{1\mu} Z^{\mu}_1 }{2} \,,
 \end{equation}
the $H\to Z_1 Z_1$ decay rate can be enhanced by the longitudinal component, which is proportional to $1/m_{Z_1}$.  This leads to a partial decay width,
  \begin{equation}
  \Gamma(H\to Z_1 Z_1) \simeq  \frac{m_H}{32 \pi} \frac{m^2_H}{v^2} \left|  \frac{  t^2_\beta -1 }{t_\beta}  s_{\beta-\alpha}\right|^2\,. \label{eq:HZ1Z1}
  \end{equation}
The original suppression factor from the gauge coupling is seen to be canceled by the longitudinal effect of $1/m^2_{Z_1}$ from each $Z_1$ boson.  With $m_H=50$~GeV and $t_\beta=20$, we obtain $\Gamma(H\to Z_1 Z_1)\approx 8.2$~GeV.  The other decay processes are subdominant.  For example, the $H\to Z_1 Z^*_2\to Z_1 f \bar f$ decay has additional suppression factors due to the phase space and $1/m^2_{Z_2}$.  An explicit estimate shows that the partial width for $H\to Z_1 Z^*_2$ is of ${\cal O}(10^{-5})$~GeV.  According to the earlier analysis, $Z_1 \to \nu \bar\nu$ is the dominant decay channel.  Consequently, $H$ predominantly decays into invisible neutrinos and becomes missing energy in the detector.

We now turn to the production of $H$ at the LHC.  First, $H$ could be singly produced according to Eq.~(\ref{eq:h_V}) via the vector boson fusion (VBF) process.  But the $W^- W^+ (Z Z) H$ coupling is suppressed by $c_{\beta-\alpha}$.  Additionally, the Yukawa coupling for the bremsstrahlung production of $H$ with the top quark is determined by $(m_t/v) (c_{\beta-\alpha}- s_{\beta-\alpha}/t_\beta)$ and is also suppressed.  However, $H$ can be pair-produced more copiously through the $hHH$ and $W^-H^+H$ couplings.  In the former case, the $H$ pair is produced by the on-shell Higgs boson, i.e., $pp\to h \to HH$.  From Eq.~(\ref{eq:hHH}), although $\Gamma(h\to HH)$ is associated with the small factor $\xi$, its BR can still be at the percent level.  This amounts to the invisible decay of the Higgs boson~\cite{ATLAS:2022yvh}.  In the latter case, the $W^-H^+H$ coupling, as given in Eq.~(\ref{eq:h_V}), is determined by the gauge coupling $g$ with $s_{\beta-\alpha}\approx 1$.  When $H^+$ is taken as an intermediate state in the t-channel scattering, $H$ pair production occurs via the VBF channel, i.e., $pp\to HH + \mbox{forward jets}$.  We may probe such an effect via the search for invisible decays of the new Higgs boson $H$~\cite{ATLAS:2022yvh}.

\section{Summary} \label{sec:summary}

A sub-GeV $Z'$ gauge boson has received much attention recently in the literature due to its distinctive characteristics, which could potentially resolve the observed anomalies, such as the muon $g-2$, and serve as a messenger between visible and dark sectors. Additionally, a light $Z'$ gauge boson can make a significant contribution to CE$\nu$NS, as recently observed by the COHERENT experiment.  Accordingly, we investigate the phenomenological impacts on flavor physics when the light $Z'$ gauge boson originates from the local $U(1)_{L_\mu-L_\tau}$ gauge symmetry. 

We have found that when a second Higgs doublet carrying the $U(1)_{L_\mu-L_\tau}$ charge is introduced to spontaneously break the $U(1)_{L_\mu-L_\tau}$ gauge symmetry, the new neutral and charged scalars can result in a larger $W$ mass. Moreover, when a scalar leptoquark $S^{\frac{1}{3}}=(\bar{3}, 1, 2/3)$ is added to the model, it would couple to the third-generation leptons in a unique way due to the $U(1)_{L_\mu-L_\tau}$ symmetry, so that the branching ratios of $B\to (D, D^*) \tau \nu_\tau$ are enhanced, thus solving the $R(D)$ and $R(D^*)$ anomalies.

With the new Higgs doublet, the mixing between the new scalar boson and the SM-like Higgs leads to new decay channels for the Higgs boson, including $h\to \mu \tau/Z_1 Z_1/Z_1 Z_2$ (and $h\to HH$ when $m_H < m_h/2$).  It is found that due to the enhancement of $1/m^2_{Z_1}$, the $\tau\to \mu Z_1$ decay strictly constrains the $\mu-\tau$ flavor mixing, resulting in a highly suppressed $h\to \mu \tau$ decay.  By assuming proper partial widths to the new Higgs decay channels, the $\tan\beta$ and $\cos(\beta-\alpha)$ parameters are limited and the large $\tan\beta$ scheme is favored.  Although the $\mu-\tau$ flavor-changing coupling is restricted to be small, the $\tau\to \mu Z_1 \to \mu e^- e^+ $ decay, induced through the $Z-Z'$ mixing, can still reach the sensitivity of ${\cal O}(10^{-9})$ at Belle II.

Taking into account all potential constraints, we have found that the cross section of CE$\nu$NS induced by the $Z'-Z$ mixing depends solely on the light gauge boson mass, $m_{Z_1}$.  The mass region of $m_{Z_1}$ that is used to fit the CE$\nu$NS cross section, measured by COHERENT using the CsI target~\cite{COHERENT:2021xmm}, can also explain the muon $g-2$ anomaly within $3\sigma$.  To demonstrate the sensitivity of new physics to CE$\nu$NS in the model, we propose to study the cross section as a function of the incident neutrino energy.  Our results show that in the low energy region, such as $E_\nu\sim 10$~MeV, the deviation from the SM can exceed $15\%$, depending on the value of $m_{Z_1}$. To compare with results from other $U(1)$ gauge symmetries, we have examined the influence on the CE$\nu$NS cross section from selected $U(1)$ gauged models, such as the universal, $B-L$, $B-3L_\mu$, and $L_\mu-L_\tau$ with kinetic mixing. It has been found that only the model with dynamical $U(1)_{L_\mu-L_\tau}$ breaking can explain the anomaly of muon $g-2$ when the $1\sigma$ upper limits of the COHERENT data are imposed.

In addition to explaining the observed excesses in $R(D)$ and $R(D^*)$ using the introduced leptoquark, we have proposed a $q^2$-dependent ratio of $d\Gamma/dq^2(B\to M \tau\nu)$ to the avreaged differential decay rate of the light leptons $d\Gamma/dq^2(B\to M \ell'\nu)$, denoted by $R_M(q^2)$.  Our results show that in the high $q^2$ region, $R_D(q^2)$ is more sensitive to the new physics effects and exhibits a significant deviation from the SM.

We have also studied the impact of the two-Higgs-doublet model on the oblique parameters and their relations to the $W$ boson mass.  With the approximation that $\cos(\beta-\alpha)\ll 1$, the parameters involved in the oblique parameters are $m_H$ and $m_{H^+}$.  We find a significant space in the $m_H$-$m_{H^+}$ plane that allows an enhancement of $m_W$ up to the value observed by CDF II.  Finally, we have discussed the possible decay channels for $Z_1$ and $H$ in the scenario where $m_{Z_1} \in(10,100)$~MeV and $m_H < m_h/2$.  The analysis shows that $Z_1\to \nu_\mu \bar\nu_\mu/\nu_\tau \bar\nu_\tau$ and $H\to Z_1 Z_1$ are the dominant decay channels.

\acknowledgments
This work was supported in part by the National Science and Technology Council, Taiwan under Grant Nos. MOST-110-2112-M-006-010-MY2 (C.~H.~Chen) and MOST-111-2112-M-002-018-MY3 (C.~W.~Chiang and C.~W.~Su).

\appendix 

\section{$\bar B \to D^{(*)}$ transition form factors} 

\subsection{Form factor parameterization}\label{app:FF}

In this section, we define the $\bar B \to D^{(*)}$ transition form factors for the $\bar B\to D^{(*)} \ell \nu$ decays. First, the transition form factors associated with the various currents mediating the $\bar  B \to D$ transitions are parametrized as:
\begin{align} 
\langle D(p_2) | q  b | \bar B(p_1) \rangle &= (m_B + m_D) F_S(q^2) 
\,,\nonumber \\
\langle D(p_2) | q \gamma^\mu b | \bar B(p_1) \rangle &= F_{+}(q^2) \left( (p_1 + p_2)^\mu -\frac{m^2_B -m^2_D}{q^2 } q^\mu \right)+ \frac{m^2_B - m^2_D}{q^2} q^\mu F_0(q^2)
\,, \nonumber \\
 \langle D(p_2) | q \sigma_{\mu\nu} b | \bar B(p_1) \rangle &= - i (p_{1\mu} p_{2\nu} - p_{1\nu} p_{2\mu} ) \frac{2F_T(q^2) }{m_B + m_D}\,, \label{eq:ffP}
\end{align}
where the momentum transfer $q=p_1 - p_2$.  For the $\bar B\to D^*$ transitions, the form factors are parametrized as:
 \begin{align}
 \langle D^*(p_2, \epsilon)| \bar q \gamma_\mu b|\bar B(p_1)\rangle 
 =& 
 i \varepsilon_{\mu \nu \rho \sigma} \epsilon^{\nu*} p^\rho_1 p^\sigma_2 \frac{2V(q^2) }{m_B + m_{D^*}}
 \,, \nonumber \\
  \langle D^*(p_2, \epsilon)| \bar q  \gamma_5 b|\bar B(p_1)\rangle 
  =& 
  - \frac{2 m_{D^*}}{m_B+m_{D^*}} F_P(q^2) \epsilon^*\cdot q 
  ~,  \nonumber \\
 \langle D^*(p_2, \epsilon)| \bar q \gamma_\mu \gamma_5 b|\bar B(p_1)\rangle 
 =& 
 2m_{D^*} A_{0}(q^2) \frac{\epsilon^*\cdot q}{q^2}q_\mu +
 (m_B + m_{D^*})A_1(q^2) \left( \epsilon^*_\mu - \frac{\epsilon^* \cdot q}{q^2}q_\mu \right) 
 \nonumber \\
 &
 - A_2(q^2) \frac{\epsilon^* \cdot q}{m_B + m_{D^*}} \left( (p_1 + p_2)_\mu - \frac{m^2_B -m^2_{D^*}}{q^2} q_\mu  \right)
 \,, \nonumber \\
\langle D^*(p_2, \epsilon)| \bar q \sigma_{\mu \nu} b|\bar B(p_1)\rangle 
=& \varepsilon_{\mu \nu \rho \sigma} \left[ 
 \epsilon^{\rho *} (p_1 + p_2)^\sigma T_1(q^2) + \epsilon^{\rho *} q^\sigma \frac{m^2_B -m^2_{D^*}}{q^2}(T_2(q^2) - T_1(q^2)) \right. \nonumber \\
 &\left. + 2 \frac{\epsilon^* \cdot q}{q^2} p^\rho_1 p^\sigma_2 \left( T_2 (q^2) - T_1(q^2) + \frac{q^2}{m^2_B -m^2_{D^*}} T_3(q^2) \right) \right]
 \,, \label{eq:ffV}
 \end{align}
where  $\epsilon^{0123} \equiv 1$, $\sigma_{\mu \nu} \gamma_5 = \frac{i}{2} \epsilon_{\mu \nu \rho \sigma} \sigma^{\rho \sigma}$, and $\epsilon^\mu$ denotes the polarization vector of the $D^*$ meson.

\subsection{Form factors in the HQET}

To numerically estimate the BRs of the $\bar B\to D^{(*)} \ell \nu$ decays, a QCD approach is necessary to evaluate the involved form factors.  In this study, we use the results presented in Ref.~\cite{Bernlochner:2017jka}, which is based on the HQET.  Since the parametrization of the form factors in the HQET differs from those in Eqs.~(\ref{eq:ffP}) and (\ref{eq:ffV}), we introduce here the HQET notation and provide the relationship between the different parametrizations.  We first define the dimensionless kinetic variables in the HQET:
  \begin{equation}
  v^\mu=\frac{p^\mu_B}{m_B}\,, \  v'^\mu=\frac{p^\mu_{D^{(*)}}}{m_{D^{(*)}}}\,, \ w= v\cdot v' = \frac{m^2_B + m^2 _{D^{(*)}} -q^2 }{2 m_B m_{D^{(*)}}}\,.
  \end{equation}
The form factors for the $\bar B\to D$ transitions are then parametrized as~\cite{Bernlochner:2017jka}:
  \begin{align}
  \langle D| \bar c b| \bar B\rangle &= \sqrt{m_B m_D} h_S (w+1)
  \,, \nonumber \\
   \langle D| \bar c \gamma^\mu  b| \bar B\rangle &= \sqrt{m_B m_D} 
   \left[ h_+ (v+v')^\mu + h_{-} (v-v')^\mu \right] 
   \,, \nonumber \\
    \langle D| \bar c \sigma^{\mu \nu}   b| \bar B\rangle &= i \sqrt{m_B m_D}  h_T \left( v'^{\mu} v^{\nu} - v'^\nu v^\mu \right)\,,
     \label{eq:HQETBP}
  \end{align}
and those for the $\bar B\to D^*$ transitions are:
  \begin{align}
   \langle D^* | \bar c \gamma^5 b| \bar B\rangle &= -\sqrt{m_B m_{D^*}} h_P \, \epsilon^* \cdot v 
   \,, \nonumber \\
    \langle D^* | \bar c \gamma^\mu  b| \bar B\rangle &= i \sqrt{m_B m_{D^*}} h_V \varepsilon^{\mu \nu \alpha \beta} \epsilon^*_\nu v'_\alpha v_\beta 
    \,, \nonumber \\
     \langle D^* | \bar c \gamma^\mu  \gamma^5 b| \bar B\rangle &= \sqrt{m_B m_{D^*}} \left[ h_{A_1} ( w+1) \epsilon^{*\mu} -h_{A_2} (\epsilon^*\cdot v) v^\mu - h_{A_3} (\epsilon^* \cdot v) v'^\mu   \right]
     \,, \nonumber \\
   \langle D^* | \bar c \sigma^{\mu\nu}   b| \bar B\rangle &=  - \sqrt{m_B m_{D^*}} \left[ h_{T_1} \epsilon^*_\alpha (v+ v')_\beta + h_{T_2} \epsilon^*_\alpha ( v-v')_\beta + h_{T_3} (\epsilon^*\cdot v ) v_\alpha v'_\beta \right]\,, \label{eq:HQETBV}
  \end{align}
where $h_{-}$, $h_{A_2}$, and $h_{T_{2,3}}$  vanish in the heavy quark limit, and the remaining form factors are equal to the leading-order Isgur-Wise function $\xi(w)$.

We take the parametrization of the leading-order Isgur-Wise function as~\cite{Caprini:1997mu}:
 \begin{equation}
 \frac{\xi(w)}{\xi(w_0)} \simeq  1 - 8 a^2 \bar\rho^2_{*}  z_{*} + \left[ V_{21} \bar\rho^2_{*} -V_{20} + \Delta(e_b,e_c,\alpha_s) \right] z^2_{*}
 \,,
 \end{equation}
where $V_{21}=57.0$, $V_{20}=7.5$, $z_*$ and  $a$ are defined as~\cite{Caprini:1997mu}:
 \begin{equation}
 z_* = \frac{\sqrt{w+1} -\sqrt{2} a}{\sqrt{w+1} + \sqrt{2} a}\,,~  a = \sqrt{\frac{1+r_D}{2\sqrt{r_D} } }\,, 
 \end{equation}
$r_D=m_D/m_B$, $w_0$ is determined from solving $z_*(w_0)=0$, $\bar \rho^2_{*}$ is the slope parameter of $\xi(w)/\xi(w_0)$, and $\Delta(e_b,e_c,\alpha_s)$ denotes the correction effects of ${\cal O}(e_{b,c})$ with $e_{b(c)}=\bar\Lambda/m_{b(c)}$ and ${\cal O}(\alpha_s)$.  For numerical estimates, we take the results from the fit scenario of ``$L_{w\geq 1}$+SR" shown in~\cite{Bernlochner:2017jka}. In addition to $\bar\rho^2_{*}=1.24\pm 0.08$, the values of sub-leading Isgur-Wise functions at $w=1$ are given in Table~\ref{tab:subIW}. Using these results, the correction of ${\cal O}(e_{b,c})$ and ${\cal O}(\alpha_s)$ can be obtained as:
  \begin{equation}
  \Delta(e_b,e_c,\alpha_s) \approx 0.582 \pm 0.298\,,
  \end{equation}
where we take the $1S$ scheme for $m_b$ and $m^{1S}_{b}=4.71 \pm 0.05$~GeV~\cite{Bernlochner:2017jka}. In addition, $\delta m_{bc}=m_b - m_c =3.40\pm 0.02$~GeV and $\bar \Lambda=0.45$~GeV are used.

 \begin{table}[htbp]
   \caption{ The results of sub-leading Isgur-Wise functions from the ``$L_{w\geq  1}$+SR" fit scenario.}
  \label{tab:subIW}
   \begin{tabular}{c|ccccc} \hline \hline
   FS &  $\hat\chi_{2}(1)$ & $\hat\chi'_2(1)$ & $\hat\chi'_3(1)$ & $\eta(1)$ & $\eta'(1)$   
   \\ \hline 
   $L_{w\geq 1}$ + SR & ~$ -0.06 \pm 0.02$~  &  ~$-0.00\pm 0.02$ & ~$0.05 \pm 0.02$~ & ~$0.30\pm 0.03$~ & ~$-0.05\pm 0.09$~  
   \\ \hline  \hline   
   \end{tabular}
\end{table}

Hence, the form factors up to ${\cal O}(e_{b,c})$ and ${\cal O}(\alpha_s)$ can be expressed by factoring out $\xi$, i.e., $ h_{i} = \hat h_{i}\, \xi$, where $\hat h_i$ for the $\bar B \to D$ transitions are given by~\cite{Bernlochner:2017jka}: 
 \begin{subequations}
 \begin{align}
 \hat h_+ & = 1 + \hat\alpha_s \left[ C_{V_1} + \frac{w + 1}{2} (C_{V_1} + C_{V_3} )\right] + ( e_c + e_b) \hat L_1
 \,,  \\
 \hat h_{-} & = \hat\alpha_s \frac{w + 1}{2} (C_{V_2} - C_{V_3} ) + ( e_c - e_b) \hat L_4
 \,, \\
 \hat h_S & = 1 + \hat\alpha_s C_S + ( e_c + e_b) \left[ \hat L_1 - \hat L_4 \frac{w-1}{w+1} \right]
 \,, \\
 \hat h_T &= 1+ \hat\alpha_s (C_{T_1} - C_{T_2} + C_{T_3}) + (e_c + e_b) (\hat L_1 - \hat L_4)\,,
 \end{align}
 \end{subequations}
and those for the $\bar B \to D^*$ transitions are given by:
  \begin{subequations}
  \begin{align}
  \hat h_V & = 1 + \alpha_s C_{V_1} + e_c (\hat L_2 -\hat L_5) + e_b (\hat L_1 -\hat L_4)
  \,, \\
   \hat h_{A_1} & = 1 + \hat \alpha_s C_{A_1} + e_c  \left(\hat L_2 - \hat L_5 \frac{w-1}{w+1} \right) + e_b \left(\hat L_1 - \hat L_4 \frac{w-1}{w+1} \right)
   \,, \\
  \hat h_{A_2} & = \hat \alpha_s C_{A_2} + e_c (\hat L_3 + \hat L_6)
  \,, \\
  \hat h_{A_3} &= 1+ \hat \alpha_s (C_{A_1}+C_{A_3}) + e_c (\hat L_2 - \hat L_3 + \hat L_6 - \hat L_5) + e_b (\hat L_1 - \hat L_4)
  \,, \\
  \hat h_P &= 1+ \hat \alpha_s C_P + e_c \left[ \hat L_2 + \hat L_3 (w-1) + \hat L_5 - \hat L_6 (w+1) \right] + e_b (\hat L_1 - \hat L_4)
  \,, \\
  \hat h_{T_1} &= 1 + \hat \alpha_s \left[ C_{T_1} + \frac{w-1}{2} (C_{T_2} - C_{T_3}) \right] + e_c \hat L_2 + e_b \hat L_1
  \,,\\
  \hat h_{T_2} &= \hat \alpha_s \frac{w+1}{2} (C_{T_2}+ C_{T_3}) + e_c \hat L_5 - e_b \hat L_4
  \,,\\
  \hat h_{T_3} &= \hat \alpha_s C_{T_2} + e_c (\hat L_6 - \hat L_3)
  \,.
  \end{align}  
  \end{subequations}
  The $w$-dependent functions $C_{\Gamma_i}$ can be found in Ref.~\cite{Neubert:1992qq}, and the sub-leading Isgur-Wise functions are~\cite{Falk:1992wt}:
   \begin{align}
   &\hat L_1 = -4(w-1) \hat\chi_2 + 12 \hat \chi_3\,, \ \hat L_2 = -4 \hat\chi_3\,, \ \hat L_3 = 4 \hat\chi_2
   \,,  \nonumber \\
   & \hat L_4 = 2 \eta -1\,, \ \hat L_5= -1\,, \  \hat L_6 = -2 \frac{1+\eta}{w+1}
   \,,
   \end{align}
   where the $w$-dependent functions $\hat\chi_i$ and $\eta$ can be approximated as:
    \begin{align}
    \hat\chi_2 (w) & \simeq \hat\chi_2(1) + \hat\chi'_2(1) (w-1)\,, \nonumber \\
    \hat\chi_3(w) & \simeq \hat\chi'_3 (1) (w-1) \,, \nonumber \\
    \eta(w) & \simeq \eta(1) + \eta'(1)(w-1)\,.
    \end{align}

The form factor parametrizations in Eqs.~(\ref{eq:ffP}) and (\ref{eq:ffV}), using which we formulate the BRs, and in Eqs.~(\ref{eq:HQETBP}) and (\ref{eq:HQETBV}), for which we evaluate within the framework of the HQET, are related as follows:
 \begin{align}
 F_S(q^2) & = \frac{\sqrt{m_B m_D}}{m_B+m_D}\left(w-1 \right) h_S(w)
 \,, \nonumber \\
 F_{+}(q^2)  & = \frac{1}{2\sqrt{m_B m_D}} \left[ (m_B + m_D) h_{+}(w) - (m_B-m_D) h_{-}(w) \right] 
 \,, \nonumber \\
 F_0(q^2) &= \frac{1}{2\sqrt{m_B m_D}} \left[ \frac{(m_B+m_D)^2 -q^2}{m_B+m_D} h_{+}(w) - \frac{(m_B-m_D)^2 -q^2}{m_B-m_D} h_{-}(w) \right] 
 \,, \nonumber \\
  F_T (q^2)& = \frac{m_B+m_D}{2\sqrt{m_B m_D}} h_T(w)\,.
 \end{align}
The relations for the form factors arising from the pseudoscalar, vector, and axial-vector currents for the $\bar B\to D^*$ transitions are found to be:
\begin{align}
F_P(q^2) & =  \frac{m_B + m_{D^*}}{2\sqrt{m_B m_{D^*}}} h_P (w) 
\,, \nonumber \\
V(q^2)  & = \frac{m_B + m_{D^*}}{2\sqrt{m_B m_{D^*}}}  h_V(w)
\,, \nonumber \\
 A_0(q^2) &= \frac{1}{2\sqrt{m_B m_{D^*}}} \left[ \frac{(m_B + m_{D^*})^2 -q^2}{2m_{D^*}} h_{A_1}(w)  \right. 
 \nonumber \\ 
 & \left. \qquad\qquad
 -  \frac{m^2_B - m^2_{D^*} +q^2}{2m_{B}} h_{A_2}(w) 
 - \frac{m^2_B - m^2_{D^*} -q^2}{2m_{D^*}}  h_{A_3}(w) \right] 
 \,, \nonumber \\
 A_1(q^2) & = \frac{(m_B + m_{D^*})^2 -q^2}{2\sqrt{m_B m_{D^*}}(m_B + m_{D^*})} h_{A_1}(w)
 \,, \nonumber \\
  A_2 (q^2)& = \frac{m_B+m_{D^*}}{2\sqrt{m_B m_D}} \left( h_{A_3}(w) + \frac{m_{D^*}}{m_B}h_{A_2}(w) \right)
  \,.
 \end{align}
Finally, the tensor form factors for the $\bar B\to D^*$ transitions are related by:
\begin{align}
T_1(q^2) & = \frac{1}{2\sqrt{m_B m_{D^*}}} \left[ (m_B+m_{D^*}) h_{T_1}(w) - (m_B - m_{D^*}) h_{T_2}(w) \right] 
\,, \nonumber \\
T_2(q^2) & = \frac{1}{2\sqrt{m_B m_{D^*}}}  \left[  \frac{(m_B + m_{D^*})^2 -q^2}{m_B+m_{D^*}} h_{T_1}(w) -  \frac{(m_B -m_{D^*})^2 -q^2}{m_B-m_{D^*}} h_{T_2}(w)\right] 
\,, \nonumber \\
T_3(q^2) & =  \frac{1}{2\sqrt{m_B m_{D^*}}}  \Bigg[ (m_B - m_{D^*}) h_{T_1}(w) - (m_B + m_{D^*}) h_{T_2}(w) 
\nonumber \\
 &  \qquad\qquad
 + \frac{m^2_B - m^2_{D^*}}{m_B} h_{T_3}(w) \Bigg]\,.
\end{align}

\section{Oblique parameters in the model}  \label{app:STU}

To calculate the $S$, $T$, and $U$ parameters in the model, we apply the results obtained in Ref.~\cite{Grimus:2008nb}. Using the mixing matrices of Goldstone and scalar bosons shown in Eqs.~(\ref{eq:Goldstone}) and (\ref{eq:H_h}), the resulting $T$ parameter subtracting the SM result is expressed as:
\begin{align}
\alpha_{\rm em}  T  
= &  
\frac{1}{16 \pi^2 v^2}  \left\{ c^2_{\beta-\alpha} F(m^2_{H^+}, m_h) +s^2_{\beta-\alpha} F(m^2_{H^+}, m_H)  \right. 
\nonumber \\
 &  \left.  + 3 c^2_{\beta-\alpha} \left[ F(m^2_Z,m^2_H)- F(m^2_W, m^2_H) -F(m^2_Z,m^2_h) + F(m^2_W, m^2_h)\right] \right\} 
 \,, \label{eq:T}
 \end{align}
where $\alpha_{\rm em}=e^2/4\pi$ is the fine structure constant of QED, and the function $F$ is defined as:
\begin{equation}
F(m^2_a, m^2_b)  = \frac{m^2_a + m^2_b}{2} - \frac{m^2_a m^2_b }{m^2_a - m^2_b} \ln \frac{m^2_a}{m^2_b}\,.  
\end{equation}
In the limit of $s_{\beta-\alpha}\to 1$, the $H^\pm$- and $H$-mediated loop effects are the most dominant.

The $S$ and $U$ parameters are respectively given by:
 \begin{align}
 S  = & 
 \frac{1}{24 \pi} \Bigg[ (c^2_W -s^2_W)^2 G(m^2_{H^+}, m^2_{H^+}, m^2_Z) + \ln\frac{m^2_{H^+} }{m^2_H} 
 \nonumber \\
 & 
 + c^2_{\beta-\alpha} \left( \hat G(m^2_{H},  m^2_Z)-\hat G( m^2_{h}, m^2_Z)\right) \Bigg]
 \,, \label{eq:S}
 \end{align}
and
 \begin{align}
 U = & \frac{1}{24\pi} \Big[ c^2_{\beta-\alpha} G(m^2_{H^+}, m^2_h, m^2_W) + s^2_{\beta-\alpha} G(m^2_{H^+}, m^2_H, m^2_W) - (2 s^2_W -1)^2 G(m^2_{H^+}, m^2_{H^+}, m^2_Z) 
 \nonumber \\
& + c^2_{\beta-\alpha} \left( \hat{G}(m^2_H,m^2_W)-\hat{G}(m^2_H,m^2_Z) - \hat{G}(m^2_h, m^2_W)+ \hat{G}(m^2_h, m^2_Z )\right) \Big]
\,,  \label{eq:U}
\end{align}
where the  functions of $G$ and $\tilde G$  are given by:
 \begin{align}
 G(m^2_a, m^2_b, m^2_c)  
 = & 
 - \frac{16}{3} + \frac{5(m^2_a +m^2_b)}{m^2_c} - \frac{2 (m^2_a -m^2_b)^2}{m^4_c} + \frac{r}{m^6_c} f(t,r) 
 \nonumber \\
 & + \frac{3}{m^2_c} \left(\frac{m^4_a+m^4_b}{m^2_a -m^2_b}  - \frac{m^4_a -m^4_b}{m^2_c} 
 + \frac{(m^2_a -m^2_b)^3}{3 m^4_c}\right) \ln \frac{m^2_a}{m^2_b}
 \,,  \\
 \tilde{G}(m^2_a, m^2_b, m^2_c) 
 =& -2 + \left( \frac{m^2_a - m^2_b}{m^2_c} - \frac{m^2_a + m^2_b}{m^2_a - m^2_b}\right) \ln\frac{m^2_a}{m^2_b}
 + \frac{f(t, r)}{m^2_c}
 \,. 
 \end{align}
and $\hat{G}$, $t$, $r$, and $f(t,r)$ are defined as:
 \begin{align}
\hat G(m^2_a, m^2_b)&=G(m^2_a, m^2_b, m^2_b) +12  \tilde G(m^2_a, m^2_b, m^2_b)\,, \nonumber \\
t & =m^2_a+m^2_b-m^2_c \,, \nonumber \\
r&=m^4_c-2 m^2_c (m^2_a + m^2_b) + (m^2_a-m^2_b)^2\,, \nonumber \\
f(t,r)&= 
\begin{cases}
\displaystyle
\sqrt{r} \ln\left| \frac{t-\sqrt{r}}{t+\sqrt{r}}\right|  & \text {for}~r > 0 
\,, \\
0  & \text{for}~r=0 
\,,  \\    
\displaystyle
2\sqrt{-r}\, \arctan\frac{\sqrt{-r}}{t} & \text{for}~ r < 0
\,.
\end{cases}
\end{align}


\begin{thebibliography}{99}



\bibitem{Freedman:1973yd}
D.~Z.~Freedman,
Phys. Rev. D \textbf{9}, 1389-1392 (1974).

\bibitem{COHERENT:2017ipa}
D.~Akimov \textit{et al.} [COHERENT],
Science \textbf{357}, no.6356, 1123-1126 (2017)
[arXiv:1708.01294 [nucl-ex]].

\bibitem{COHERENT:2021xmm}
D.~Akimov \textit{et al.} [COHERENT],
Phys. Rev. Lett. \textbf{129}, no.8, 081801 (2022)
[arXiv:2110.07730 [hep-ex]].

\bibitem{COHERENT:2020iec}
D.~Akimov \textit{et al.} [COHERENT],
Phys. Rev. Lett. \textbf{126}, no.1, 012002 (2021)
[arXiv:2003.10630 [nucl-ex]].

\bibitem{Coloma:2017ncl}
P.~Coloma, M.~C.~Gonzalez-Garcia, M.~Maltoni and T.~Schwetz,
Phys. Rev. D \textbf{96}, no.11, 115007 (2017)
[arXiv:1708.02899 [hep-ph]].

\bibitem{Liao:2017uzy}
J.~Liao and D.~Marfatia,
Phys. Lett. B \textbf{775}, 54-57 (2017)
[arXiv:1708.04255 [hep-ph]].


\bibitem{Giunti:2019xpr}
C.~Giunti,
Phys. Rev. D \textbf{101}, no.3, 035039 (2020)
[arXiv:1909.00466 [hep-ph]].

\bibitem{Coloma:2019mbs}
P.~Coloma, I.~Esteban, M.~C.~Gonzalez-Garcia and M.~Maltoni,
JHEP \textbf{02}, 023 (2020)
[arXiv:1911.09109 [hep-ph]].

\bibitem{Denton:2020hop}
P.~B.~Denton and J.~Gehrlein,
JHEP \textbf{04}, 266 (2021)
[arXiv:2008.06062 [hep-ph]].

\bibitem{Khan:2021wzy}
A.~N.~Khan, D.~W.~McKay and W.~Rodejohann,
Phys. Rev. D \textbf{104}, no.1, 015019 (2021)
[arXiv:2104.00425 [hep-ph]].


\bibitem{Hoferichter:2020osn}
M.~Hoferichter, J.~Men\'endez and A.~Schwenk,
Phys. Rev. D \textbf{102}, no.7, 074018 (2020)
[arXiv:2007.08529 [hep-ph]].

\bibitem{Liao:2022hno}
J.~Liao, H.~Liu and D.~Marfatia,
Phys. Rev. D \textbf{106}, no.3, L031702 (2022)
[arXiv:2202.10622 [hep-ph]].

\bibitem{Abdullah:2022zue}
M.~Abdullah, H.~Abele, D.~Akimov, G.~Angloher, D.~Aristizabal Sierra, C.~Augier, A.~B.~Balantekin, L.~Balogh, P.~S.~Barbeau and L.~Baudis, \textit{et al.}
[arXiv:2203.07361 [hep-ph]].


\bibitem{Calabrese:2022mnp}
R.~Calabrese, J.~Gunn, G.~Miele, S.~Morisi, S.~Roy and P.~Santorelli,
Phys. Rev. D \textbf{107}, no.5, 055039 (2023)
[arXiv:2212.11210 [hep-ph]].



\bibitem{Papoulias:2017qdn}
D.~K.~Papoulias and T.~S.~Kosmas,
Phys. Rev. D \textbf{97}, no.3, 033003 (2018)
[arXiv:1711.09773 [hep-ph]].

\bibitem{Abdullah:2018ykz}
M.~Abdullah, J.~B.~Dent, B.~Dutta, G.~L.~Kane, S.~Liao and L.~E.~Strigari,
Phys. Rev. D \textbf{98}, no.1, 015005 (2018)
[arXiv:1803.01224 [hep-ph]].

\bibitem{Denton:2018xmq}
P.~B.~Denton, Y.~Farzan and I.~M.~Shoemaker,
JHEP \textbf{07}, 037 (2018)
[arXiv:1804.03660 [hep-ph]].

\bibitem{CONNIE:2019xid}
A.~Aguilar-Arevalo \textit{et al.} [CONNIE],
JHEP \textbf{04}, 054 (2020)
[arXiv:1910.04951 [hep-ex]].

\bibitem{Miranda:2020tif}
O.~G.~Miranda, D.~K.~Papoulias, G.~Sanchez Garcia, O.~Sanders, M.~T\'ortola and J.~W.~F.~Valle,
JHEP \textbf{05}, 130 (2020)
[erratum: JHEP \textbf{01}, 067 (2021)]
[arXiv:2003.12050 [hep-ph]].

\bibitem{Cadeddu:2020nbr}
M.~Cadeddu, N.~Cargioli, F.~Dordei, C.~Giunti, Y.~F.~Li, E.~Picciau and Y.~Y.~Zhang,
JHEP \textbf{01}, 116 (2021)
[arXiv:2008.05022 [hep-ph]].

\bibitem{Coloma:2020gfv}
P.~Coloma, M.~C.~Gonzalez-Garcia and M.~Maltoni,
JHEP \textbf{01}, 114 (2021)
[erratum: JHEP \textbf{11}, 115 (2022)]
[arXiv:2009.14220 [hep-ph]].

\bibitem{delaVega:2021wpx}
L.~M.~G.~de la Vega, L.~J.~Flores, N.~Nath and E.~Peinado,
JHEP \textbf{09}, 146 (2021)
[arXiv:2107.04037 [hep-ph]].

\bibitem{CONUS:2021dwh}
H.~Bonet \textit{et al.} [CONUS],
JHEP \textbf{05}, 085 (2022)
[arXiv:2110.02174 [hep-ph]].

\bibitem{Coloma:2022avw}
P.~Coloma, I.~Esteban, M.~C.~Gonzalez-Garcia, L.~Larizgoitia, F.~Monrabal and S.~Palomares-Ruiz,
JHEP \textbf{05}, 037 (2022)
[arXiv:2202.10829 [hep-ph]].


\bibitem{AtzoriCorona:2022moj}
M.~Atzori Corona, M.~Cadeddu, N.~Cargioli, F.~Dordei, C.~Giunti, Y.~F.~Li, E.~Picciau, C.~A.~Ternes and Y.~Y.~Zhang,
JHEP \textbf{05}, 109 (2022)
[arXiv:2202.11002 [hep-ph]].





\bibitem{He:1990pn}
X.~G.~He, G.~C.~Joshi, H.~Lew and R.~R.~Volkas,
Phys. Rev. D \textbf{43}, 22-24 (1991)

\bibitem{He:1991qd}
X.~G.~He, G.~C.~Joshi, H.~Lew and R.~R.~Volkas,
Phys. Rev. D \textbf{44}, 2118-2132 (1991).

\bibitem{Chen:2017cic}
C.~H.~Chen and T.~Nomura,
Phys. Rev. D \textbf{96}, no.9, 095023 (2017)
[arXiv:1704.04407 [hep-ph]].



\bibitem{Aoyama:2020ynm}
T.~Aoyama, N.~Asmussen, M.~Benayoun, J.~Bijnens, T.~Blum, M.~Bruno, I.~Caprini, C.~M.~Carloni Calame, M.~C\`e and G.~Colangelo, \textit{et al.}
Phys. Rept. \textbf{887}, 1-166 (2020)
[arXiv:2006.04822 [hep-ph]].

\bibitem{Heeck:2011wj}
J.~Heeck and W.~Rodejohann,
Phys. Rev. D \textbf{84}, 075007 (2011)
[arXiv:1107.5238 [hep-ph]].











\bibitem{CDF:2022hxs}
T.~Aaltonen \textit{et al.} [CDF],
Science \textbf{376}, no.6589, 170-176 (2022).


\bibitem{CDF:2013dpa}
T.~A.~Aaltonen \textit{et al.} [CDF and D0],
Phys. Rev. D \textbf{88}, no.5, 052018 (2013)
[arXiv:1307.7627 [hep-ex]].

\bibitem{ATLAS:2023fsi}
 [ATLAS],
ATLAS-CONF-2023-004.

\bibitem{Heinemeyer:2013dia}
S.~Heinemeyer, W.~Hollik, G.~Weiglein and L.~Zeune,
JHEP \textbf{12}, 084 (2013)
[arXiv:1311.1663 [hep-ph]].




\bibitem{Fan:2022dck}
Y.~Z.~Fan, T.~P.~Tang, Y.~L.~S.~Tsai and L.~Wu,
Phys. Rev. Lett. \textbf{129}, no.9, 091802 (2022)
[arXiv:2204.03693 [hep-ph]].


\bibitem{Strumia:2022qkt}
A.~Strumia,
JHEP \textbf{08}, 248 (2022)
[arXiv:2204.04191 [hep-ph]].



\bibitem{Bagnaschi:2022whn}
E.~Bagnaschi, J.~Ellis, M.~Madigan, K.~Mimasu, V.~Sanz and T.~You,
JHEP \textbf{08}, 308 (2022)
[arXiv:2204.05260 [hep-ph]].

\bibitem{Bahl:2022xzi}
H.~Bahl, J.~Braathen and G.~Weiglein,
Phys. Lett. B \textbf{833}, 137295 (2022)
[arXiv:2204.05269 [hep-ph]].

\bibitem{Cheng:2022jyi}
Y.~Cheng, X.~G.~He, Z.~L.~Huang and M.~W.~Li,
Phys. Lett. B \textbf{831}, 137218 (2022)
[arXiv:2204.05031 [hep-ph]].

\bibitem{Asadi:2022xiy}
P.~Asadi, C.~Cesarotti, K.~Fraser, S.~Homiller and A.~Parikh,
[arXiv:2204.05283 [hep-ph]].

\bibitem{Heckman:2022the}
J.~J.~Heckman,
Phys. Lett. B \textbf{833}, 137387 (2022)
[arXiv:2204.05302 [hep-ph]].

\bibitem{Crivellin:2022fdf}
A.~Crivellin, M.~Kirk, T.~Kitahara and F.~Mescia,
Phys. Rev. D \textbf{106}, no.3, L031704 (2022)
[arXiv:2204.05962 [hep-ph]].

\bibitem{FileviezPerez:2022lxp}
P.~Fileviez Perez, H.~H.~Patel and A.~D.~Plascencia,
Phys. Lett. B \textbf{833}, 137371 (2022)
[arXiv:2204.07144 [hep-ph]].

\bibitem{Kanemura:2022ahw}
S.~Kanemura and K.~Yagyu,
Phys. Lett. B \textbf{831}, 137217 (2022)
[arXiv:2204.07511 [hep-ph]].

\bibitem{Kim:2022hvh}
J.~Kim, S.~Lee, P.~Sanyal and J.~Song,
Phys. Rev. D \textbf{106}, no.3, 035002 (2022)
[arXiv:2205.01701 [hep-ph]].

\bibitem{Li:2022gwc}
X.~Q.~Li, Z.~J.~Xie, Y.~D.~Yang and X.~B.~Yuan,
[arXiv:2205.02205 [hep-ph]].

\bibitem{Dcruz:2022dao}
R.~Dcruz and A.~Thapa,
[arXiv:2205.02217 [hep-ph]].

\bibitem{Chowdhury:2022dps}
T.~A.~Chowdhury and S.~Saad,
Phys. Rev. D \textbf{106}, no.5, 055017 (2022)
[arXiv:2205.03917 [hep-ph]].

\bibitem{Gao:2022wxk}
J.~Gao, D.~Liu and K.~Xie,
[arXiv:2205.03942 [hep-ph]].

\bibitem{Han:2022juu}
X.~F.~Han, F.~Wang, L.~Wang, J.~M.~Yang and Y.~Zhang,
Chin. Phys. C \textbf{46}, no.10, 103105 (2022)
[arXiv:2204.06505 [hep-ph]].

\bibitem{Cheng:2022hbo}
Y.~Cheng, X.~G.~He, F.~Huang, J.~Sun and Z.~P.~Xing,
[arXiv:2208.06760 [hep-ph]].


\bibitem{Bandyopadhyay:2022bgx}
T.~Bandyopadhyay, A.~Budhraja, S.~Mukherjee and T.~S.~Roy,
[arXiv:2212.02534 [hep-ph]].

\bibitem{Chen:2023eof}
C.~H.~Chen, C.~W.~Chiang and C.~W.~Su,
[arXiv:2301.07070 [hep-ph]].

\bibitem{Heeck:2014qea}
J.~Heeck, M.~Holthausen, W.~Rodejohann and Y.~Shimizu,
Nucl. Phys. B \textbf{896}, 281-310 (2015)
[arXiv:1412.3671 [hep-ph]].

\bibitem{Heeck:2016xkh}
J.~Heeck,
Phys. Lett. B \textbf{758}, 101-105 (2016)
[arXiv:1602.03810 [hep-ph]].






\bibitem{MILC:2015uhg}
J.~A.~Bailey \textit{et al.} [MILC],
Phys. Rev. D \textbf{92}, no.3, 034506 (2015)
[arXiv:1503.07237 [hep-lat]].

\bibitem{Na:2015kha}
H.~Na \textit{et al.} [HPQCD],
Phys. Rev. D \textbf{92}, no.5, 054510 (2015)
[erratum: Phys. Rev. D \textbf{93}, no.11, 119906 (2016)]
[arXiv:1505.03925 [hep-lat]].


  
\bibitem{Bigi:2016mdz}
D.~Bigi and P.~Gambino,
Phys. Rev. D \textbf{94}, no.9, 094008 (2016)
[arXiv:1606.08030 [hep-ph]].

\bibitem{Bernlochner:2017jka}
F.~U.~Bernlochner, Z.~Ligeti, M.~Papucci and D.~J.~Robinson,
Phys. Rev. D \textbf{95}, no.11, 115008 (2017)
[erratum: Phys. Rev. D \textbf{97}, no.5, 059902 (2018)]
[arXiv:1703.05330 [hep-ph]].



\bibitem{Jaiswal:2017rve}
S.~Jaiswal, S.~Nandi and S.~K.~Patra,
JHEP \textbf{12}, 060 (2017)
[arXiv:1707.09977 [hep-ph]].

\bibitem{BaBar:2019vpl}
J.~P.~Lees \textit{et al.} [BaBar],
Phys. Rev. Lett. \textbf{123}, no.9, 091801 (2019)
doi:10.1103/PhysRevLett.123.091801
[arXiv:1903.10002 [hep-ex]].

\bibitem{Bordone:2019vic}
M.~Bordone, M.~Jung and D.~van Dyk,
Eur. Phys. J. C \textbf{80}, no.2, 74 (2020)
[arXiv:1908.09398 [hep-ph]].

\bibitem{Martinelli:2021onb}
G.~Martinelli, S.~Simula and L.~Vittorio,
Phys. Rev. D \textbf{105}, no.3, 034503 (2022)
[arXiv:2105.08674 [hep-ph]].



\bibitem{HeavyFlavorAveragingGroup:2022wzx}
Y.~S.~Amhis \textit{et al.} [Heavy Flavor Averaging Group and HFLAV],
Phys. Rev. D \textbf{107}, no.5, 052008 (2023)
[arXiv:2206.07501 [hep-ex]].

\bibitem{LHCb:2023zxo}
R.~Aaij \textit{et al.} [LHCb],
[arXiv:2302.02886 [hep-ex]].


\bibitem{LHCb:2023cjr}
R.~Aaij \textit{et al.} [LHCb],
[arXiv:2305.01463 [hep-ex]].

\bibitem{LHCb:2017vlu}
R.~Aaij \textit{et al.} [LHCb],
Phys. Rev. Lett. \textbf{120}, no.12, 121801 (2018)
[arXiv:1711.05623 [hep-ex]].

\bibitem{LHCb:2022piu}
R.~Aaij \textit{et al.} [LHCb],
Phys. Rev. Lett. \textbf{128}, no.19, 191803 (2022)
[arXiv:2201.03497 [hep-ex]].

\bibitem{Fedele:2022iib}
M.~Fedele, M.~Blanke, A.~Crivellin, S.~Iguro, T.~Kitahara, U.~Nierste and R.~Watanabe,
Phys. Rev. D \textbf{107}, no.5, 055005 (2023)
[arXiv:2211.14172 [hep-ph]].


\bibitem{Crivellin:2012ye}
A.~Crivellin, C.~Greub and A.~Kokulu,
Phys. Rev. D \textbf{86}, 054014 (2012)
[arXiv:1206.2634 [hep-ph]].

\bibitem{Crivellin:2013wna}
A.~Crivellin, A.~Kokulu and C.~Greub,
Phys. Rev. D \textbf{87}, no.9, 094031 (2013)
[arXiv:1303.5877 [hep-ph]].

\bibitem{Crivellin:2015hha}
A.~Crivellin, J.~Heeck and P.~Stoffer,
Phys. Rev. Lett. \textbf{116}, no.8, 081801 (2016)
[arXiv:1507.07567 [hep-ph]].

\bibitem{Chen:2017eby}
C.~H.~Chen and T.~Nomura,
Eur. Phys. J. C \textbf{77}, no.9, 631 (2017)
[arXiv:1703.03646 [hep-ph]].



\bibitem{Akeroyd:2017mhr}
A.~G.~Akeroyd and C.~H.~Chen,
Phys. Rev. D \textbf{96}, no.7, 075011 (2017)
[arXiv:1708.04072 [hep-ph]].

\bibitem{Chen:2018hqy}
C.~H.~Chen and T.~Nomura,
Phys. Rev. D \textbf{98}, no.9, 095007 (2018)
[arXiv:1803.00171 [hep-ph]].


\bibitem{Becirevic:2016yqi}
D.~Be\v{c}irevi\'c, S.~Fajfer, N.~Ko\v{s}nik and O.~Sumensari,
Phys. Rev. D \textbf{94}, no.11, 115021 (2016)
[arXiv:1608.08501 [hep-ph]].

\bibitem{Bhattacharya:2016mcc}
B.~Bhattacharya, A.~Datta, J.~P.~Gu\'evin, D.~London and R.~Watanabe,
JHEP \textbf{01}, 015 (2017)
[arXiv:1609.09078 [hep-ph]].

\bibitem{Crivellin:2016ejn}
A.~Crivellin, J.~Fuentes-Martin, A.~Greljo and G.~Isidori,
Phys. Lett. B \textbf{766}, 77-85 (2017)
[arXiv:1611.02703 [hep-ph]].

\bibitem{Crivellin:2017zlb}
A.~Crivellin, D.~M\"uller and T.~Ota,
JHEP \textbf{09}, 040 (2017)
[arXiv:1703.09226 [hep-ph]].

\bibitem{Chen:2017hir}
C.~H.~Chen, T.~Nomura and H.~Okada,
Phys. Lett. B \textbf{774}, 456-464 (2017)
[arXiv:1703.03251 [hep-ph]].

\bibitem{Chen:2017usq}
C.~H.~Chen and T.~Nomura,
Phys. Lett. B \textbf{777}, 420-427 (2018)
[arXiv:1707.03249 [hep-ph]].

\bibitem{Crivellin:2019dwb}
A.~Crivellin, D.~M\"uller and F.~Saturnino,
JHEP \textbf{06}, 020 (2020)
[arXiv:1912.04224 [hep-ph]].


\bibitem{Heeck:2022znj}
J.~Heeck and A.~Thapa,
Eur. Phys. J. C \textbf{82}, no.5, 480 (2022)
[arXiv:2202.08854 [hep-ph]].


\bibitem{Belle-II:2023esi}
I.~Adachi \textit{et al.} [Belle-II],
[arXiv:2311.14647 [hep-ex]].

\bibitem{Chen:2023wpb}
C.~H.~Chen and C.~W.~Chiang,
[arXiv:2309.12904 [hep-ph]].


\bibitem{Klimenko:1984qx}
K.~G.~Klimenko,
Theor. Math. Phys. \textbf{62}, 58-65 (1985).
  
\bibitem{Kannike:2012pe}
K.~Kannike,
Eur. Phys. J. C \textbf{72}, 2093 (2012)
[arXiv:1205.3781 [hep-ph]].

\bibitem{Baur:2002rb}
U.~Baur, T.~Plehn and D.~L.~Rainwater,
Phys. Rev. Lett. \textbf{89}, 151801 (2002)
[arXiv:hep-ph/0206024 [hep-ph]].

\bibitem{Branco:2011iw}
G.~C.~Branco, P.~M.~Ferreira, L.~Lavoura, M.~N.~Rebelo, M.~Sher and J.~P.~Silva,
Phys. Rept. \textbf{516}, 1-102 (2012)
[arXiv:1106.0034 [hep-ph]].


\bibitem{Klein:1999qj}
S.~Klein and J.~Nystrand,
Phys. Rev. C \textbf{60}, 014903 (1999)
[arXiv:hep-ph/9902259 [hep-ph]].

\bibitem{Barbeau:2021exu}
P.~S.~Barbeau, Y.~Efremenko and K.~Scholberg,
[arXiv:2111.07033 [hep-ex]].


\bibitem{Bertuzzo:2021opb}
E.~Bertuzzo, G.~Grilli di Cortona and L.~M.~D.~Ramos,
JHEP \textbf{06}, 075 (2022)
[arXiv:2112.04020 [hep-ph]].


\bibitem{AristizabalSierra:2019zmy}
D.~Aristizabal Sierra, J.~Liao and D.~Marfatia,
JHEP \textbf{06}, 141 (2019)
[arXiv:1902.07398 [hep-ph]].


\bibitem{Scholberg:2018vwg}
K.~Scholberg [COHERENT],
PoS \textbf{NuFact2017}, 020 (2018)
[arXiv:1801.05546 [hep-ex]]. 



\bibitem{Dorsner:2013tla}
I.~Dor\v{s}ner, S.~Fajfer, N.~Ko\v{s}nik and I.~Ni\v{s}and\v{z}i\'c,
JHEP \textbf{11}, 084 (2013)
[arXiv:1306.6493 [hep-ph]].

 \bibitem{Peskin:1990zt}
M.~E.~Peskin and T.~Takeuchi,
Phys. Rev. Lett. \textbf{65} (1990), 964-967.


\bibitem{Peskin:1991sw}
M.~E.~Peskin and T.~Takeuchi,
Phys. Rev. D \textbf{46} (1992), 381-409.

\bibitem{Grimus:2008nb}
W.~Grimus, L.~Lavoura, O.~M.~Ogreid and P.~Osland,
Nucl. Phys. B \textbf{801}, 81-96 (2008)
[arXiv:0802.4353 [hep-ph]].


\bibitem{Maksymyk:1993zm}
I.~Maksymyk, C.~P.~Burgess and D.~London,
Phys. Rev. D \textbf{50} (1994), 529-535
[arXiv:hep-ph/9306267 [hep-ph]].

\bibitem{Burgess:1993vc}
C.~P.~Burgess, S.~Godfrey, H.~Konig, D.~London and I.~Maksymyk,
Phys. Rev. D \textbf{49}, 6115-6147 (1994)
[arXiv:hep-ph/9312291 [hep-ph]].

\bibitem{Altmannshofer:2014pba}
W.~Altmannshofer, S.~Gori, M.~Pospelov and I.~Yavin,
Phys. Rev. Lett. \textbf{113}, 091801 (2014)
[arXiv:1406.2332 [hep-ph]].

\bibitem{CCFR:1991lpl}
S.~R.~Mishra \textit{et al.} [CCFR],
Phys. Rev. Lett. \textbf{66}, 3117-3120 (1991)

\bibitem{BaBar:2016sci}
J.~P.~Lees \textit{et al.} [BaBar],
Phys. Rev. D \textbf{94}, no.1, 011102 (2016)
[arXiv:1606.03501 [hep-ex]].

\bibitem{LHCHiggsCrossSectionWorkingGroup:2016ypw}
D.~de Florian \textit{et al.} [LHC Higgs Cross Section Working Group],
[arXiv:1610.07922 [hep-ph]].



\bibitem{PDG2022}
R.~L.~Workman \textit{et al.} [Particle Data Group],
PTEP \textbf{2022} (2022) no.8, 083C01

\bibitem{Holdom:1985ag}
B.~Holdom,
Phys. Lett. B \textbf{166}, 196-198 (1986)


\bibitem{Belle-II:2022heu}
I.~Adachi \textit{et al.} [Belle-II],
Phys. Rev. Lett. \textbf{130}, no.18, 181803 (2023)
[arXiv:2212.03634 [hep-ex]].



\bibitem{CMS:2020wzx}
A.~M.~Sirunyan \textit{et al.} [CMS],
Phys. Lett. B \textbf{819}, 136446 (2021)
[arXiv:2012.04178 [hep-ex]].

\bibitem{ATLAS:2021oiz}
G.~Aad \textit{et al.} [ATLAS],
JHEP \textbf{06}, 179 (2021)
[arXiv:2101.11582 [hep-ex]].




\bibitem{deBlas:2022hdk}
J.~de Blas, M.~Pierini, L.~Reina and L.~Silvestrini,
Phys. Rev. Lett. \textbf{129}, no.27, 271801 (2022)
[arXiv:2204.04204 [hep-ph]].


\bibitem{Banerjee:2022vdd}
S.~Banerjee,
Universe \textbf{8}, no.9, 480 (2022)
[arXiv:2209.11639 [hep-ex]].

\bibitem{ATLAS:2022yvh}
G.~Aad \textit{et al.} [ATLAS],
JHEP \textbf{08}, 104 (2022)
[arXiv:2202.07953 [hep-ex]].


\bibitem{Caprini:1997mu} 
  I.~Caprini, L.~Lellouch and M.~Neubert,
  Nucl.\ Phys.\ B {\bf 530}, 153 (1998)
  [hep-ph/9712417].

\bibitem{Neubert:1992qq} 
  M.~Neubert,
  Nucl.\ Phys.\ B {\bf 371}, 149 (1992).

\bibitem{Falk:1992wt} 
  A.~F.~Falk and M.~Neubert,
  Phys.\ Rev.\ D {\bf 47}, 2965 (1993)
  [hep-ph/9209268].
  

\end{thebibliography}
\end{document}